\documentclass[letterpaper,10pt,titlepage,linenumber]{article}
\usepackage[english]{babel}
\usepackage[utf8x]{inputenc}
\usepackage[T1]{fontenc}
\usepackage{amsmath}
\usepackage{amsfonts}
\usepackage{amssymb}
\usepackage{graphicx}
\usepackage[letterpaper,left=3cm,right=3cm,top=3cm,bottom=3cm]{geometry}
\usepackage{epstopdf}
\usepackage{lineno}
\usepackage{hyphenat}

\title{\textbf{Muon Tomography sites for Colombian volcanoes}}
\author{
\textbf{A. Vesga-Ram\'{\i}rez} \\
\textit{Centro Internacional para Estudios de la Tierra,} \\
\textit{Comisi\'on Nacional de Energ\'{\i}a At\'omica Buenos Aires-Argentina.} \\
\textbf{D. Sierra-Porta\thanks{\small Corresponding author} } \\ 
\textit{Escuela de F\'{\i}sica, Universidad Industrial de Santander, Bucaramanga-Colombia and }\\
\textit{Centro de Modelado Cient\'{\i}fico, Universidad del Zulia,  Maracaibo-Venezuela,} \\
\textbf{J. Pe\~na-Rodr\'{\i}guez, J.D. Sanabria-G\'omez, M. Valencia-Otero} \\
\textit{Escuela de F\'{\i}sica, Universidad Industrial de Santander, Bucaramanga-Colombia.} \\
\textbf{C. Sarmiento-Cano}\\ 
\textit{Instituto de Tecnolog\'{\i}as en Detecci\'on y Astropart\'{\i}culas, 1650, Buenos Aires-Argentina.} , \\ 
\textbf{M. Su\'arez-Dur\'an} \\
\textit{Departamento de F\'{\i}sica y Geolog\'{\i}a, Universidad de Pamplona, Pamplona-Colombia} \\
\textbf{H. Asorey} \\ 
\textit{Laboratorio Detecci\'on de Part\'{\i}culas y Radiaci\'on, Instituto Balseiro} \\ 
\textit{Centro At\'omico Bariloche, Comisi\'on Nacional de Energ\'{\i}a At\'omica, Bariloche-Argentina;} \\
\textit{Universidad Nacional de R\'{\i}o Negro, 8400, Bariloche-Argentina and} \\
\textit{Instituto de Tecnolog\'{\i}as en Detecci\'on y Astropart\'{\i}culas, 1650, Buenos Aires-Argentina.} \\
\textbf{L. A. N\'u\~nez } \\ 
\textit{Escuela de F\'{\i}sica, Universidad Industrial de Santander, Bucaramanga-Colombia and }\\
\textit{Departamento de F\'{\i}sica, Universidad de Los Andes,  M\'erida-Venezuela.}
}

\begin{document}
\maketitle
\begin{abstract}
By using a very detailed simulation scheme, we have calculated the cosmic ray background flux at 13 active Colombian volcanoes and developed a methodology to identify the most convenient places for a muon telescope to study their inner structure. 

Our simulation scheme considers three critical factors with different spatial and time scales: the geomagnetic effects, the development of extensive air showers in the atmosphere, and the detector response at ground level. The muon energy dissipation along the path crossing the geological structure is modeled considering the losses due to ionization, and also contributions from radiative Brem\ss trahlung, nuclear interactions, and pair production.

By considering each particular volcano topography and assuming reasonable statistics for different instrument acceptances, we obtained the muon flux crossing each structure and estimated the exposure time for our hybrid muon telescope at several points around each geological edifice.  After a detailed study from the topography, we have identified the best volcano to be studied, spotted the finest points to place a muon telescope and estimated its time exposures for a significant statistics of muon flux.  

We have devised a mix of technical and logistic criteria --the ``rule of thumb'' criteria-- and found that only Cerro Mach\'{\i}n, located at the Cordillera Central (4$^{\circ}$29'N 75$^{\circ}$22'W), can be feasibly studied today through muography. Cerro Negro and Chiles could be good candidates shortly.  
\end{abstract}

\section{Introduction}
Muography is a non-invasive astroparticle technique, introduced several decades ago for imaging anthropic and geologic structures, benefiting from the high penetration atmospheric muons produced in cosmic rays showers. This method offers a spatial resolution in the order of tens of meters, up to a kilometer of penetration, and measures the out-coming muon flux for different directions by means of a hodoscope. The flux variance between trajectories allows us to extract information about the inner density distribution of the scanned object.

In the beginning, the study of cosmic rays --and particularly their detection after crossing geological structures-- were motivated by the need to understand the background noise in particle detectors inside anthropic/natural structures \cite{George1955,ZichichiEtal2000}. Luis \'Alvarez \cite{AlvarezEtal1970} was the first to apply muon radiography, with no results, to the pyramid of Cheops; but this initiative led to the successful \textit{ScanPyramids project}, which recently discovered cavities in the Pyramid of Khufu (Cheops)\cite{MorishimaEtal2017}. 

The interest in muon radiography for Earth sciences studies arose after the discovery of the significant penetrating power of some high energy secondary particles produced by the interaction of cosmic rays with the atmosphere (hereafter, primaries). These particles can cross hundreds of meters of rock with an attenuation related to the amount of matter traversed along its trajectory\,\cite{Nagamine2003}. This technique uses the same basic principles as a standard medical radiography: it measures, with a sensitive device, the attenuation of cosmic muons when crossing geological structures. Although there are limitations with muography in detecting deep structures beneath the volcano (such as a magma chamber), it is particularly useful when it is applied to shallow volcano phenomena such as the conduit dynamics. Thus, volcano muography constitutes an unique method to obtain direct information on the density distribution inside geological objects with a better spatial resolution than other geophysical techniques (see \cite{ThompsonEtal2019,Kaiser2019,TanakaOlah2019, CarboneEtal2014,JourdeEtal2013, Tanaka2013, Tanaka2016, CarloganuEtal2013, PortalEtal2013, MarteauEtal2012, Okubo2012, LesparreEtal2010, TanakaEtal2007},  and references therein). 

Colombia, located in the Pacific Belt, has more than a dozen active volcanoes (see figure \ref{ColombiaVolcanoesMap})  clustered in three main groups along the Cordillera Central, the highest of the three branches of the Colombian Andes.  Most of these volcanoes represent a significant risk to the nearby population in towns and/or cities\cite{Cortes2016,Agudelo2016,Munoz2017} and have caused major disasters. The most recent, the Armero tragedy occurred in November 13, 1985, when pyroclastics of the Nevado del Ruiz  fused about 10\% of the mountain glacier, sending  lahars with the terrible, devastating result of 20,000 casualties \cite{PiersonEtal1990}.

Therefore, to determine and to model the inner volcano structure is crucial in evaluating its potential risks. Muon tomography is a powerful technique, which measures the cosmic muon flux attenuation by rock volumes of different densities, allowing the projection of images of volcanic conduits at the top of the volcanic edifice. It constitutes an attractive way to infer density distributions inside different geological structures, which is critical in the study of possible eruption dynamics associated with specific eruptive styles. Nowadays, astroparticle research groups in Colombia have started to explore this technique to estimate the density distribution within geological edifices by recording the variation of the atmospheric muon flux crossing the structure\cite{TapiaEtal2016,SierraPortaEtal2018, PenarodriguezEtal2018B, PenarodriguezEtal2019, GuerreroEtal2019, TorresEtal2019, ParraAvila2019}. 

The objective of the present work is twofold: first, to identify possible muon telescope observation sites to study Colombian active volcanoes by estimating the muon flux emerging from the geological edifices and second, to determine the time exposure for our hybrid muon telescope at those selected sites. Section \ref{ColombiaVolcanoes} describes four of the main active Colombian volcanoes. In Section \ref{SimulationMuonFlux}, we discuss the detailed simulation scheme used to estimate the incoming atmospheric muon flux and, we also investigate the energy loss of muons crossing the volcano edifice. For completeness, our hybrid muon telescope is briefly described in section \ref{subsec:BriefMuonTelescope}. In  \ref{Results}, we analyze muon propagation across the Mach\'{\i}n volcano and calculate the exposure times for four observation points. We also estimate the outgoing muon flux from the geological structures that could be detected by our telescope. Besides, this Section discusses a ray-tracing analysis for muon propagation at three other active volcanoes: Chiles, Cerro Negro, and Galeras. From the previous results we devise in Section \ref{sec:Criteria} ``rule of thumb'' criteria that should be fulfilled by the potential sites (listed below) and apply them to 13 active Colombian volcanoes.  Finally, in Section \ref{sec:disc} conclusions are presented along with possible future works.

\section{Volcanoes in Colombia}
\label{ColombiaVolcanoes}
\begin{figure}[!ht]
\begin{center}
\includegraphics[width=0.75\textwidth]{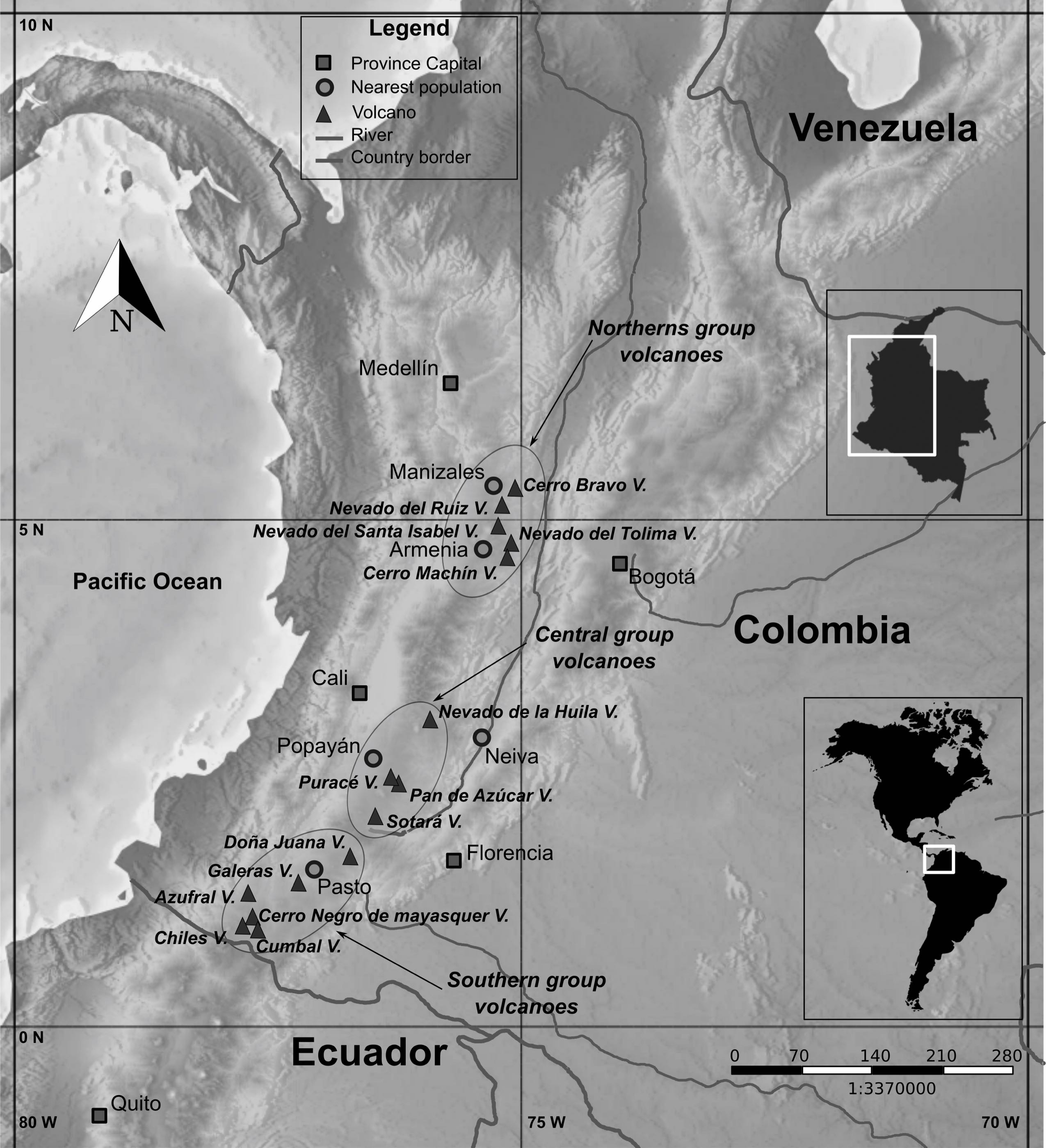}
\caption{Artistic representation of 13 Colombian active volcanoes --Azufral, Cerro Negro, Chiles, Cumbal, Do\~na Juana, Galeras, Mach\'{\i}n,  Nevado del Huila,  Nevado del Ru\'{\i}z,  Nevado Santa Isabel,  Nevado del Tolima,  Purac\'e, and Sotar\'a-- are displayed in three disperse clusters through the Cordillera Central. Because of their social significance and eruptive history, we shall briefly focus on four of them: Galeras, Nevado del Ruiz, Cerro Mach\'{\i}n and Cerro Negro-Chiles. Galeras, Cerro Negro-Chiles volcanoes are found in the southern cluster, while Nevado del Ruiz and Cerro Mach\'{\i}n are located within the most northern one.}
\label{ColombiaVolcanoesMap}
\end{center}
\end{figure}

In this work, we have considered 13 Colombian active volcanoes: Azufral, Cerro Negro, Chiles, Cumbal, Do\~na Juana, Galeras, Mach\'{\i}n,  Nevado del Huila,  Nevado del Ru\'{\i}z,  Nevado Santa Isabel,  Nevado del Tolima,  Purac\'e, and Sotar\'a. Figure \ref{ColombiaVolcanoesMap} displays an artistic representation of their geographic distribution. Because of their social significance and eruptive history, we shall briefly describe some of the characteristics of four of them: Galeras, Nevado del Ruiz, Cerro Mach\'{\i}n, and the Cerro Negro-Chiles complex.

\subsection{Cerro Mach\'{\i}n}
\label{subsec:Machin}
The Mach\'{\i}n volcano is often overlooked as a minor edifice in the Cerro Bravo-Cerro Mach\'{\i}n volcanic belt but, considering its high explosive potential, dacitic composition and magnitude of past eruptions, it must be considered one of the most dangerous active volcanoes in Colombia \cite{Cortes2001}. It is also located at the Cordillera Central (4$^{\circ}$29'N 75$^{\circ}$22'W), between Cajamarca and Mach\'{\i}n \cite{CepedaEtal1995}. 

Over the last 5,000 years, Mach\'{\i}n has had six eruptions --the last one occurred about 850 years ago-- generating pyroclastic flows, depositing several tens of centimeters of ashlayers, throwing eruptive columns (several tens of kilometers) and flows of volcanic mud. In recent times, some of the manifestations of Mach\'{\i}n's volcanic activity are the presence of fumaroles, permanent micro-seismicity, thermal waters flowing in the vicinity of the crater, geoforms of the well-preserved volcanic building and a more significant presence of Radon gas in the sector \cite{Cortes2001, Rueda2005}.

The Mach\'{\i}n volcano, with a crater $2.4$ km diameter and $450$m high, polygenetic, has a dome --developed through hundreds of thousands or million years--, with sedimentary and morphologic characteristics that suggest a Toba cone structure, built during the most recent phreatomagmatic \cite{AguilarPiedrahitaENG2017}. According to studies related to the geological history of this volcano, a future pyroclastic eruptive episode would be deposited mainly in an area of $10$ km$^2$ around the volcano edifice \cite{MurciaEtal2010}.
The similarity in the morphology of two volcanic structures and the activity of one of them must be taken into account, as one more element, in the evaluation of the danger of an eventual eruption. That could be the case with the similitude of the Cerro Mach\'{\i}n Volcano and the Chich\'on or Chichonal Volcano in southern M\'exico in the state of Chiapas ($ 17^{o}$ 21'N; $93^{o}$ 41'W; 1100\,m.a.s.l)
\footnote{\texttt{https://en.wikipedia.org/wiki/Chichonal}} which should alert about the potential danger of Mach\'{\i}n volcano.

\subsection{Cerro Negro - Chiles}
\label{subsec:CerroNegro}
The Chiles and Cerro Negro are located on the Colombia-Ecuador border, 86\, km from the city of San Juan de Pasto, at geographic coordinates 0$^{\circ}$49'N 77$^{\circ}$56'W and 0$^{\circ}$46'N 77$^{\circ}$57'W, respectively. 

This volcanic complex lies at the intersections of three faults: Chiles-Cerro Negro, Chiles-Cumbal and Cerro Negro-Nasta. The volcanic domes reach 4748\,m a.s.l (Chiles), 4470\,m a.s.l (Cerro Negro), and their craters have diameters of 1.0\,km and 1.8\,km, respectively. These two adjacent volcanic cones are collapsed towards the north (Chiles) and west (Cerro Negro), with the presence of geoforms of an already extinct glacier action. Their buildings are formed mainly by several episodes of lava and pyroclastic, with main volcanic products classified as andesites of two pyroxenes and olivines. 

Although there are no historical records of eruptive activity, there is evidence of highly explosive stages, and the current activity is displayed by the presence of hot springs and solfataras. On the Ecuadorian side of Chiles, there is a seismological station which detects various activity.

\subsection{Galeras}
\label{subsec:Galeras}
The Galeras volcanic complex --located in southwest Colombia: 1$^{\circ}$13'18.58"N, 77$^{\circ}$21'33.86"W-- is the most active volcano in Colombia with the highest social risk due to its regular activity and the populated area that surrounds it. 

Surpassing 5,000 years of antiquity, this volcanic complex has a base diameter of 20\,km, a summit elevation of 4,276\,m a.s.l., and a primary crater diameter of 320\,m. The active cone, called Galeras Volcano, rises 1600 m above and approximately 9\, km away from the city of San Juan de Pasto (capital of Nari\~no department) with a population of 313,000 inhabitants.

Galeras is characterized by andesite lava and pyroclastic with significant fallout deposits, displaying a conical shape with a large caldera at the top. After a long period of inactivity of more than 40 years, it awoke again in 1987, experimenting mainly minor eruptions, some with explosive character: fumarolic formation and enlargement, strong tremors, shockwaves and emission of pyroclasts and ashes \cite{CalvacheCortesWilliams1997, CruzChouet1997}. Since 2009, the activity has been considerably reduced to the expulsion of ashes --columns that have reached $10$\,km-- and shockwaves \cite{CortesRaigosa1997}.

\subsection{Nevado del Ru\'{\i}z}
\label{subsec:NevadoRuiz}
The ice-covered volcano Nevado del Ru\'{\i}z (NRV), found at the Cordillera Central --4$^{\circ}$53'N and 75$^{\circ}$19'W-- has an altitude of approximately $5390$\,m.\, a.s.l., covering an area of more than $200$ square kilometers. Its main crater (Arenas) is one kilometer diameter and 240\,m deep. La Pira\~na and La Olleta are two small parasitic edifices, and there are four U-shaped amphitheaters produced by flank collapse and fault activity \cite{Sennert2016}.

This enormous volcano is located at the junction of two fault systems: the N75$^{\circ}$W normal Villa Mar\'{\i}a-Termales fault system and the N20$^{\circ}$E right-lateral strike-slip Palestina fault system \cite{BorreroEtal2009}. The north and northwest borders show uneven geometries caused by the location of large amphitheaters on the upper part of the volcano, the southern and southwestern fringes are marked by sharp regular slopes while the east and southeastern fringes present moderate-to-strong declivities and a significant thickness of glacier deposits.

The eruptive history of the Nevado del Ru\'{\i}z  runs from the Pleistocene to the present, and its stratigraphy has three main stages related to the alternate construction-destruction of its edifice: Ancestral Ruiz (2-1 mega-years), Older Ruiz (0.8-0.2 mega-years) and Present Ruiz (<0.15 mega-years) \cite{ThouretEtal1990}. The present emplacement of lava domes is made of andesite and dacite inside older calderas \cite{HuggelEtal2007}. During the past 11,000 years, this volcano has passed through at least 12 eruption stages, which included multiple slope failures (rock avalanches), pyroclastic and lahars, leading to the partial destruction of the summit domes \cite{ThouretEtal1990, HuggelEtal2007}. The eruptions of the last thousand years have mostly been small, excluding some like the phreatic-magmatic eruption on November 13, 1985, which involved the partial melting of the glacier cap and consequent lahars, which reached and destroyed the municipality of Armero-Tolima and caused a large number of casualties.


\section{Muon flux simulation chain and rock opacity}
\label{SimulationMuonFlux}

\subsection{Muon flux simulation chain}
\label{SimulationChain} 
Particles measured at the ground level (secondaries from now on) come from a chain of interactions and reactions, started by the primaries impinging upon the outer atmosphere. The modulation of secondaries needs to be monitored and carefully corrected by taking into account atmospheric factors that could modify its flux. Thus, a complete and detailed simulation chain --considering factors such as geomagnetic conditions, atmospheric reaction and detector response-- is needed to characterize the expected flux at the detector level.  Any attempt to estimate the expected flux of secondaries at the detector level should consider a detailed simulation that takes into account all possible sources of flux variations of processes occurring at different spatial and time scales. We can illustrate this conceptual scheme as \cite{AsoreyEtal2015B, SuarezENG2015, AsoreyNunezSuarez2018}:
\begin{center}
    \begin{tabular}{llll}
        Cosmic Ray Flux & $\rightarrow$Heliosphere Modulated Flux & $\rightarrow$ Magnetosphere &  $\rightarrow$ \\
        $\cdots \rightarrow$ Primaries & $\rightarrow$Atmosphere Secondaries & $\rightarrow$ Detector response &$\rightarrow$ Signals.
    \end{tabular}
\end{center}

We start our simulation chain by characterizing the effects of the geomagnetic field (GMF) on the propagation of charged particles contributing to the background radiation at ground level. This is included through the calculation of directional rigidity cut-off, $R_c$, at the detector site, determined by using the {\sc{Magnetocosmics}} code, which implements the backtracking technique \cite{Desorgher2003,MasiasDasso2014}. GMF at any point of Earth is calculated by using the International Geomagnetic Field Reference \cite{IGRF11}, for modeling the near-Earth magnetic field ($r\lesssim5 R_{\oplus}$) and by the Tsyganenko Magnetic Field model version 2001  to describe the outer GMF ($r \gg 5 R_{\oplus}$) \cite{Tsyganenko2002}. 

The second link in this chain corresponds to a detailed simulation of extensive air showers produced during the interaction of the flux of primaries --from protons to irons, i.e., $1\leq Z \leq 26$-- through the atmosphere.  To obtain this very comprehensive secondary flux at ground level we use the {\sc Corsika} code \cite{HeckEtal1998}, with a specially tuned set of simulation parameters adapted to a particular geographical site. We will display these parameters in Section \ref{MachinStudy} for the particular case of the Mach\'{\i}n volcano.  For the hadronic model of interaction in the distribution of secondaries at the level of the detector, we use  QGSJET-II-04\cite{Ostapchenko2011} as a model, and for low energies, we choose the default option, GHEISHA-2002 \cite{Fesefeldt1985}.

To identify possible muon observation sites in Colombia, we shall implement the previously described calculation scheme only at secular geomagnetic conditions --i.e., static geomagnetic corrections-- focusing on the detailed calculation of the crossing muon flux and the stopping power of the volcano edifice at different volcano sites. Although we have implemented this general scheme, we found that geomagnetic corrections seem not to be significant for the geographical location of Colombian volcanoes, because it mainly affects low energy primaries\cite{AsoreyNunezSuarez2018}. 

Finally, the detector response to the different types of secondary particles at ground level is simulated using a {\sc{Geant4}} model \cite{AgostinelliEtal2003, CalderonAsoreyNunez2015} for both the scintillator panels and the water Cherenkov detector, but this last step of the simulation chain will not be considered here but will be detailed in a coming  work\cite{JaimesMottaENG2018, VasquezramirezENG2019, VasquezEtal2019, PenarodriguezEtal2020prep}.

\subsection{Directional muon rock opacity}
\label{MuonOpacity}
The open sky flux estimations at the ground level, influenced by various environmental parameters --altitude, geomagnetic corrections, solar modulation, and atmospheric variations-- induce some critical features in the instrument design. 

A comparison of the open sky particle flux $\Phi_{os}$ with the flux, $\Phi$, emerging after crossing the target provides information about its density gradient. To estimate this target density gradient we define the directional rock opacity as the mass density $\rho$ integrated along the muon path $L$ as:
\begin{equation}
\varrho \left( L \right) = \int_{L} \rho \left( \xi \right) d \xi=\bar{\rho} \times L\, , 
\label{opacity}
\end{equation}
where $\xi$ is a characteristic longitudinal coordinate through the volcano, $L$ is the total distance traveled by muons in the rock, $\bar\rho$ is the average density within the volcano. In our approach, we are assuming that the trajectories of muons are straight lines which are not affected by Coulomb scattering processes; therefore density distributions within volcano edifices can be inferred from the variation of its flux, $\Delta =  \Phi -\Phi_{os}$. A more detailed calculation, including second-order effects, is being carried out and will be included in future characterizations of the selected places.

The muon energy loss along each path can be modeled as 
\begin{equation}
-\frac{dE}{d\varrho}=a(E)+b(E) E \, ,
\label{lostenergy}
\end{equation}
for each muon arrival direction considering a uniform density distribution. Here $E$ is the muon energy; $a(E)$ and $b(E)$ are functional parameters depending on the rock composition/properties and $\varrho(L)$ is the density integrated along the trajectory of the muons (the opacity defined by eq. (\ref{opacity})). The coefficient $a(E)$ represents the energy loss due to ionization, while $b(E)$ takes into account the contribution of radiative losses, mainly Brem\ss trahlung, nuclear interactions, and pair production. The main parameters to estimate the coefficients $a(E)$ and $b(E)$ are the average ratio $<Z/A>$ between the atomic and mass numbers of the material \cite{OliveEtal2014,ValenciaoteroENG2017}.
 
\section{MuTe: Colombian Muon Telescope}
\label{subsec:BriefMuonTelescope}
There are three main types of detectors implemented for volcano muography: (a) nuclear emulsion detectors, (b) scintillation detectors, and (c) gaseous detectors. Each one has its \textit{pros} and \textit{cons} as described in \cite{TanakaOlah2019, Tanaka2016}. The main issue in muography of large-size objects is the flux overestimation due to the background sources such as the soft component of Extended Air Showers (EAS) \cite{OlahVarga2017, NishiyamaMiyamotoNaganawa2014} and upward-going muons \cite{JourdeEtal2013}. Here, we shall present a new hybrid Muon Telescope (MuTe), combining the facilities of a hodoscope with a water Cherenkov detector (WCD) in order to solve that drawback. The hodoscope estimates the event flux per trajectory depending on the fired pixels in the front and the rear panel. A Time-of-Flight, ToF, system measures the time taken by the crossing particles subtracting the nano-scale time stamp recorded in both panels. The WCD senses the Cherenkov photons generated in the water due to the interacting charged particle. The recorded photon yield is equivalent for the energy loss.

\subsection{The instrument}
Just for completeness, we briefly describe here the muon telescope its acceptance and noise reduction capabilities, later we employ these characteristics to estimate the time exposure of the instrument at the observational sites. A detailed description of the instrument capabilities will be discussed shortly elsewhere \cite{VasquezEtal2019, PenarodriguezEtal2020prep}.

\begin{figure}
\centering
{\includegraphics[width=4in]{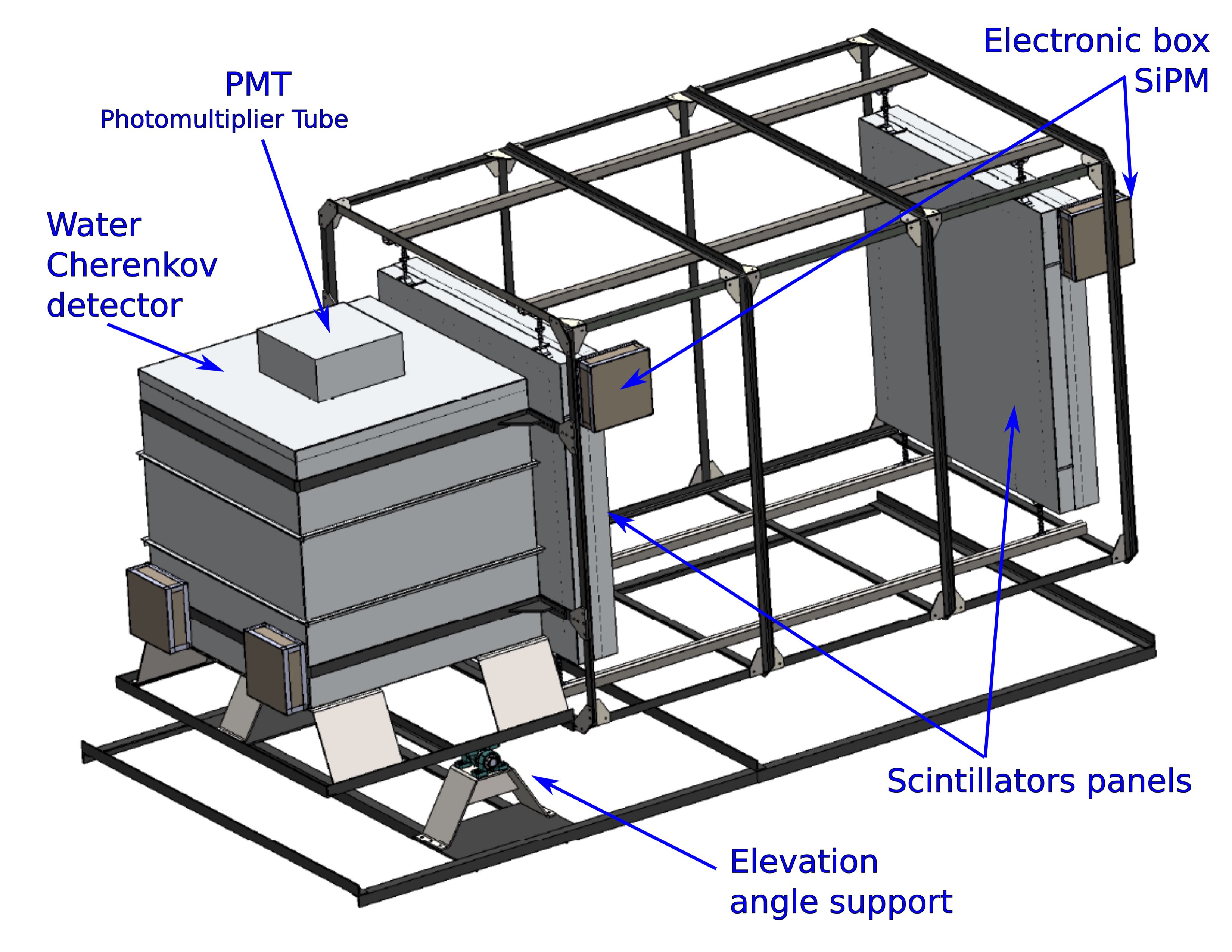}}
\caption{A sketch of our Colombian Muon Telescope (MuTe), which combines the facilities of a two-panel-hodoscope ($900$ pixels) and a 1.73 m$^3$ water Cherenkov detector. }
\label{MuTeTelescope}
\end{figure}

Figure \ref{MuTeTelescope} illustrates MuTe design which combines two detection techniques: a hodoscope formed by two detection planes of plastic scintillator bars, and a WCD, in an innovative setup which differentiates it from some previous detectors. 

The features of the two subdetectors merged in MuTe are:
\begin{itemize}
    \item \textbf{Scintillators panels:} Inspired by the experiences of other volcano muography experiments \cite{UchidaTanakaTanaka2009,GibertEtal2010}, we have designed two X-Y  arrays of $30 \times 30$ plastic scintillating strips ($120$cm$\times 4$cm$\times 1$cm), made with Styron$^{\textrm{TM}}$ 665-W polystyrene doped with a mixture of liquid organic scintillators: 1\% of 2,5-diphenyloxazole (PPO) and 0.03\% of 1,4-bis (5-phenyloxazol-2-yl) benzene (POPOP). Each array has $900$ pixels of $4$ cm$\times 4$ cm = $16$ cm$^2$, which sums up $14,400$ cm$^2$ of detection surface which can be separated up to $D=250$ cm (see figure \ref{MuTeTelescope}).
    \item \textbf{Water Cherenkov Detector:} The WCD is a purified water cube of $120$\, cm side, located behind the rear scintillator panel (again, see figure \ref{MuTeTelescope}), which acts as an absorbing element and as a third active coincidence detector is capable of isolating the muonic component of the incident particle flux. From the charge histogram, obtained by time integration of the individual pulses measured in the WCD, it is possible to separate two components of the incident flux: electromagnetic part (photon, electron \& positron) and the $\mu-$component \cite{AsoreyEtal2015B}. Additionally, due to its dimensions and location, it filters most of the electrons backward noise while protons/muons moving backward --which could cause overestimation in the hodoscope counts \cite{NishiyamaEtal2016}-- are rejected by a Time-of-flight system.
\end{itemize}

Muon generated events deposited in an energy range of ($144$MeV$ < dE/dx < 400$MeV), represent only about $40\%$ of the WCD-hodoscope acquired events. The other 60$\%$ of data is composed by ($e^{\pm}$) events under $144$MeV and multiparticle events above $400$ MeV. Subsequently, low-momentum muons ($< 1$ GeV/c), which are scattered by the volcano surface, are also rejected.

The Colombian MuTe combines particle identification techniques to discriminate noise background from data. It filters the primary noise sources for muography, i.e., the soft-component ($e^{\pm}$) of Extensive Air Showers (EAS) and scattered/upward-coming muons. Particle deposited energy identifies Electrons/positrons events in the WCD, while scattered and backward muons are rejected using a pico-second Time-of-Flight system.

\subsection{Telescope acceptance}
\label{subsec:acceptance}
The acceptance of the instrument is a convolution of the telescope geometry (number of pixels, pixel size, and panel separation) and it is obtained multiplying the detection area by the angular resolution, i.e.
\begin{equation}
\mathcal{T}(r_{mn})=S(r_{mn})\times \delta\Omega(r_{mn}),
\end{equation}  
where $r_{mn}$ represents each discrete muon incoming direction, which for an array of two panels with $N_x\times N_y$ pixels  we can identify $(2N_x-1)(2N_y-1)$ different particle trajectories\cite{LesparreEtal2010}.  Moreover, the number of incident particles $N(\varrho)$ can be expressed as
 \begin{equation}
N(\varrho)=\Delta T \times \mathcal{T}\times I(\varrho)\, , \label{Nmuons}
\end{equation}
and $I(\varrho)$ is the integrated flux (measured in cm$^{-2}$\,sr$^{-1}$\,s$^{-1}$), $\mathcal{T}$ represents the acceptance (measured in cm$^{2}$\,sr) and $\Delta T$ designates the time exposure (in seconds). It is possible to obtain a simple relationship between the exposure time and the desired opacity resolution as
\begin{equation}
\Delta T\times \mathcal{T}\times \frac{\Delta I^2(\varrho_0,\delta\varrho)}{I(\varrho_0)}>c, \label{feasibility}
\end{equation}
with $\Delta I(\varrho_0,\delta\varrho)$ the flux variation due to the different opacities $\varrho_0$ and $\varrho_0+\delta\varrho$, and $c$ is a parameter measuring the confidence level in terms of the standard deviation of the measurement. The above expression gives a bound for the minimum exposure time needed to distinguish opacity differences across the geological object\cite{LesparreEtal2010}.

\begin{figure}
\centering
{\includegraphics[width=0.49\textwidth]{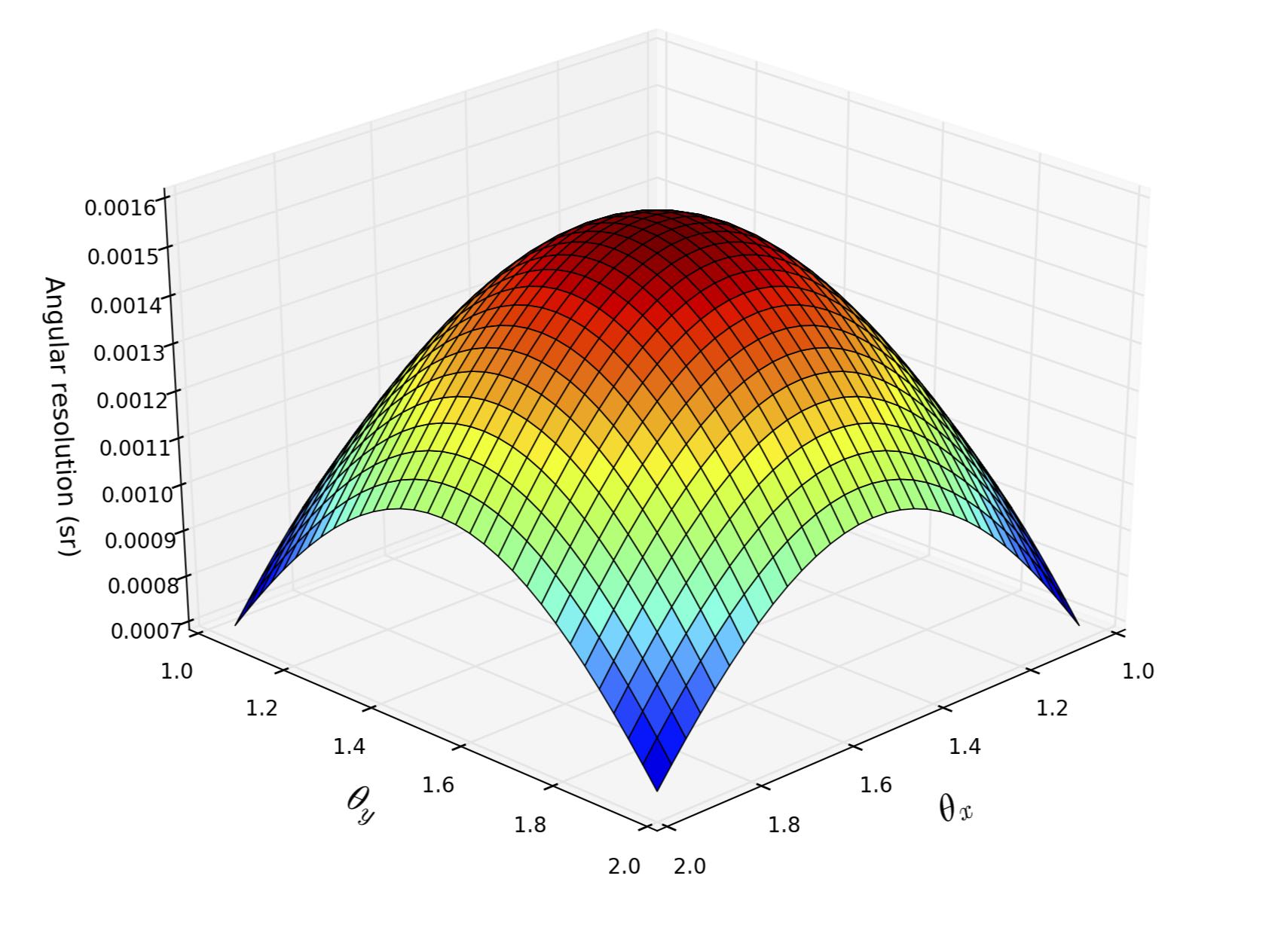}}
{\includegraphics[width=0.47\textwidth]{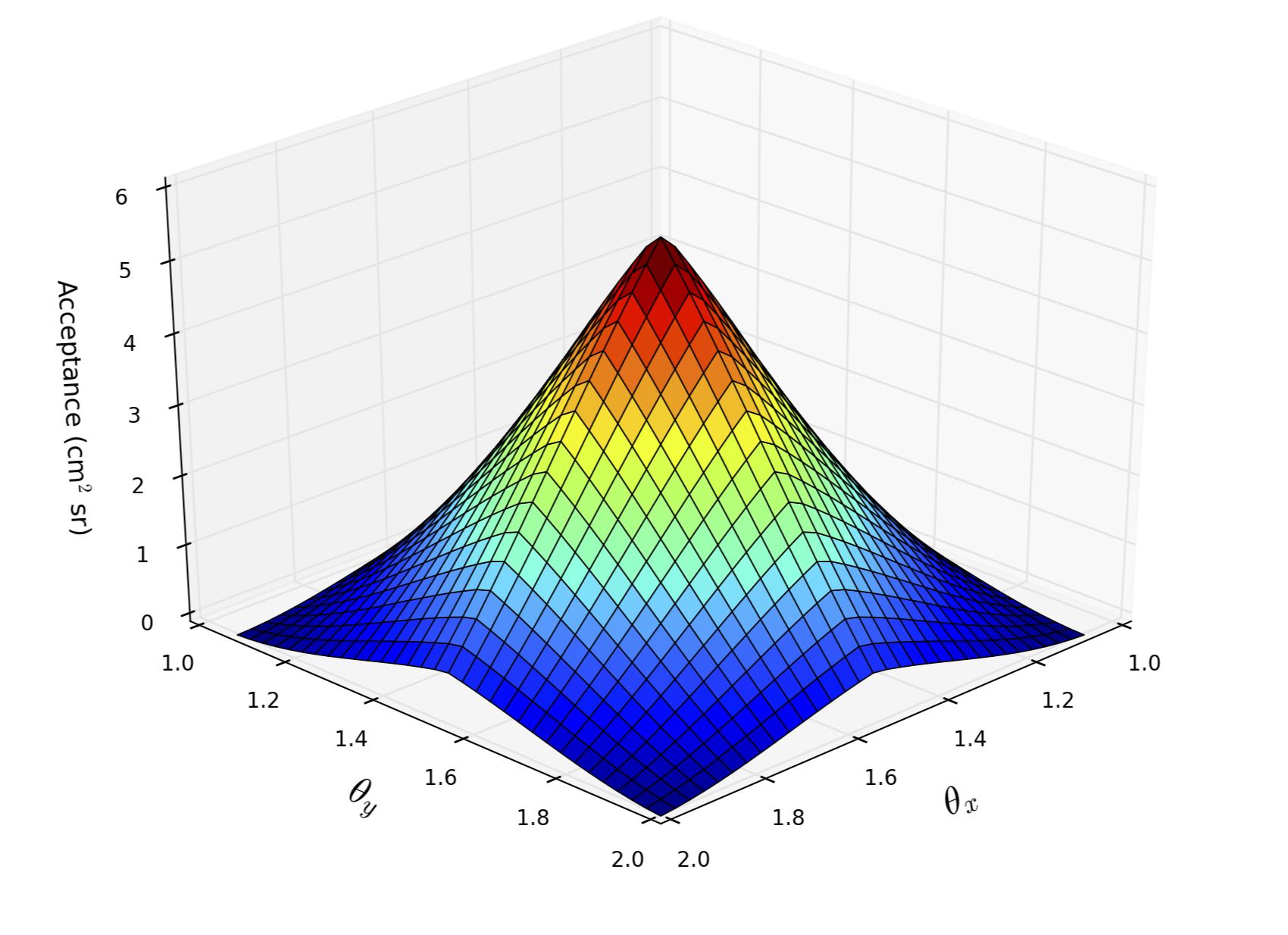}}
  \caption{Angular resolution (sr) and acceptance function (cm$^{2}$\,sr) of MuTe. Each detection panel has  $N_x=N_y=30$, 4-cm wide scintillator bars, shaping $900$ pixels of $16$\,cm$^2$ of detecting area. For this number of pixels there are  3481 discrete $r_{mn}$ possible incoming directions.}
  \label{acceptance}
\end{figure}

In figure \ref{acceptance} the angular resolution and the acceptance function for our telescope (with 900 pixels and 3481 discrete $r_{mn}$ directions) is shown. The total angular aperture of our telescope is roughly 582 mrad with the maximum point at $1.6\times 10^{-3}$sr and, as expected, the largest detection surface corresponds to the normal direction $r_{00}$, reaching $\approx 6$\,cm$^{2}$\, sr.


\section{Mach\'{\i}n, Chiles, Cerro Negro and Galeras Volcanoes}
\label{Results}
 In this Section we shall first present a detailed study of the muon propagation across the Mach\'{\i}n volcano and calculate the corresponding exposure times for four observation points. We shall also display estimations of the outgoing muon flux from the Mach\'{\i}n dome that could be detected by our muon telescope. Secondly, we also display a ray-tracing analysis for muon propagation for three other active Colombia volcanoes: Chiles, Cerro Negro, and Galeras. 
 
 From the information gathered, we identify some critical parameters that could limit the application of muography in Colombian inland volcanoes and devise a ``rule of thumb'' criteria to identify possible muography volcano candidates.

\subsection{Detailed study of the flux from the Mach\'{\i}n volcano}
\label{MachinStudy}
We start our study with Mach\'{\i}n volcano because it is one of the most dangerous active volcanoes in Colombia. It has an average height of $2750$m.a.s.l; a crater of $2.4$km of diameter blended into the landscape of its nearby topography, which makes it practically invisible. This fact increases the risk of the Mach\'{\i}n volcano for the surrounding population.
Additionally, the morphological similarities between Chichonal and Mach\'{\i}n volcanoes --both are considered stratovolcanoes and went unnoticed as volcanoes for decades \cite{Rubio1985, Macias2006}-- is impressive. The resemblances of these two volcanic buildings amaze and, the chipping of the eruption of the Chichonal volcano  --the largest in the history of M\'exico, with an affectation of almost $100$km around-- should alert about the potential danger of Mach\'{\i}n.

\subsubsection{Cosmic ray spectral composition at the Mach\'{\i}n volcano}
\begin{figure}
\centering
{\includegraphics[width=0.7\textwidth]{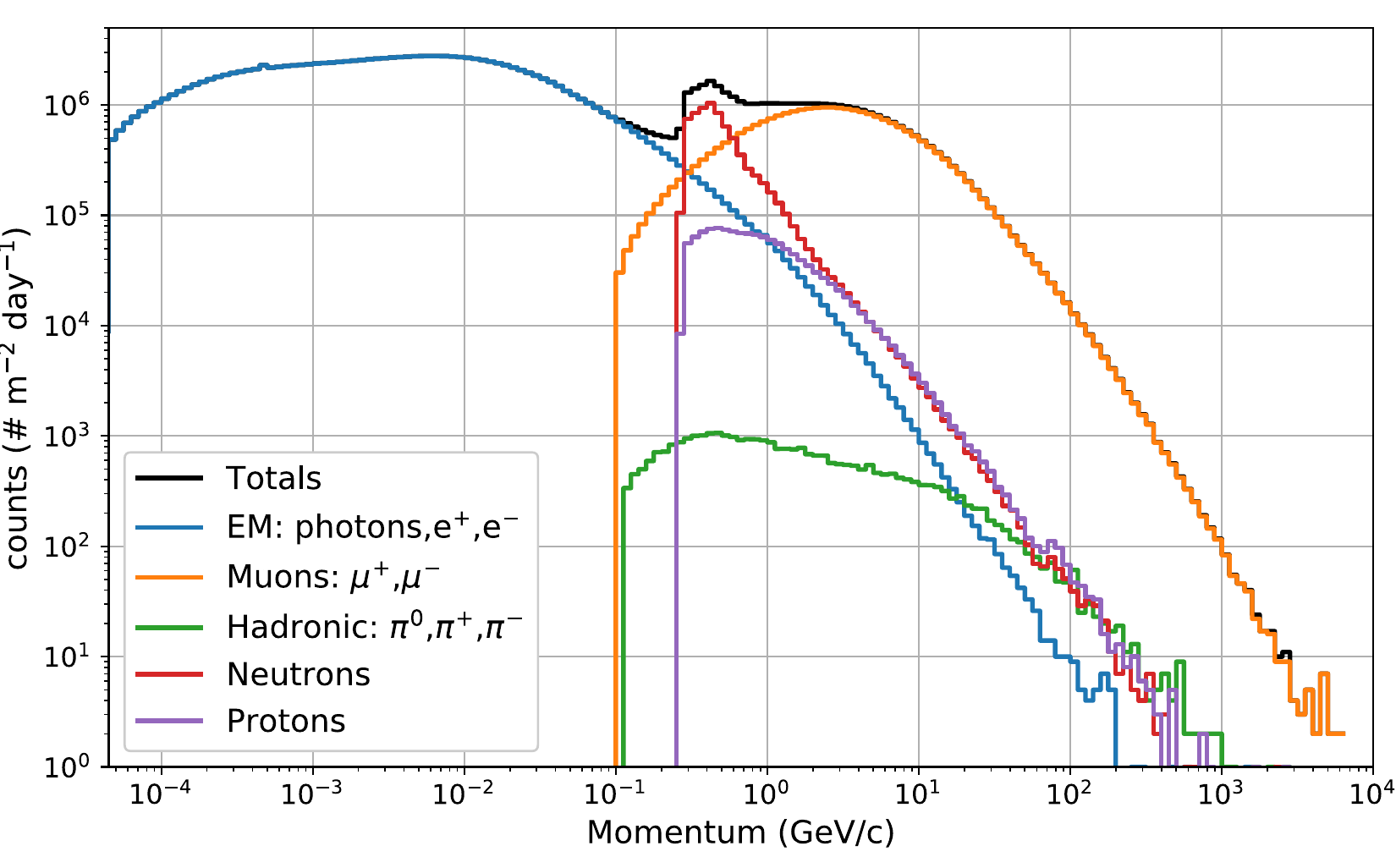}}
	\caption{The integrated spectrum of secondaries at the top of Cerro Mach\'{\i}n. At the highest momentum of the background, the flux is dominated by muons. It is noticeable that muons could reach momentums up to $10$\, TeV/c but with low occurrence, while the most probable muons arrive on average with an energy of $4$GeV/c.}
  \label{SpectralParticleComposition}
\end{figure}

We implement the simulation chain described in section \ref{SimulationChain} to estimate the spectral composition of the open sky particle flux, $\Phi_{os}$, at the top of Mach\'{\i}n volcano ($4^{\circ}$29'N 75$^{\circ}$22'W). Figure \ref{SpectralParticleComposition} displays the expected momentum spectra for the open sky flux of secondaries at the geographic coordinates of Cerro Mach\'{\i}n. Notice that the angular integrated flux is dominated by muons, which could reach momentums up to $10$\, TeV/c but with the low occurrence, while the most probable muons arrive on average with an energy of $4$GeV/c.

\begin{figure}[!ht]
\centering
{\includegraphics[width=0.49 \textwidth]{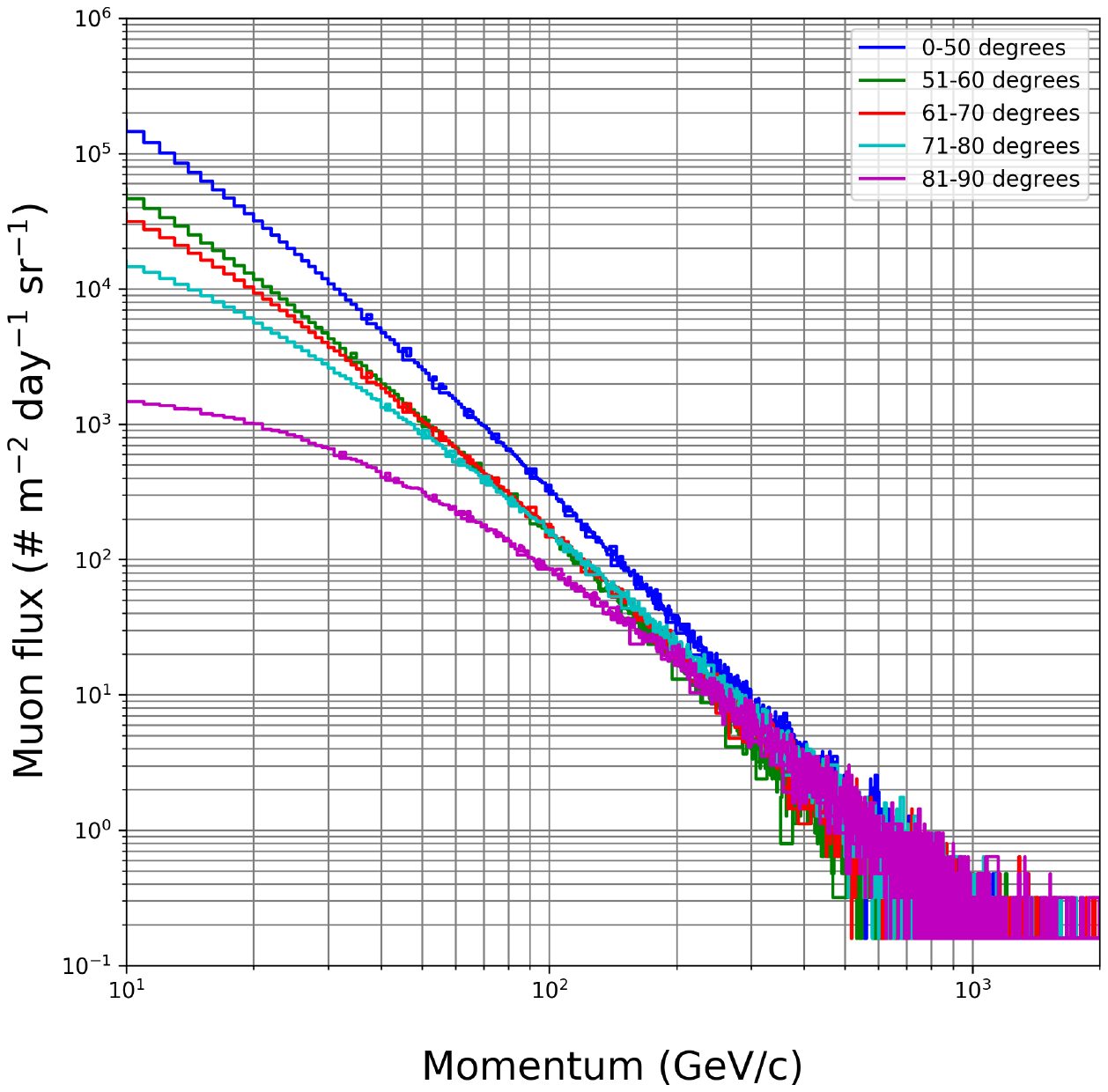}}
{\includegraphics[width=0.49 \textwidth]{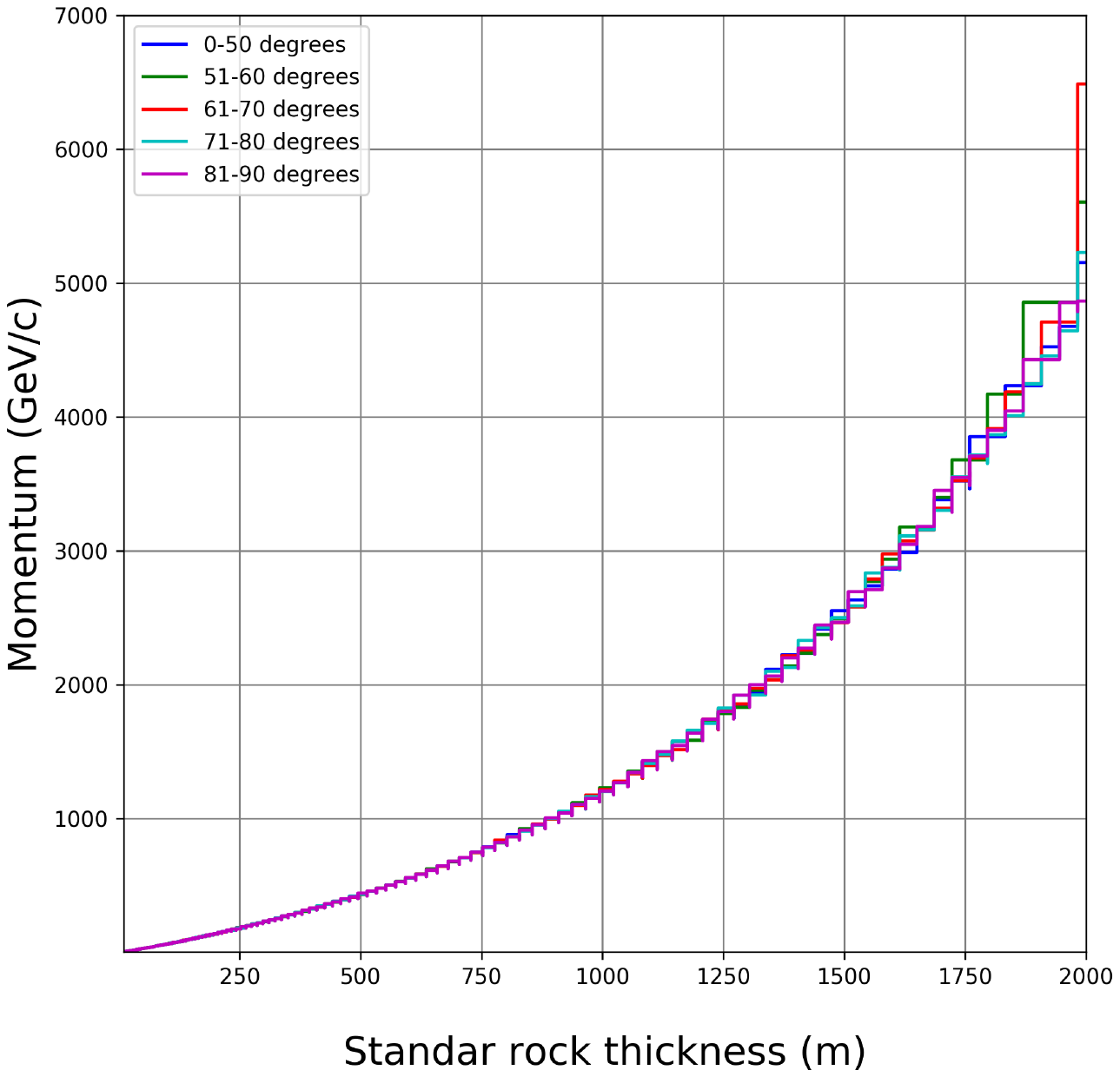}}
	\caption{Left plate displays the momentum spectra for muons  emerging in five angular bins after crossing a variable standard rock with density $2.65$ g/cm$^3$, measured at the point {\bf P$_{1M}$} of Cerro Mach\'{\i}n volcano. At $p_\mu \simeq 500$\,GeV/c, the integrated flux is $1$\,cm$^{-2}$\,sr$^{-1}$\,day$^{-1}$ and, as expected, muons with zenith angles close to the horizontal decrease by a factor of $100$ with respect to those with more vertical angles of incidence. Right plate illustrates the muon energy needed to cross standard rock ($2.65$g/cm$^3$) thickness.} 
  \label{fluxvsp}
\end{figure}

The {\sc Corsika} simulations executed at the top of Mach\'{\i}n volcano, have the following set of parameters: Latitude: 4.48N, Longitude: -75.39W; Magnetic Field: North $B_{X} = 26.91\, \mu$T  and vertical $B_{Z} = 14.37 \, \mu$T,  ( \texttt{https://www.ngdc.noaa.gov/geomag/calculators/magcalc.shtml\#igrfwmm}); Flux time: $24$h = $86400$s; Arriving altitude: $2450$m.a.s.l.; total number of simulations corresponds to $57.75$ days. The zenith incidence of primaries are between 0$^{\circ}$ and 90$^{\circ}$ (all range) and the primary energies ($90\%$ of protons and $10\%$ other heavier nuclei) are in the range of $5\,$GeV and 10$^6\,$GeV. We also select the ``tropical US Standard'' atmospheric model, a volumetric detector for flux calculations, standard energy cuts and rigidity cutoff.

\subsubsection{Muon propagation through the Mach\'{\i}n volcano}
Next, we calculate the differential flux of muons at the above mentioned four observational points as a function of the direction of arrival and determine the maximum possible depth that can be observed from each point. This is shown in the left plate of figure \ref{fluxvsp}, where we display the momentum spectra for muons emerging in five angular bins after crossing a standard variable rock with density 2.65 g/cm$^3$, measured at the point {\bf P$_{1M}$} of Mach\'{\i}n volcano.  As it can be clearly appreciated, the muon flux decreases considerately reaching $1$\,cm$^{-2}$\,sr$^{-1}$\,day$^{-1}$ for $p_\mu \simeq 500$\,GeV/c. 

\begin{figure}[!ht]
\centering
\includegraphics[width=0.45\textwidth]{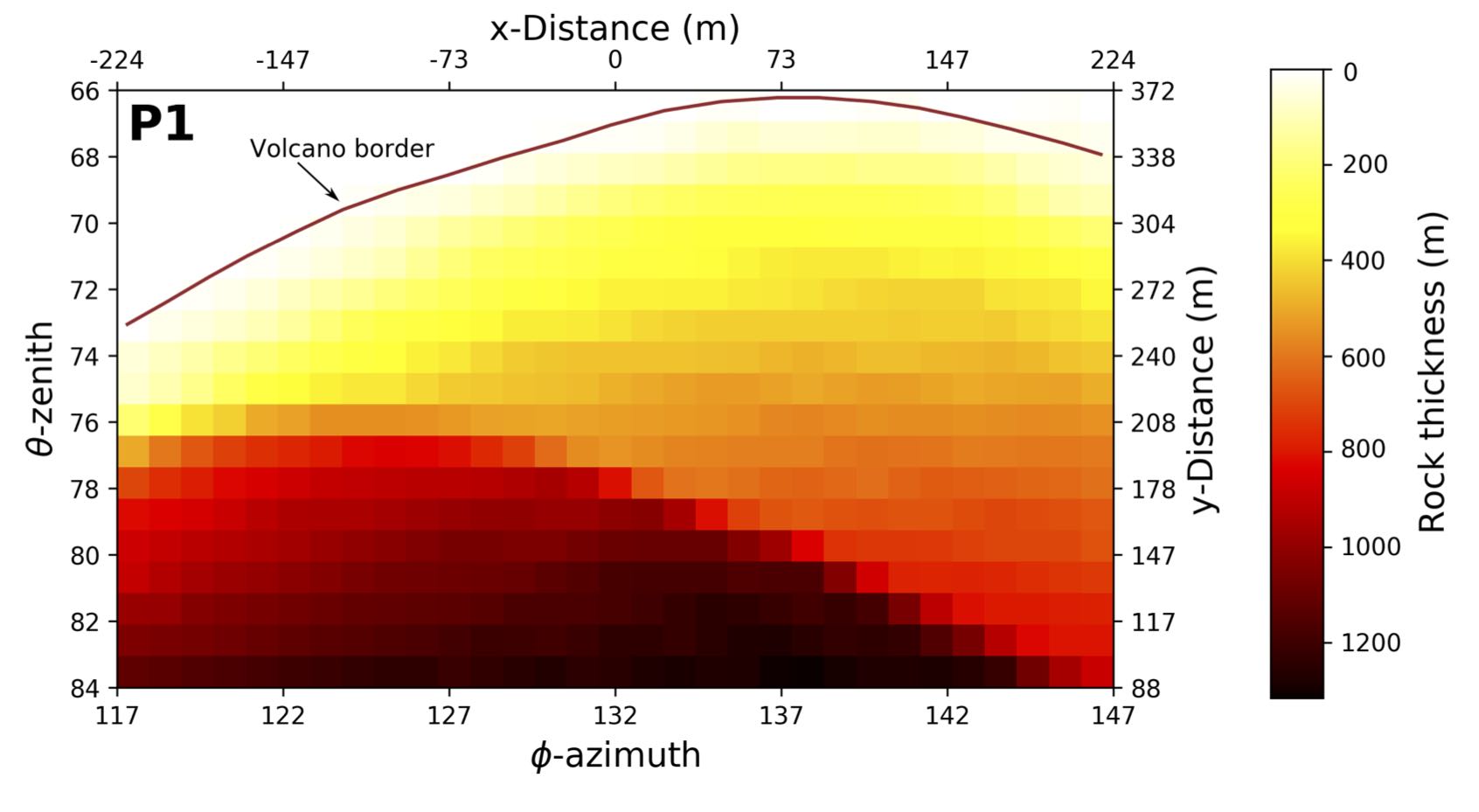}
\includegraphics[width=0.45\textwidth]{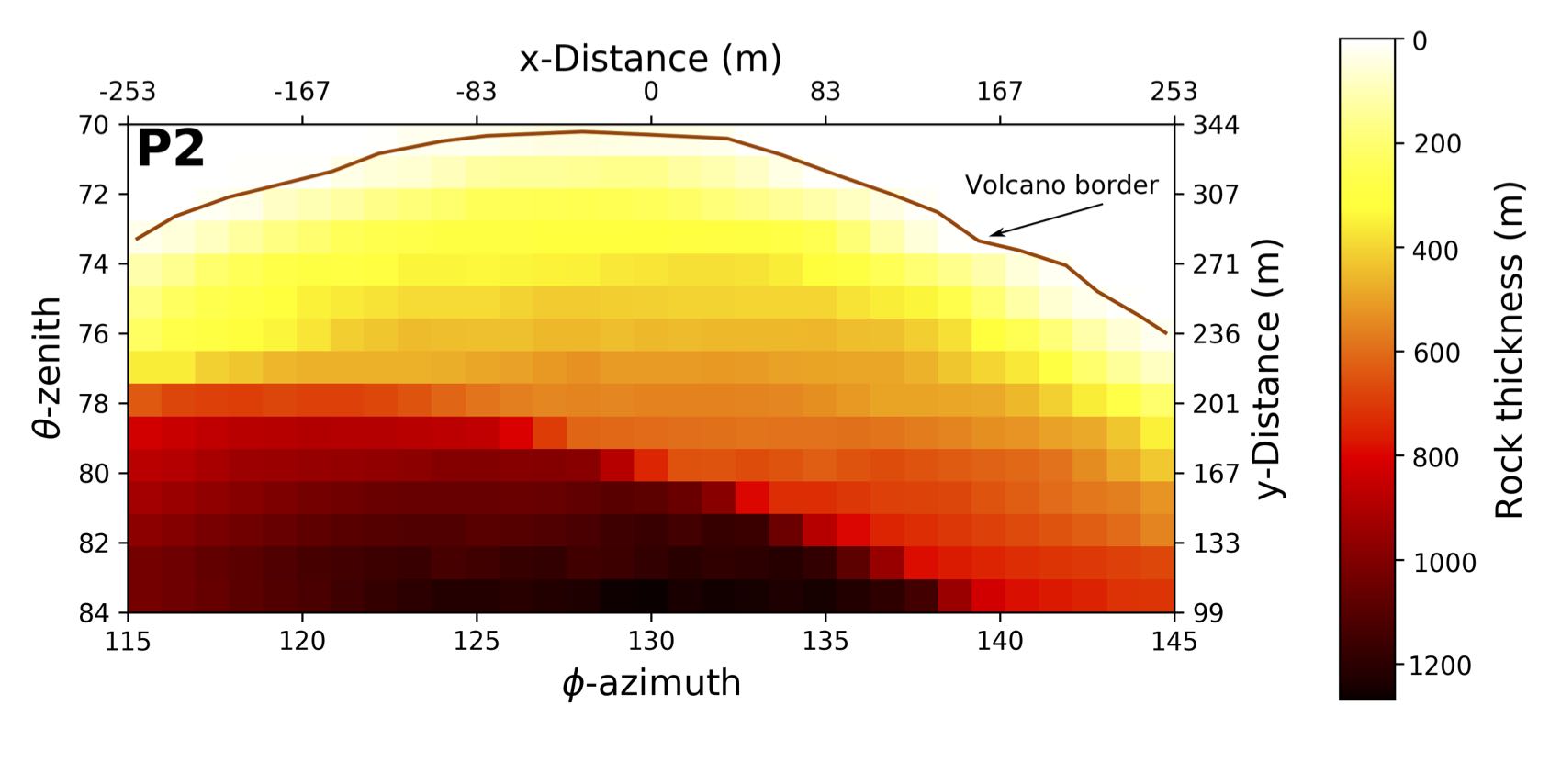}\\
\includegraphics[width=0.45\textwidth]{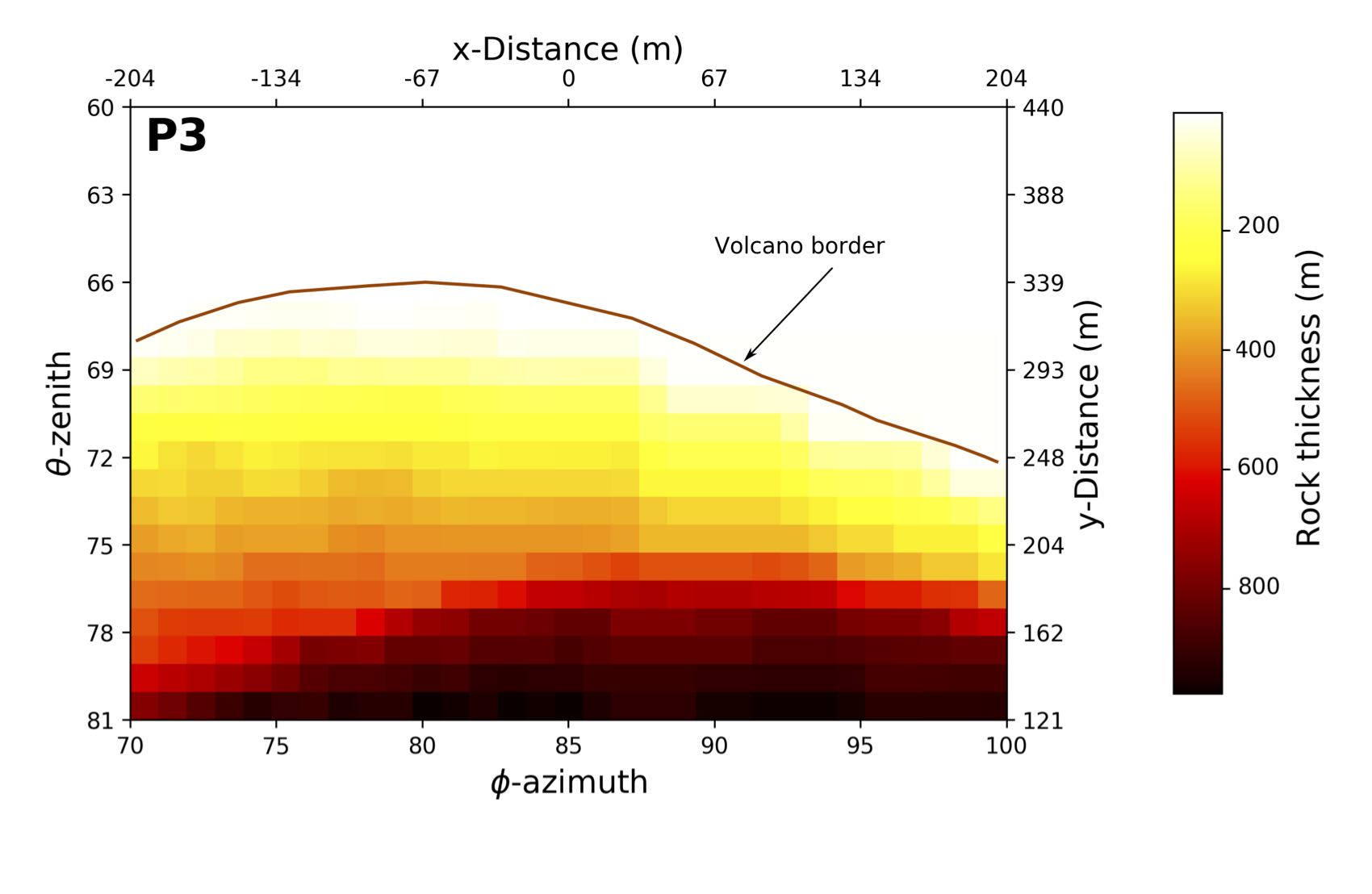}
\includegraphics[width=0.45\textwidth]{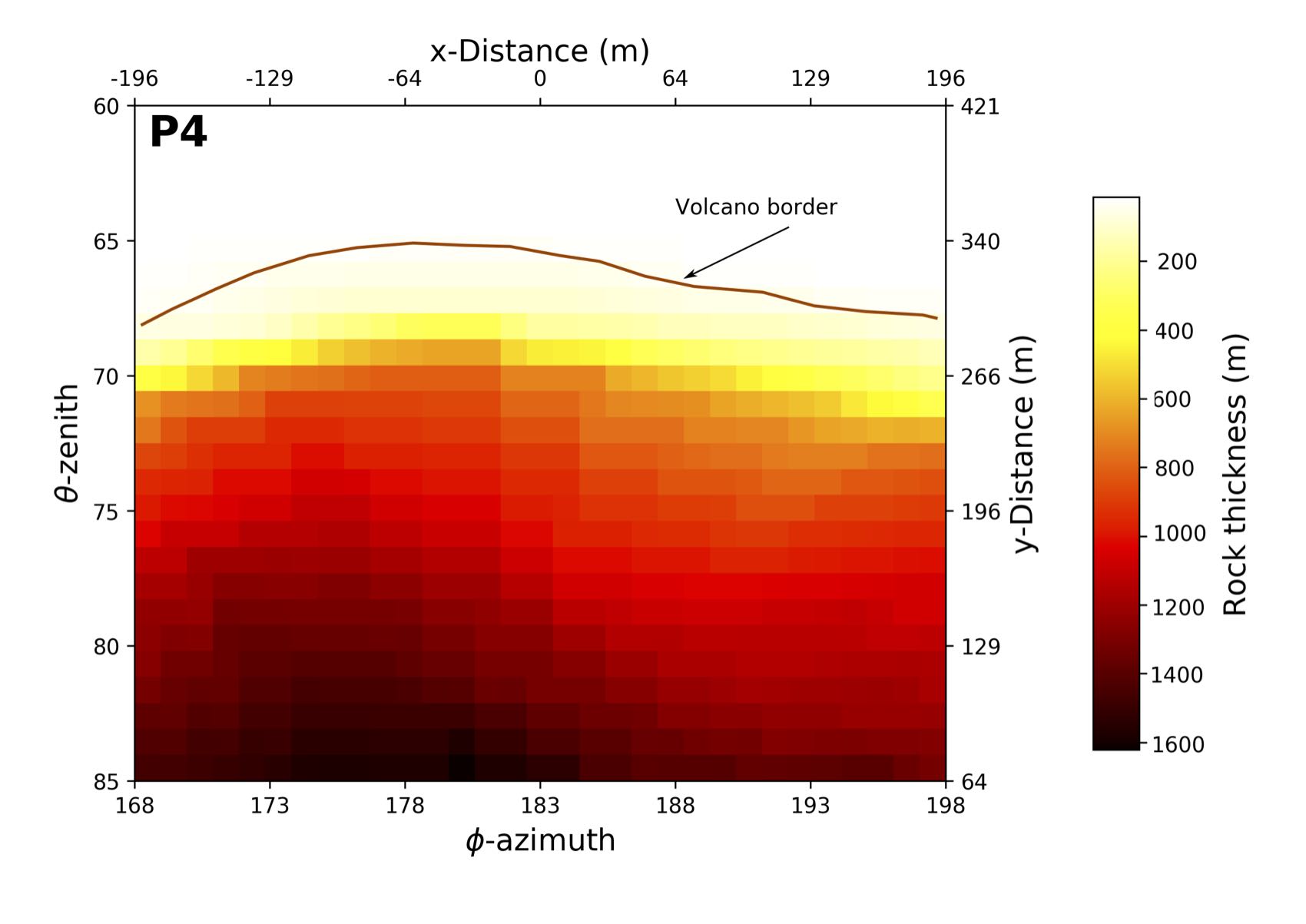}
	\caption{Particle trajectories crossing Cerro Mach\'{\i}n volcanic structure to the observation points {\bf P$_{1M}$}, {\bf P$_{2M}$}, {\bf P$_{3M}$} and {\bf P$_{4M}$}. Notice, for example, that for {\bf P$_{1M}$} observation point, muons with zenith angles $\theta>70^\circ$ travel distances exceeding $900$ meters. The topography was obtained from NASA Shuttle Radar Topography Mission global digital elevation model of Earth, with SRTM3 resolution $90$m$\times 90$m.}
  \label{ParticleTrajectories}
\end{figure}

We have employed the topography from the NASA Shuttle Radar Topography Mission global digital elevation model of the Earth\footnote{See: http://www2.jpl.nasa.gov/srtm/}, with resolution of $90$m$\times 90$m (see left plate in figure \ref{ParticleTrajectories} for the particular case of the Mach\'{\i}n volcano). Next,  we calculate all the possible distances crossed by each muon path integrating equation (\ref{lostenergy})  and (\ref{opacity}) for standard-rock-model dome, with the coefficients $a(E)$ and $b(E)$ obtained from the Particle Data Group\cite{OliveEtal2014} \footnote{Tables on: http://pdg.lbl.gov/2011/AtomicNuclearProperties/}.

At Cerro Mach\'{\i}n we have identified four observation points: {\bf P$_{1M}$}, {\bf P$_{2M}$}, {\bf P$_{3M}$} and {\bf P$_{4M}$}  around it  (see Table \ref{TableMachin}).  Figure \ref{ParticleTrajectories} displays the ray-tracing technique implemented for those  points with the corresponding  muon propagation distance through the topography, as well as the angular distribution of these distances around the upper part of the volcano. Emerging muons with these trajectories have crossed about 600 meters of rock within the geological structure.
 
\begin{table}[!ht]
\centering
\begin{tabular}{lllll}
\hline
\textbf{Cerro Mach\'{\i}n points}        & \textbf{P$_{1M}$}    & \textbf{P$_{2M}$} & \textbf{P$_{3M}$} & \textbf{P$_{4M}$} \\ \hline
\textbf{Latitude  ($^{\circ}$N)}        & 4.492298             & 4.491984       & 4.487338       & 4.494946     \\
\textbf{Longitude ($^{\circ}$W)}        & -75.381092            & -75.380085      & -75.379510      & -75.388110    \\
\textbf{Distance to edifice center (m)} & 836               & 946         & 762         & 730       \\ 
\textbf{Maximum observed depth (m)}     & 208               & 228         & 250         & 190       \\ 
\hline
\end{tabular}
	\caption{Feasible observation points at Cerro Mach\'{\i}n volcano (4$^{\circ}$29'23.08"N,$\;$ 75$^{\circ}$23'15.39"W). The maximum observed depth are those points where the emerging muon flux is less than 10$^{-2}$ muons per cm$^{2}$ per day, corresponding to zenith angles $\theta \approx 82^\circ$. }
\label{TableMachin}
\end{table}

\begin{figure}[!ht]
\centering
\includegraphics[width=0.45\textwidth]{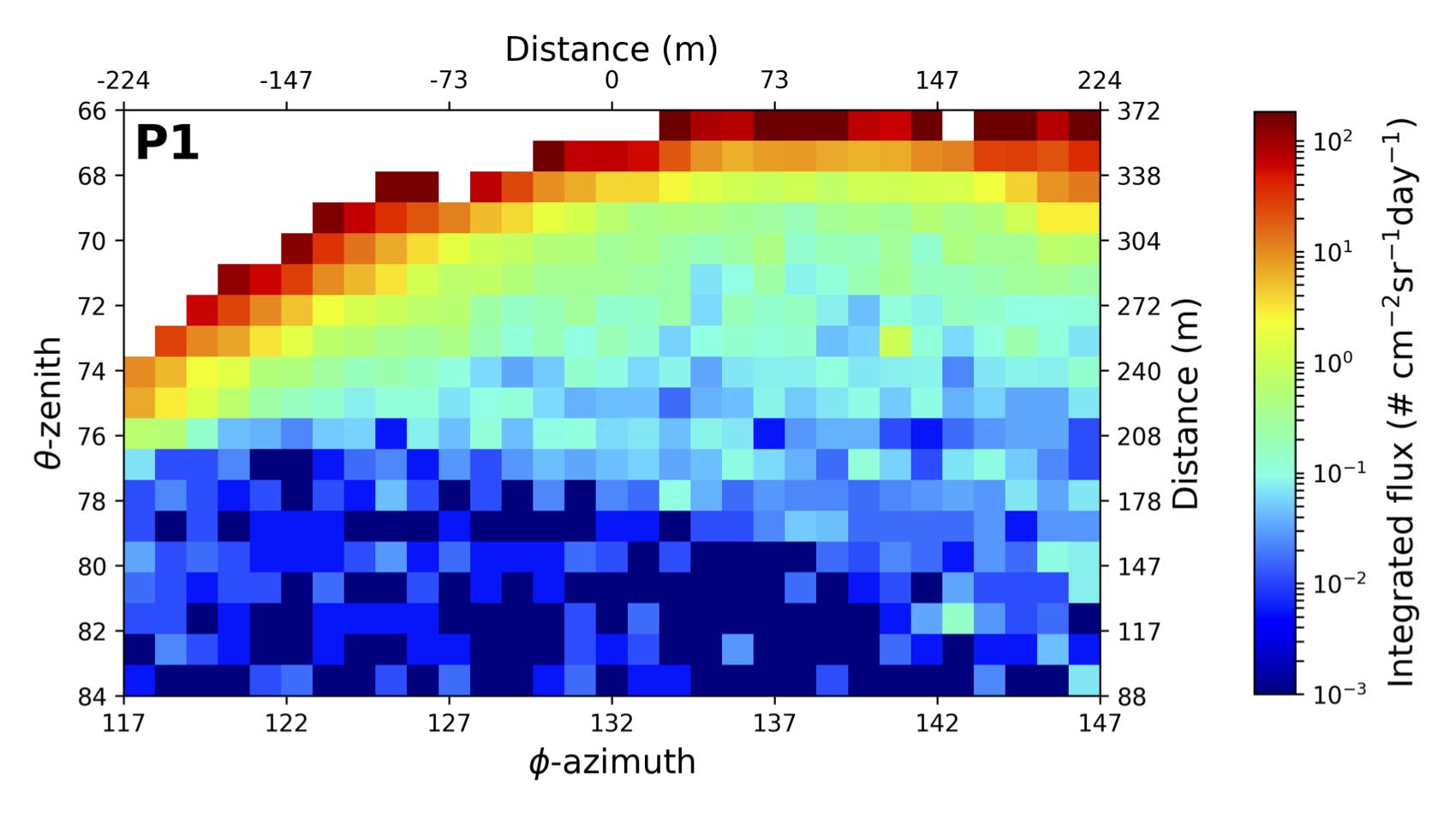}
\includegraphics[width=0.45\textwidth]{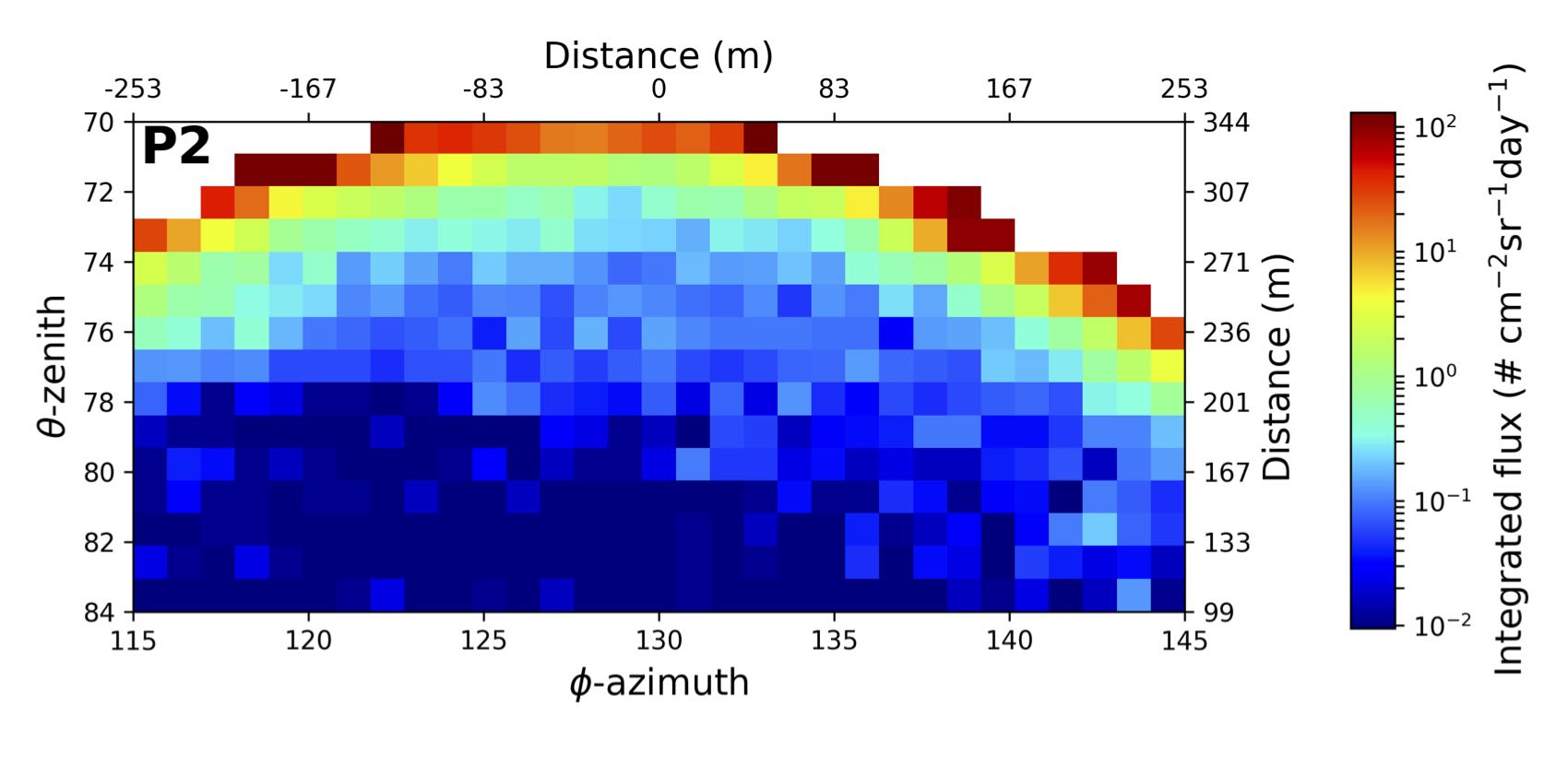}\\
\includegraphics[width=0.45\textwidth]{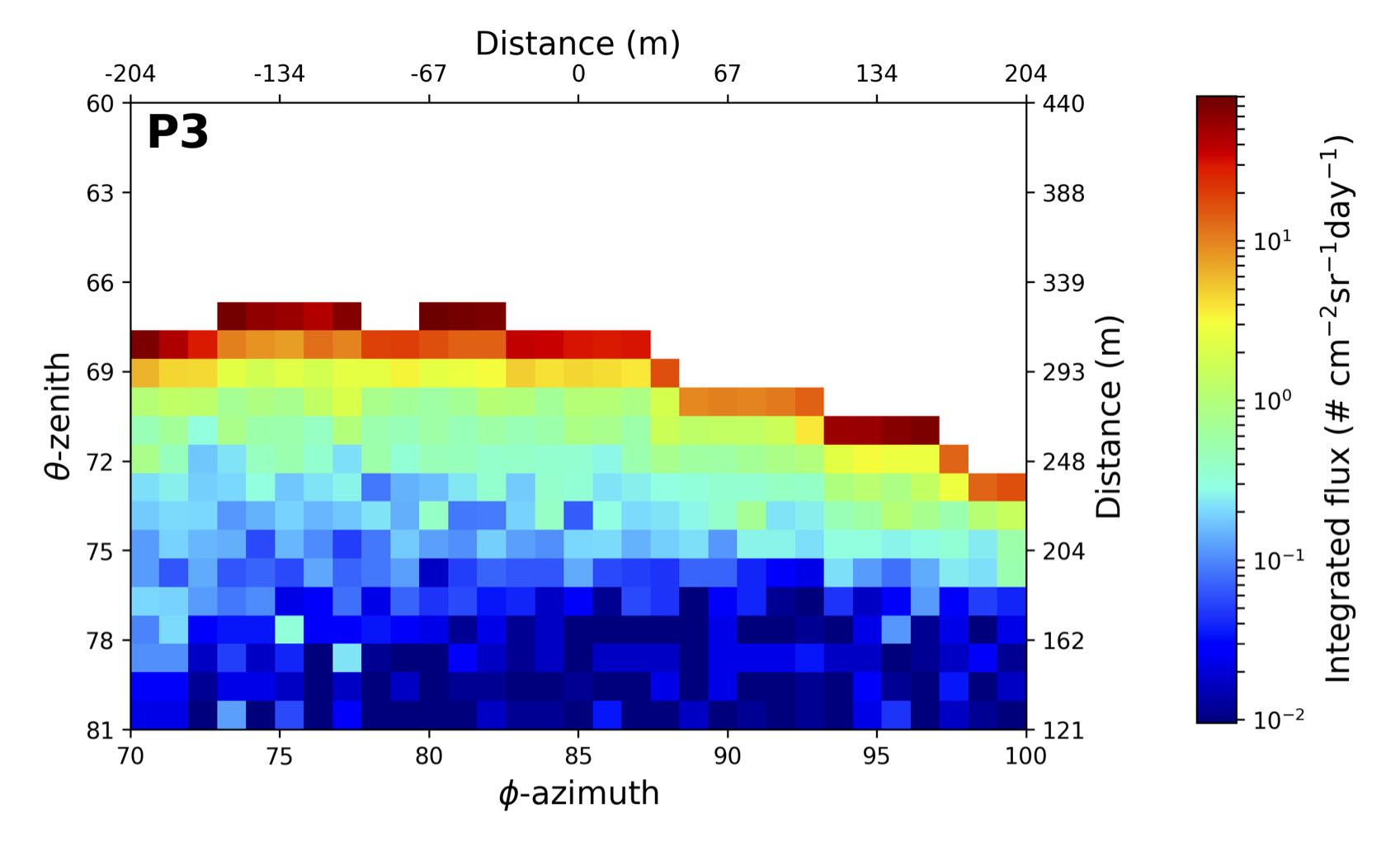}
\includegraphics[width=0.45\textwidth]{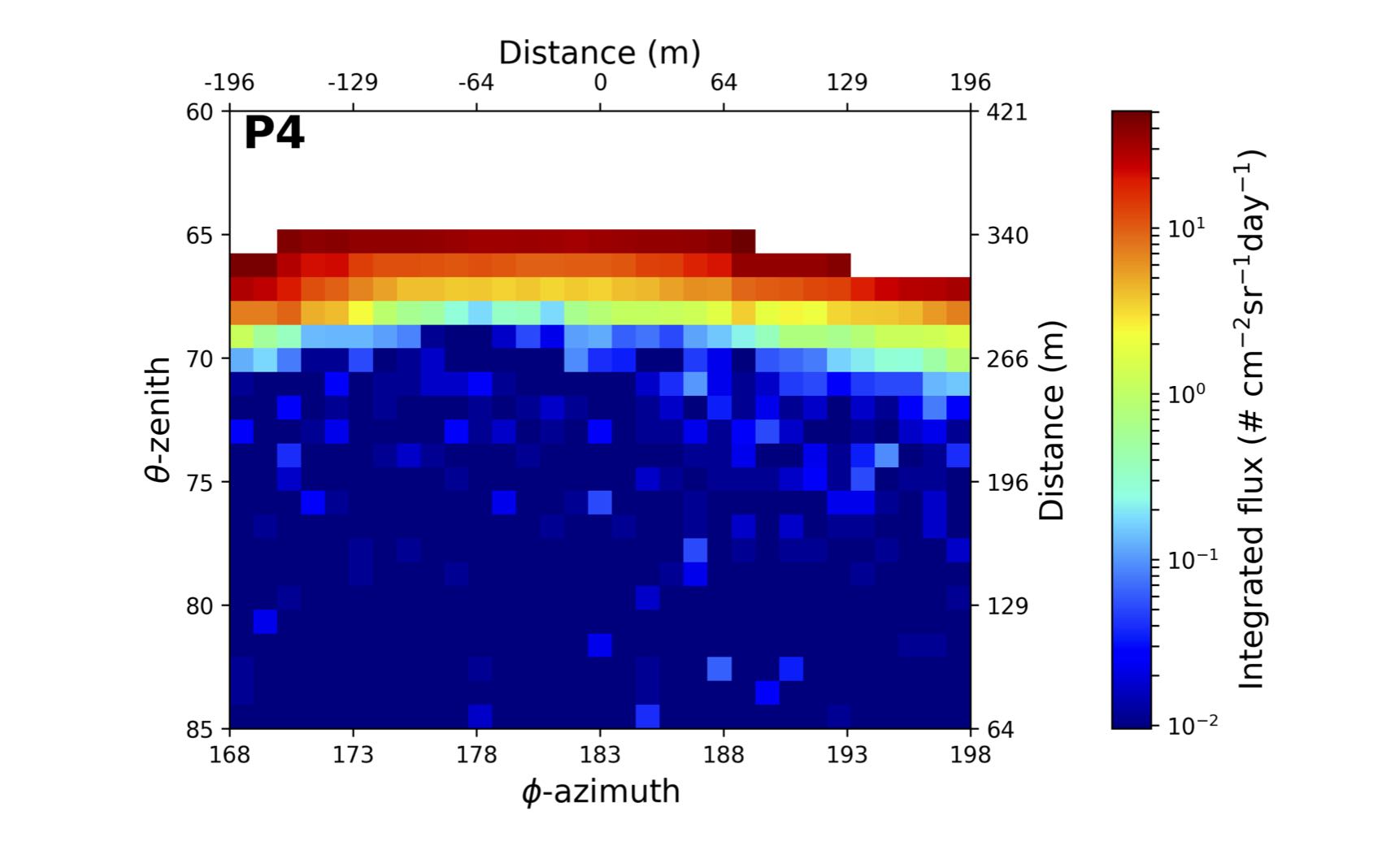}
    \caption{Expected muon flux at {\bf P$_{1M}$}, {\bf P$_{2M}$}, {\bf P$_{3M}$} and {\bf P$_{4M}$} observation points in Cerro Mach\'{\i}n, as a function of the direction of incidence. Integrating equation (\ref{lostenergy}) we obtain that muons with energies from 0.1 GeV/c to 10 TeV, generate feeble flux:$\approx$ 10$^{-2}$ muon per square centimeter per day at the maximum possible observed depth at zenith angles $\theta \approx 81^\circ-84^\circ$. White pixels represent open sky muon flux; other colors illustrate the emerging muon flux from the volcano edifice. Comparing this with the previous figure, we can observe that very few muons can cross the structure with traveling distances higher than $1000$m.}
  \label{flux_volcano}
\end{figure}

Figure \ref{flux_volcano}  displays results of $57.75$ days muon flux, emerging from the volcano, and measured at the observation points by our $30\times30$ pixels telescope with an inter-panel distance of $200$\, cm. We have set a minimum threshold count of $100$ muons/pixel. Comparing figures \ref{ParticleTrajectories} and \ref{flux_volcano}, we appreciate that there are regions where the incoming muon flux is highly absorbed due to the volcano geometry and the long distances traveled by these muons within the volcano could easily exceed  $900$ meters. Recently, we performed these simulations implementing MUSIC (for MUon SImulation Code \cite{Kudryavtsev2009}) a precise Monte Carlo muon transport code and obtained similar results\cite{MossEtal2018}.

We have found that very few muons --the most penetrant component of the particle shower, ranging from tenths to few thousands of GeV/c--, $\approx10^{-1}$ muons per square meter, per day, (see figure \ref{fluxvsp}, left plate)-- can cross almost $1,000$m of standard rock (see figure \ref{fluxvsp}, right plate or figures \ref{ParticleTrajectories} ). Therefore we set $1,500$m as the upper bound for distances that can be traveled by the most energetic horizontal  (Zenith angles $\gtrsim$ 70 degrees) muons at any Colombian Volcano. Our analysis concludes that starting from a few GeV/c there is no significant effect of the geomagnetic correction on the muon flux at any geographical zone in Colombia, but this correction is, in general, essential to determine all the possible particle background flux at other sites with different latitudes \cite{AsoreyNunezSuarez2018}.

\begin{figure}[!ht]
\centering
\includegraphics[width=0.45\textwidth]{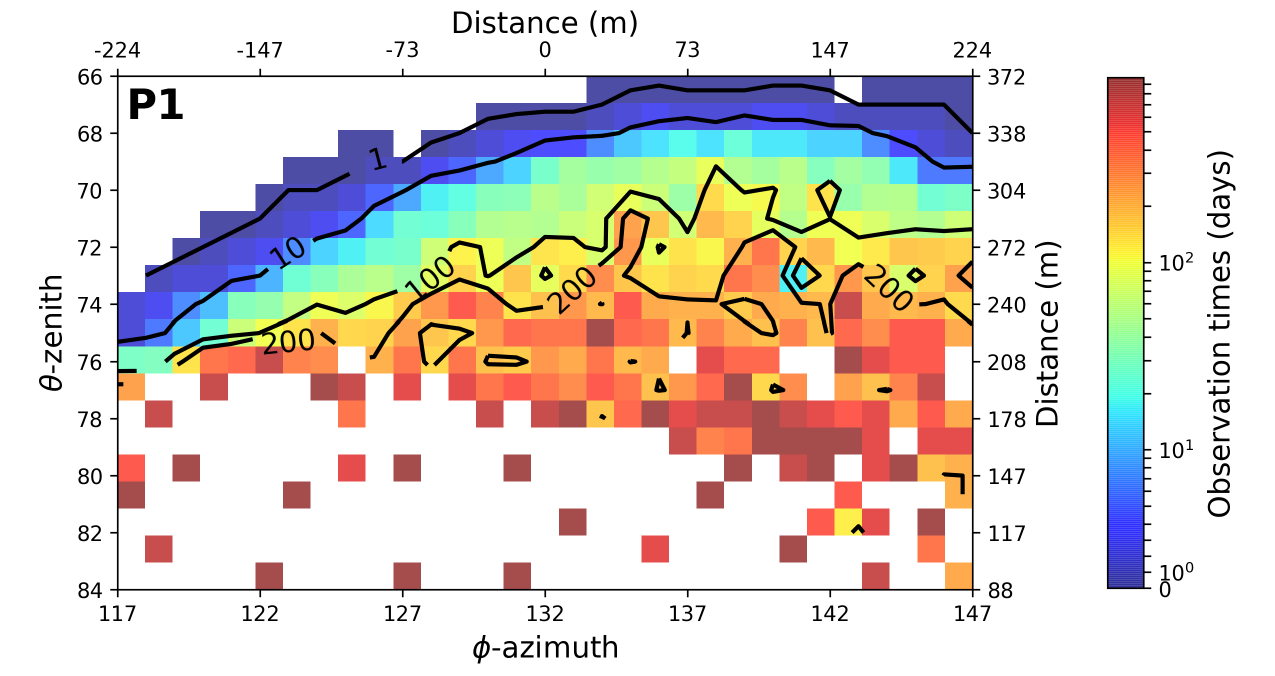}
\includegraphics[width=0.45\textwidth]{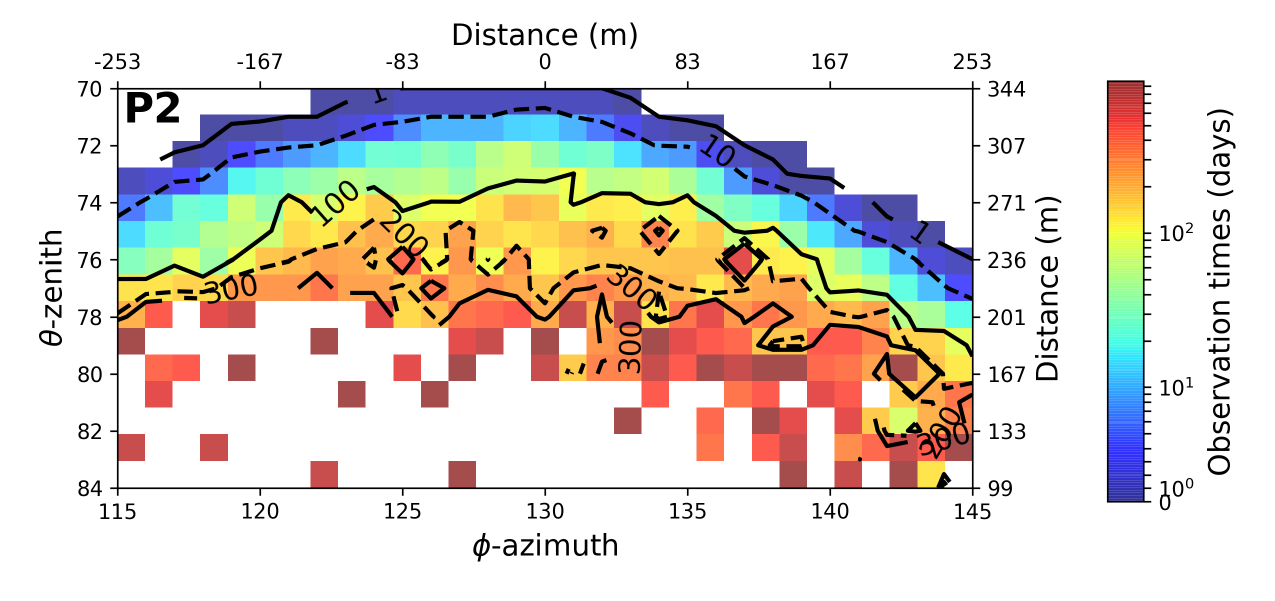}\\
\includegraphics[width=0.45\textwidth]{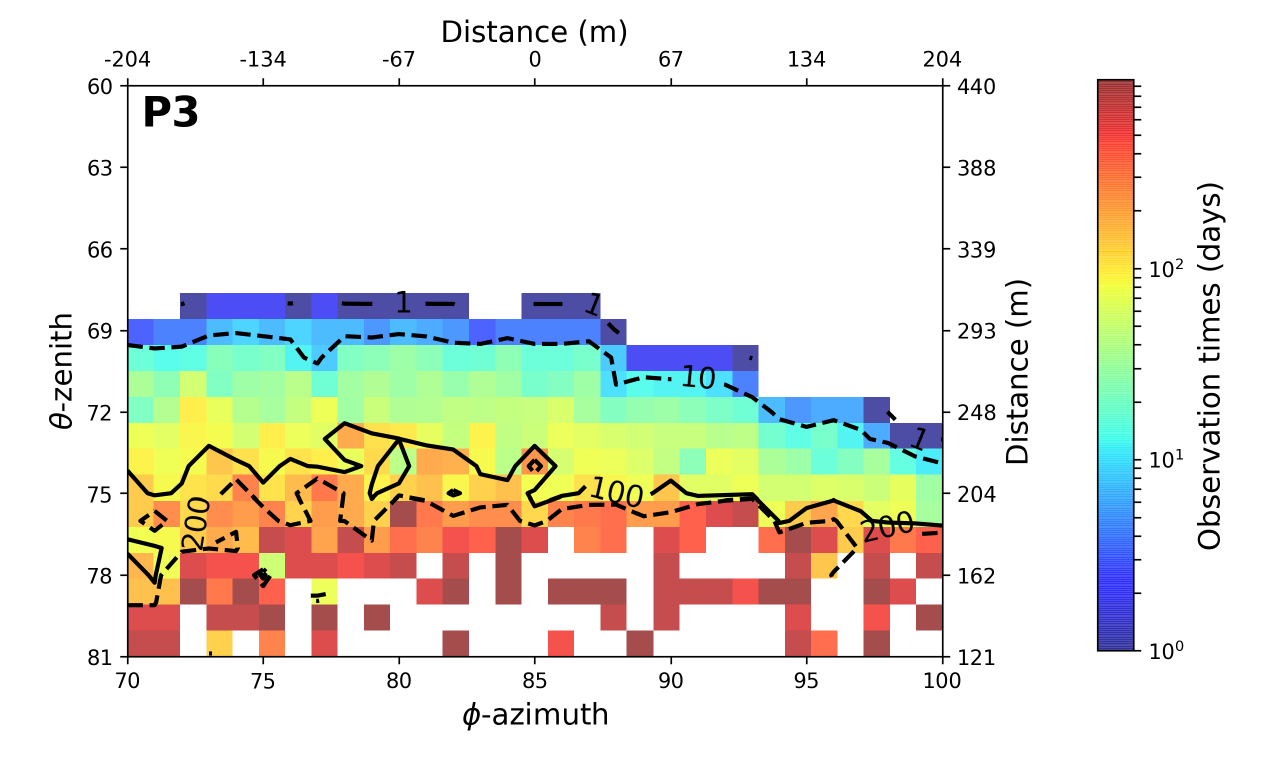}
\includegraphics[width=0.45\textwidth]{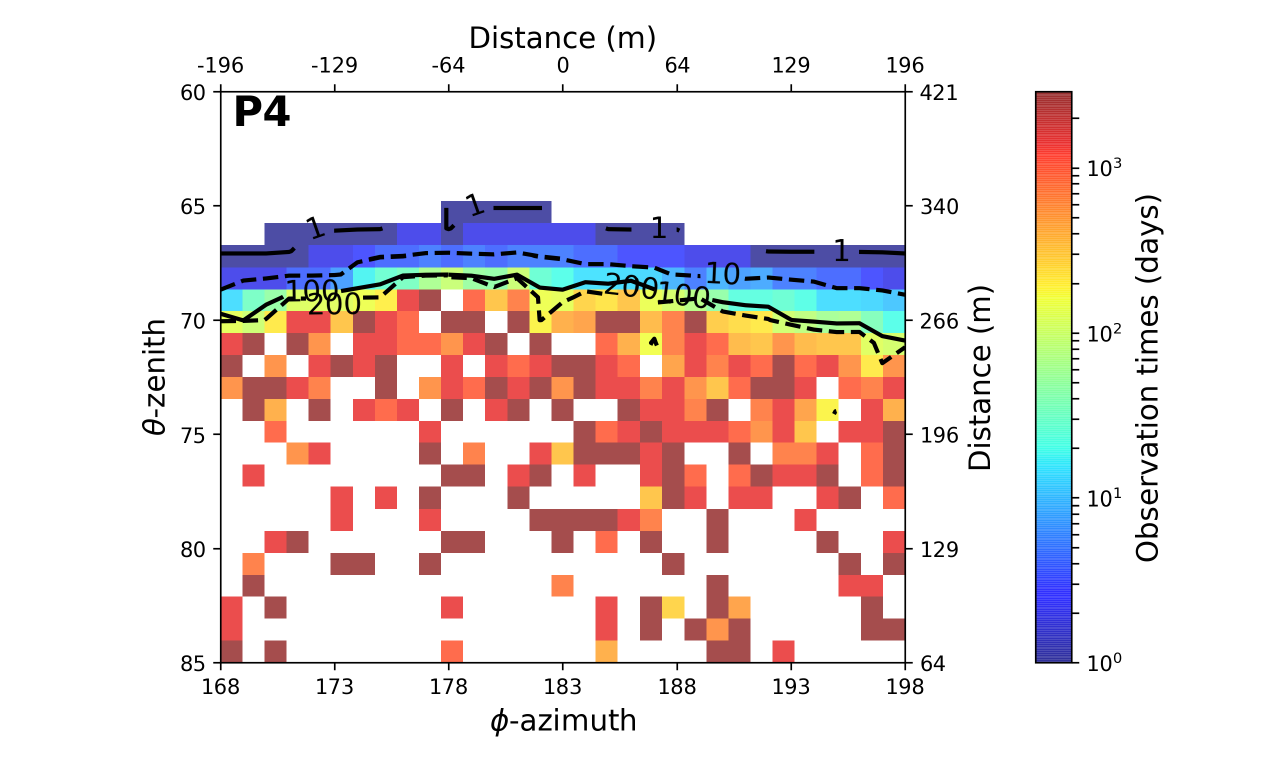}
    \caption{Expected observation times for $100$ muons at observation points {\bf P$_{1M}$}, {\bf P$_{2M}$}, {\bf P$_{3M}$} and {\bf P$_{4M}$} in Cerro Mach\'{\i}n. In this calculation we assume $6$cm$^{-2}$sr acceptance. The results agree that in order to detect $100$ muons in all directions an exposure time of $100$ days ($\sim$ 3 months) is necessary, achieving a definition at a depth of $150-160$ meters.}
  \label{time_volcano}
\end{figure}

As we have mentioned above, we set a minimum flux of $100\, \mu/$pixel and by using equation (\ref{Nmuons}) we estimate the minimum time exposure needed to examine the inner structures of the volcano edifice.  In figure \ref{time_volcano} we sketch contour lines representing the exposure times required, depending on the point of observation. If we assume an acceptance of $6$ cm$^{-2}$ sr., we will need at least $100$ days ($\sim$ three months) to explore a depth of $150-160$ meters.

\begin{figure}[!ht]
\centering
{\includegraphics[width=0.40\textwidth]{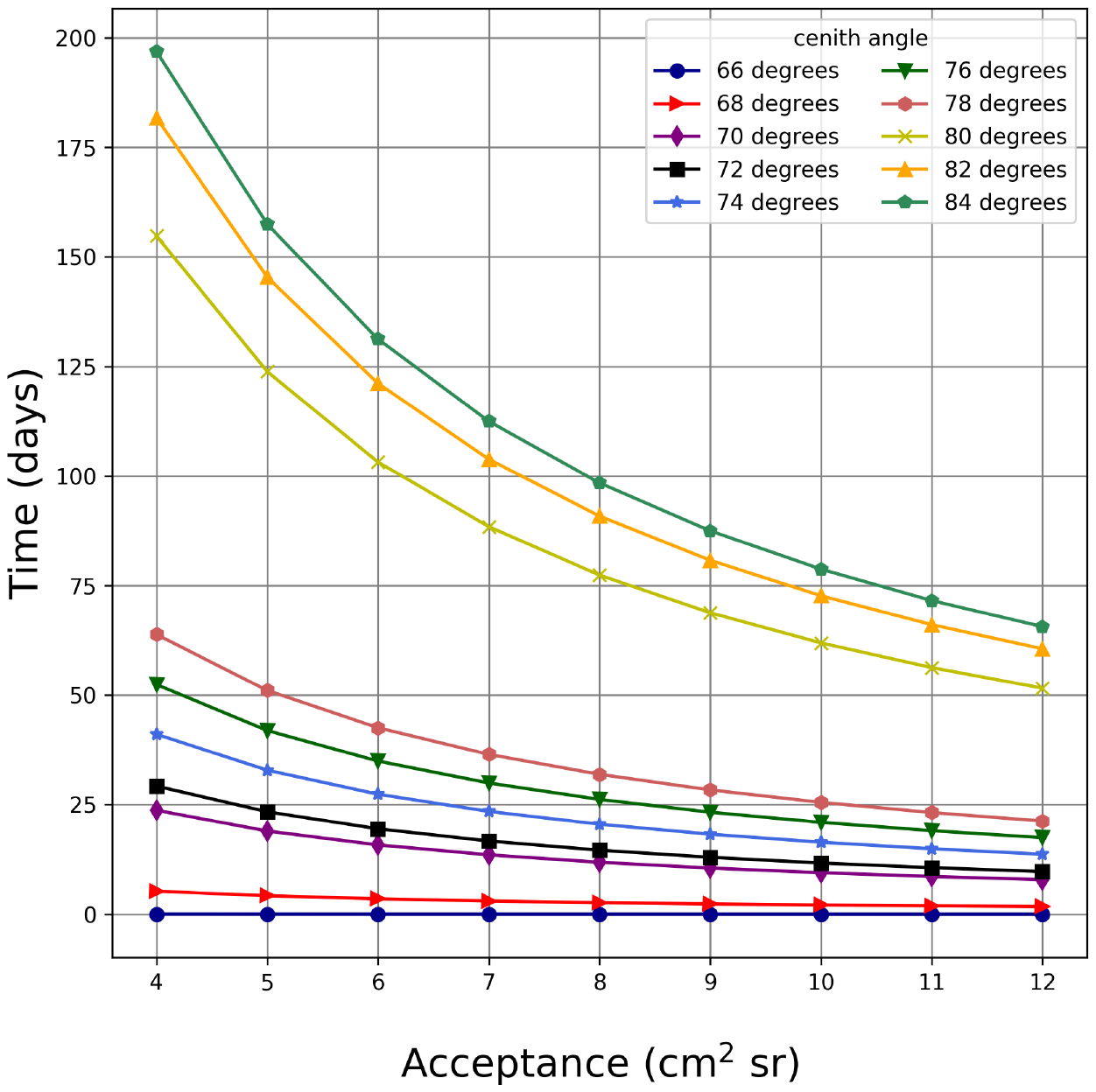}}
{\includegraphics[width=0.40\textwidth]{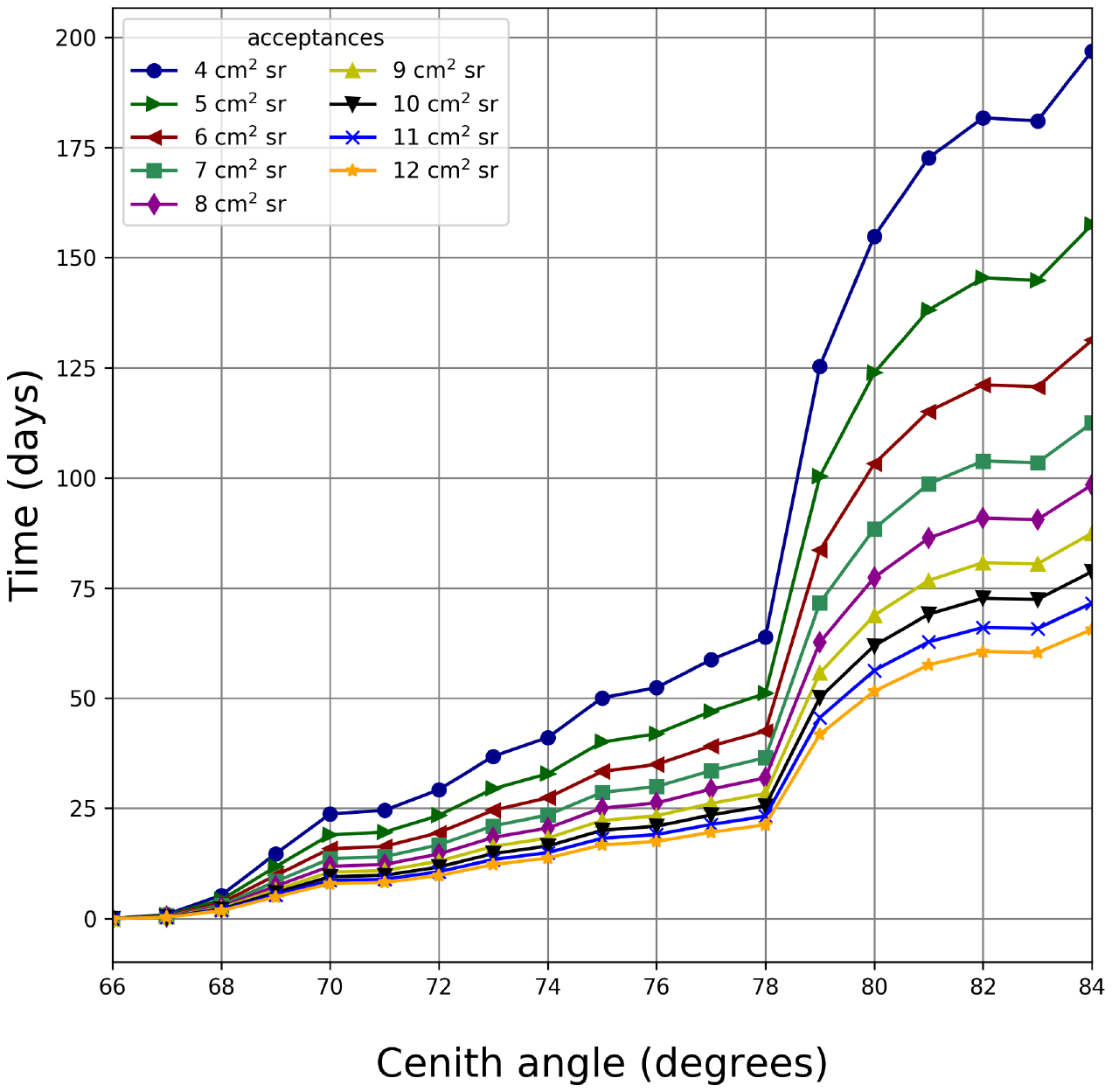}}
\caption{Exposure times for observation point \textbf{P$_{1M}$} at Cerro Mach\'{\i}n needed to identify differences of $-10\%$ in the averaged directional density for different zenith angles and telescope acceptance. We obtained exposure time lapses between two days and up to more than six months to achieve the desired density resolution, at different zenith angles.}
  \label{times}
\end{figure}

As explained before, exposure times, opacity (directional average density) and instrument resolution are related through equation (\ref{feasibility})  \cite{LesparreEtal2010}. This is another variable: the expected exposure times needed to resolve average density differences of $\approx 10$\%, are shown in figure \ref{times},  
for the zenithal range $66^\circ<\theta<84^\circ$. 

\subsubsection{Coulomb dispersion the Mach\'{\i}n volcano}

\subsection{Ray-tracing for Chiles, Cerro Negro and Galeras}
\label{RayTracing}

The idea in this Section is to look for potential observation points where the muon flux travel distance, inside the geological structures, is less than $1,500$m. It should be emphasized that almost all Colombian inland volcanoes are surrounded by other geological structures that screen the atmospheric muon flux. Thus only a few potential observation points are available.     

We examine the other three critical Colombian volcanoes using the ray-tracing technique based on the topography, already mentioned, above of the NASA global digital elevation model. First, the code returns the latitude and longitude coordinates in vectors within a matrix of elevation values and redefines the reference system (decimal degrees) to a local reference system (meters). Then we trace the muon trajectory considering in detail the topography around each volcano.

\subsubsection{Observation points at Chiles and Cerro negro volcanoes}
\begin{figure}[!h]
\centering
{\includegraphics[width=0.245\textwidth]{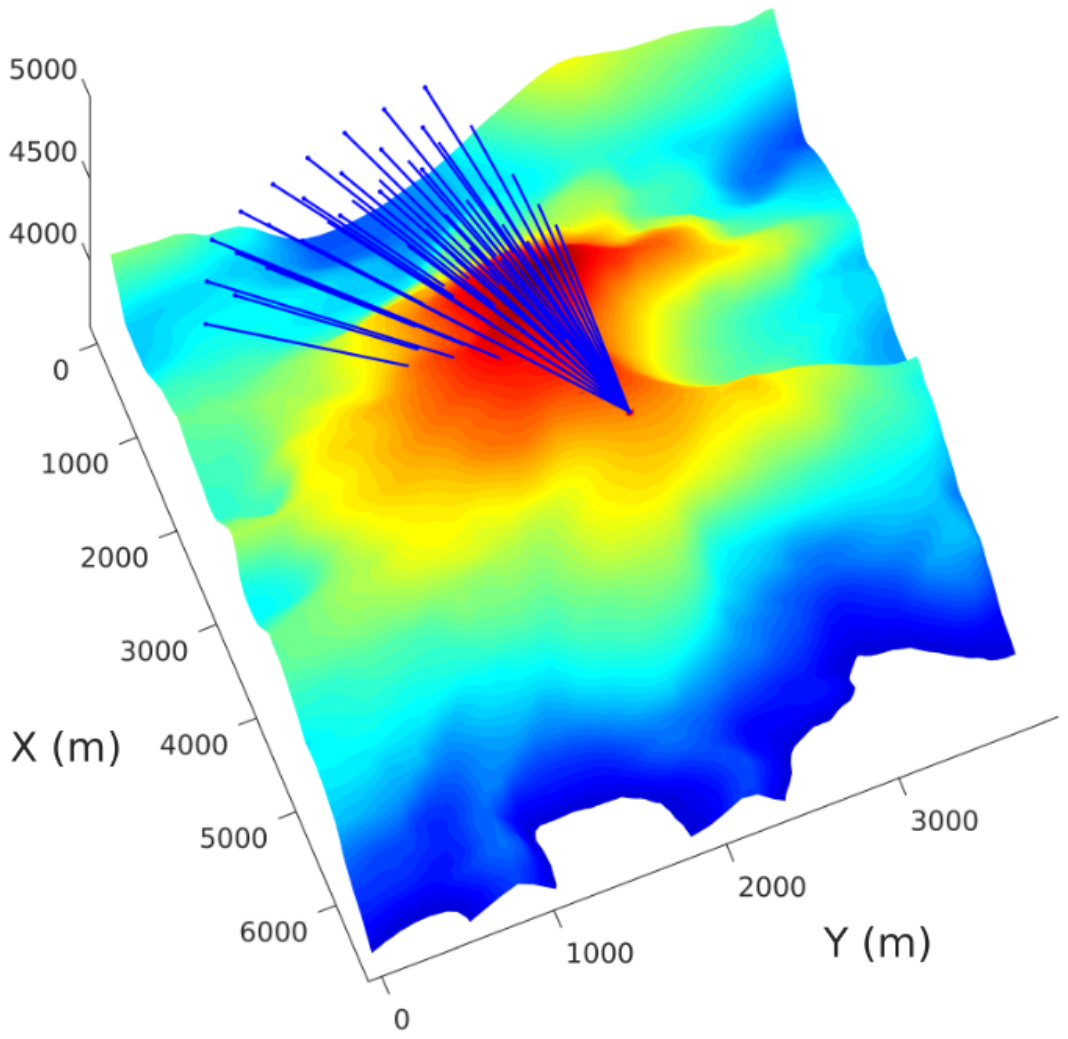}}
{\includegraphics[width=0.245\textwidth]{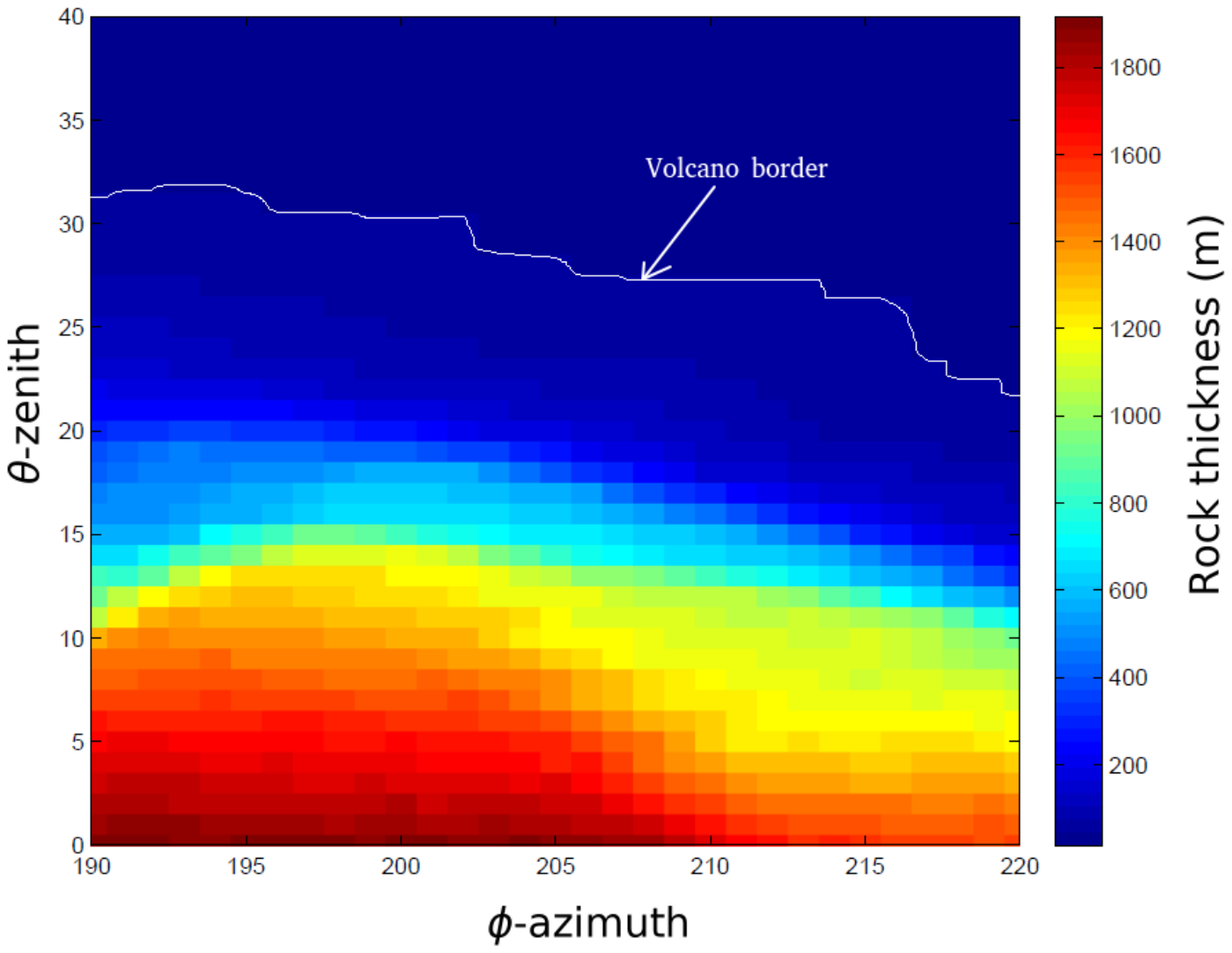}}
{\includegraphics[width=0.245\textwidth]{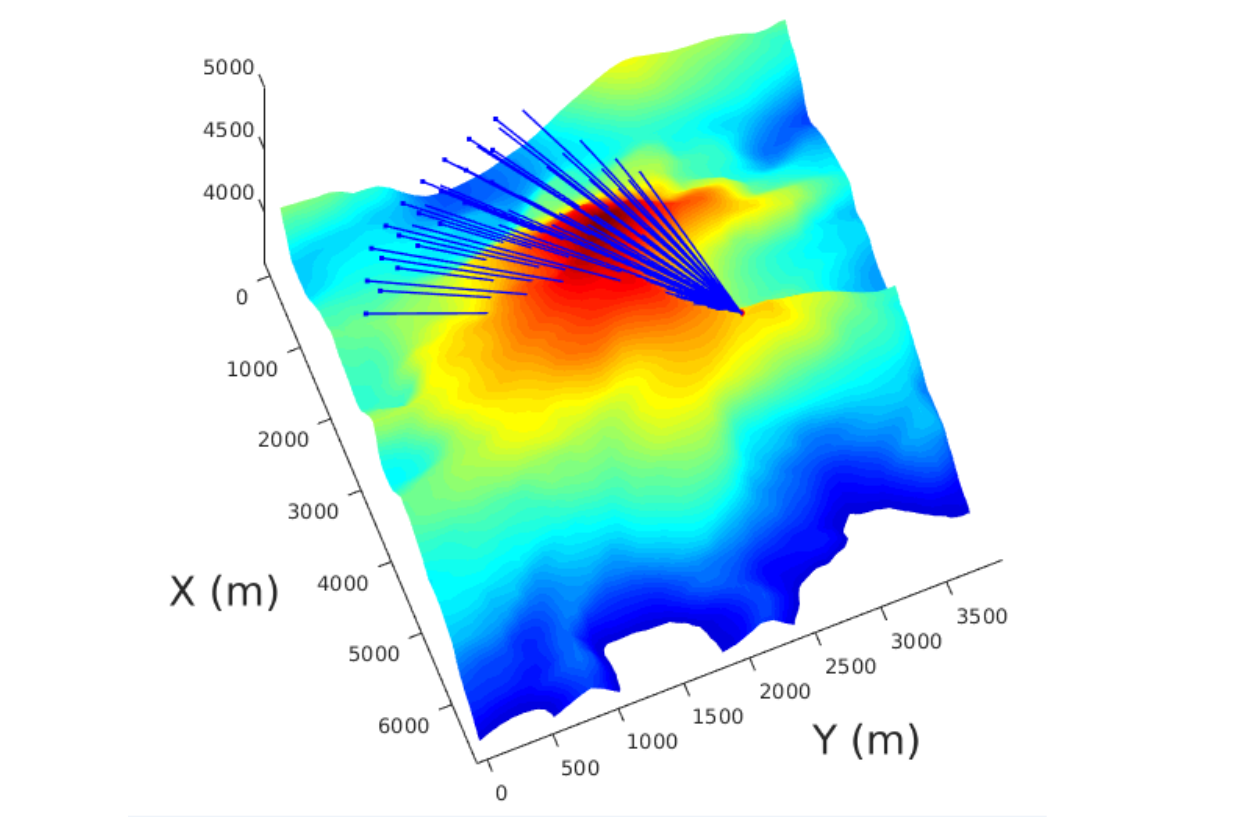}}
{\includegraphics[width=0.245\textwidth]{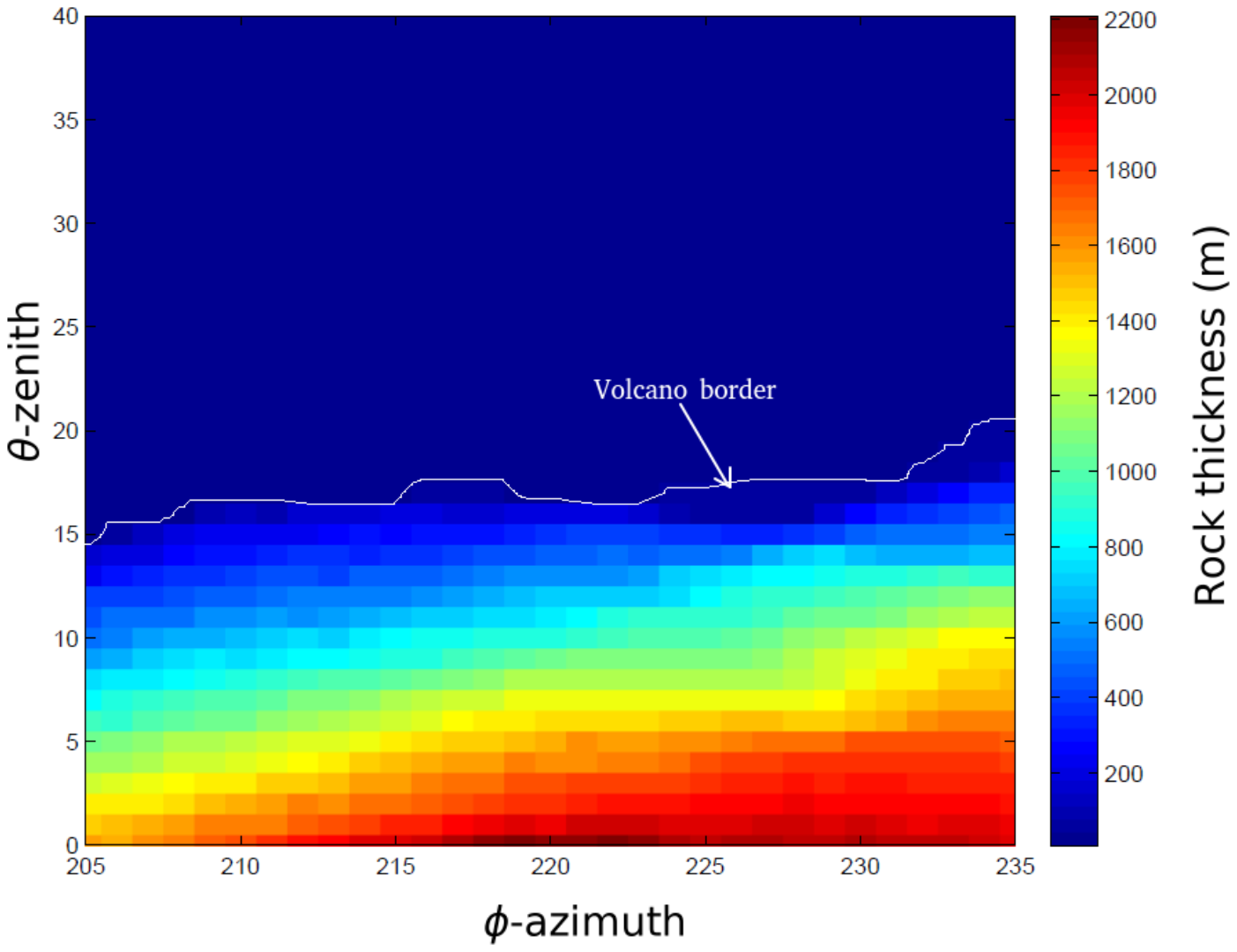}}\\
{\includegraphics[width=0.245\textwidth]{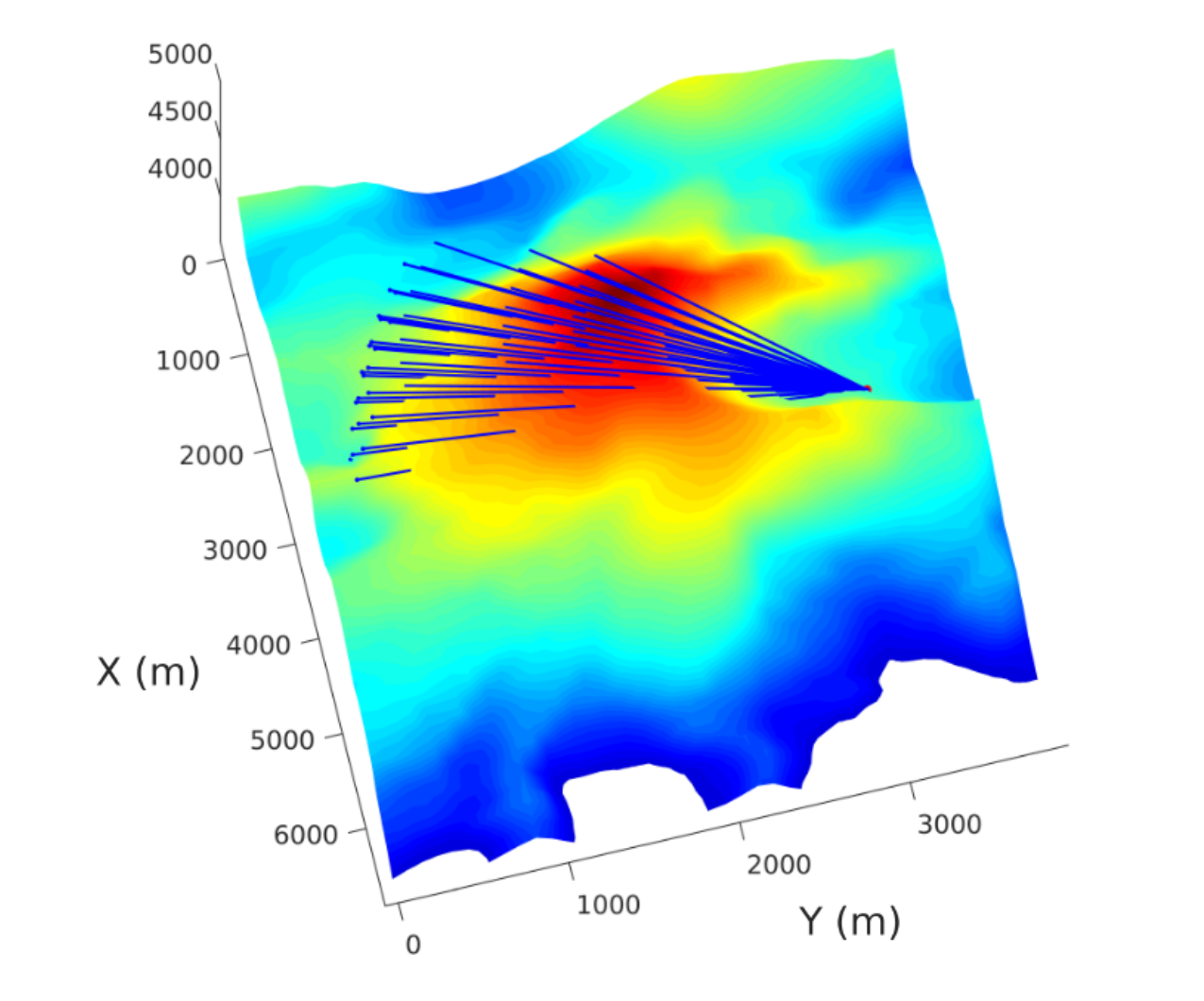}}
{\includegraphics[width=0.245\textwidth]{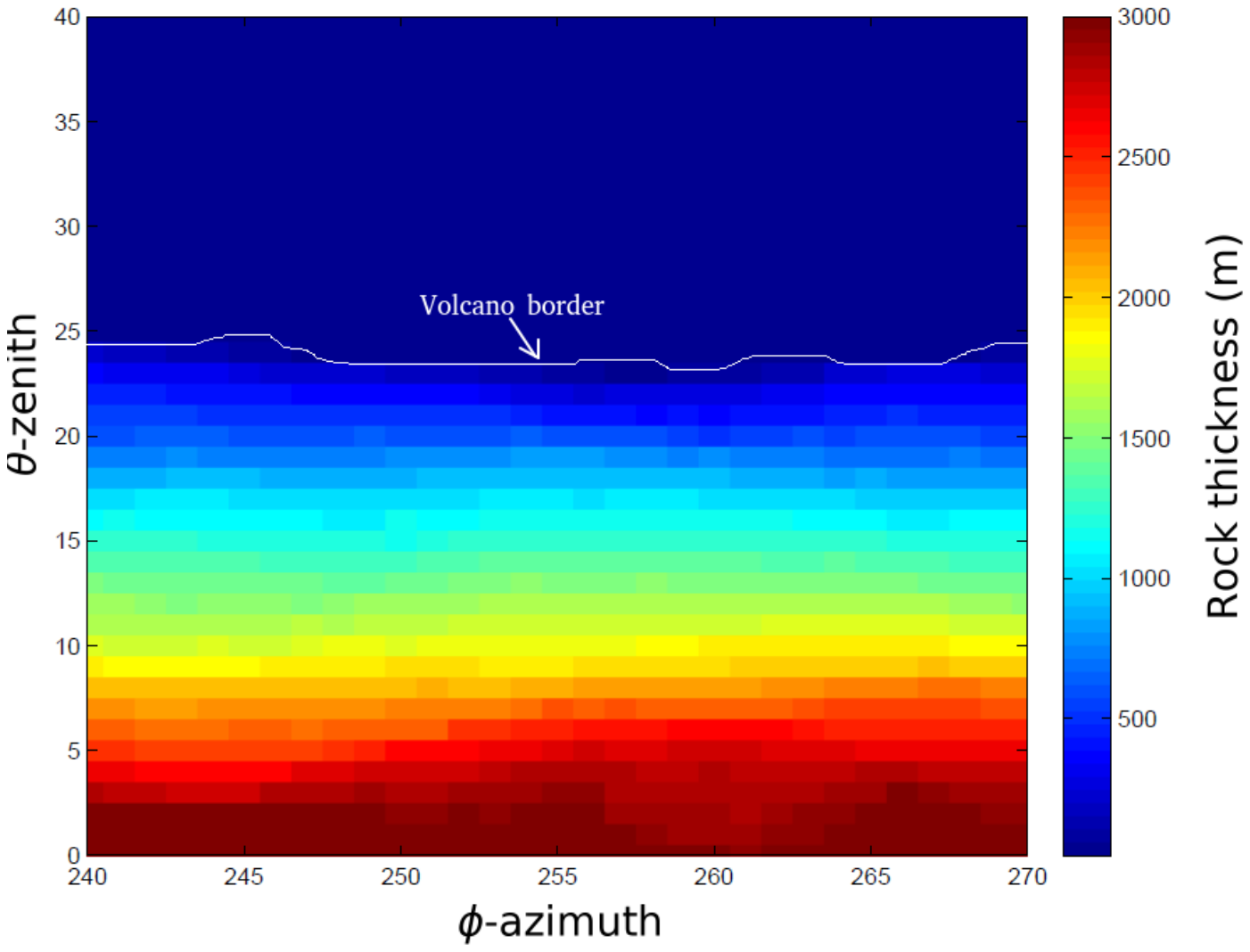}} 
{\includegraphics[width=0.245\textwidth]{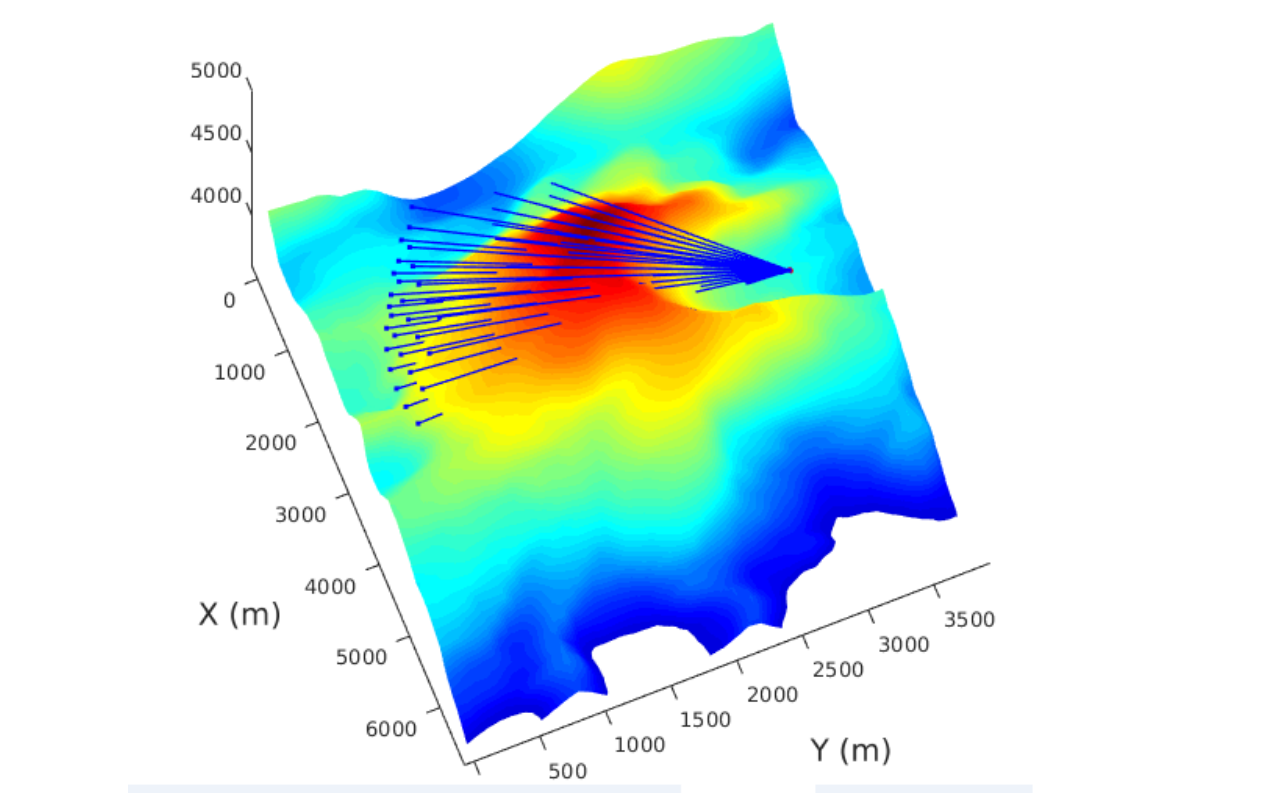}}
{\includegraphics[width=0.245\textwidth]{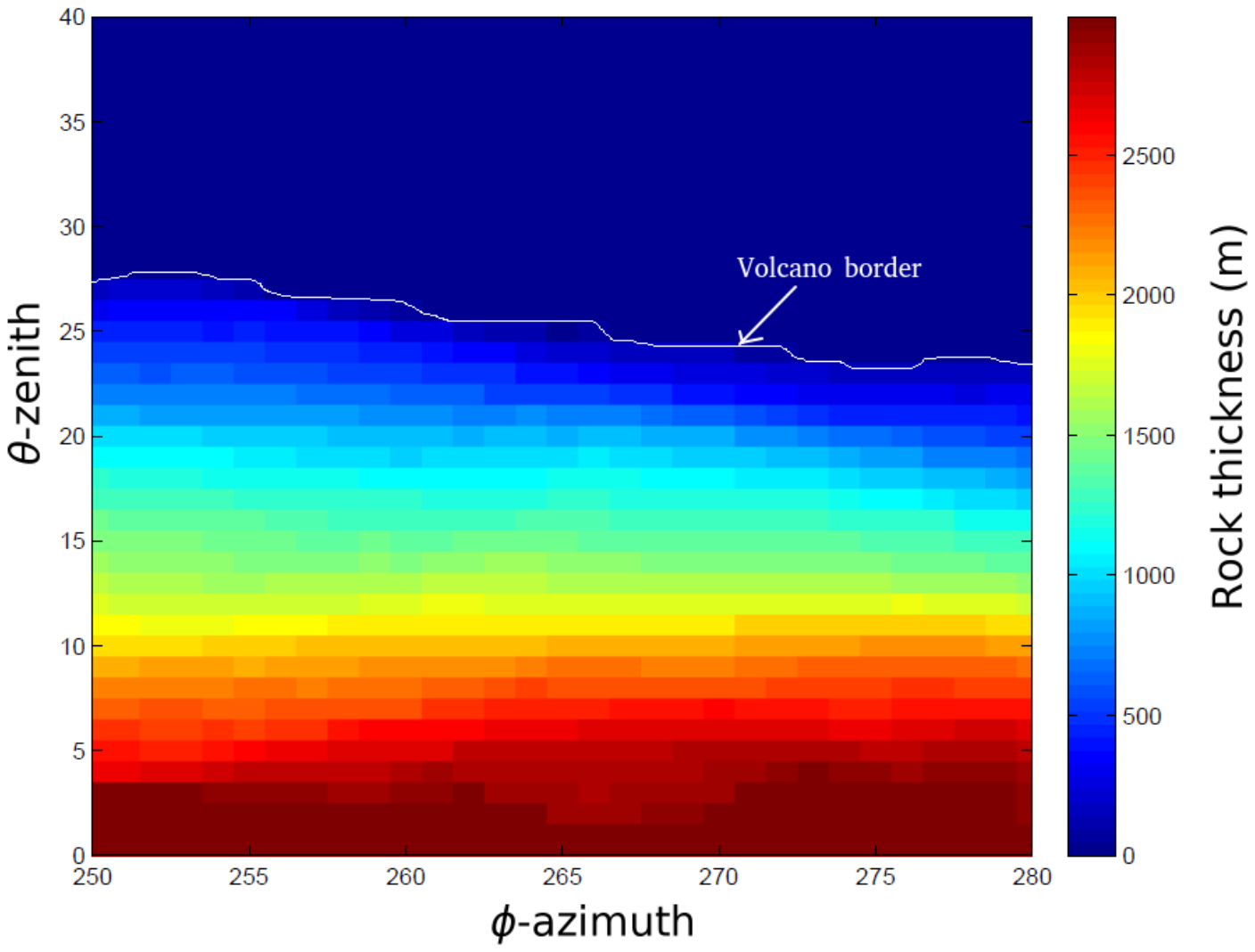}}
	\caption{Particle trajectories crossing  the Chiles volcano structure to observation point {\bf P$_{1Ch}$}, {\bf P$_{2Ch}$}, {\bf P$_{3Ch}$} and {\bf P$_{4Ch}$}. Points P$_{1Ch}$ and P$_{2Ch}$ are in areas that are difficult to access and points P$_{3Ch}$ and P$_{4Ch}$ in high risk areas}
  \label{DistanceTrajectoriesChiles}
\end{figure}

At Chiles we have identified four points  (see Table \ref{TableChiles}) and in figure \ref{DistanceTrajectoriesChiles} we can see the ray-tracing technique implemented for {\bf P$_{1Ch}$}, {\bf P$_{2Ch}$}, {\bf P$_{3Ch}$} and {\bf P$_{4Ch}$} to determine the distances of muon propagation through the volcano, as well as the angular distribution of these rays around the upper part of the volcano. Although the distances traveled by the muons to the four observation points are less than $1,500$m, the volcano was discarded due to the difficult access to these points. 

\label{ChilesObservationPoints}
\begin{table}[!ht]
\centering
\begin{tabular}{lllll}
\hline
\textbf{Chiles points}        & \textbf{P$_{1Ch}$}& \textbf{P$_{2Ch}$} & \textbf{P$_{3Ch}$} & \textbf{P$_{4Ch}$} \\ \hline
\textbf{Latitude  ($^{\circ}$N)}        & 0.819431           & 0.823917         & 0.829098        & 0.829642     \\
\textbf{Longitude ($^{\circ}$W)}        & -77.926883         & -77.927034       & -77.930539       & -77.933769    \\
\textbf{Distance to edifice center (m)} & 1050               & 1250              & 1444              & 1450       \\ 
\hline
\end{tabular}
	\caption{Feasible observation points at Chiles volcano (0$^{\circ}$49'16.32"N,$\;$ -77$^{\circ}$56'6.13"W). }
\label{TableChiles}
\end{table}

\begin{figure}[!ht]
\centering
\includegraphics[width=0.245\textwidth]{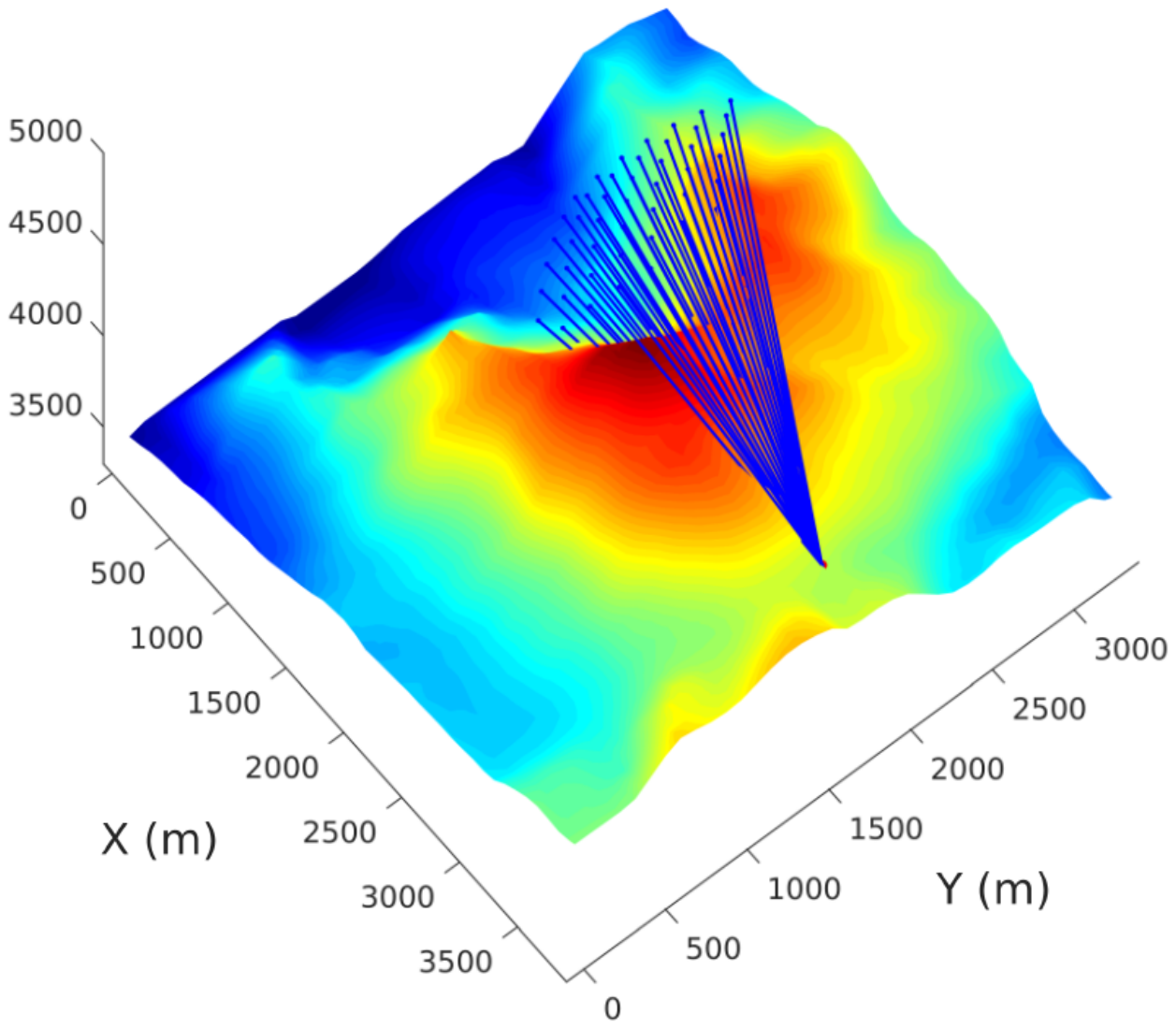}
\includegraphics[width=0.245\textwidth]{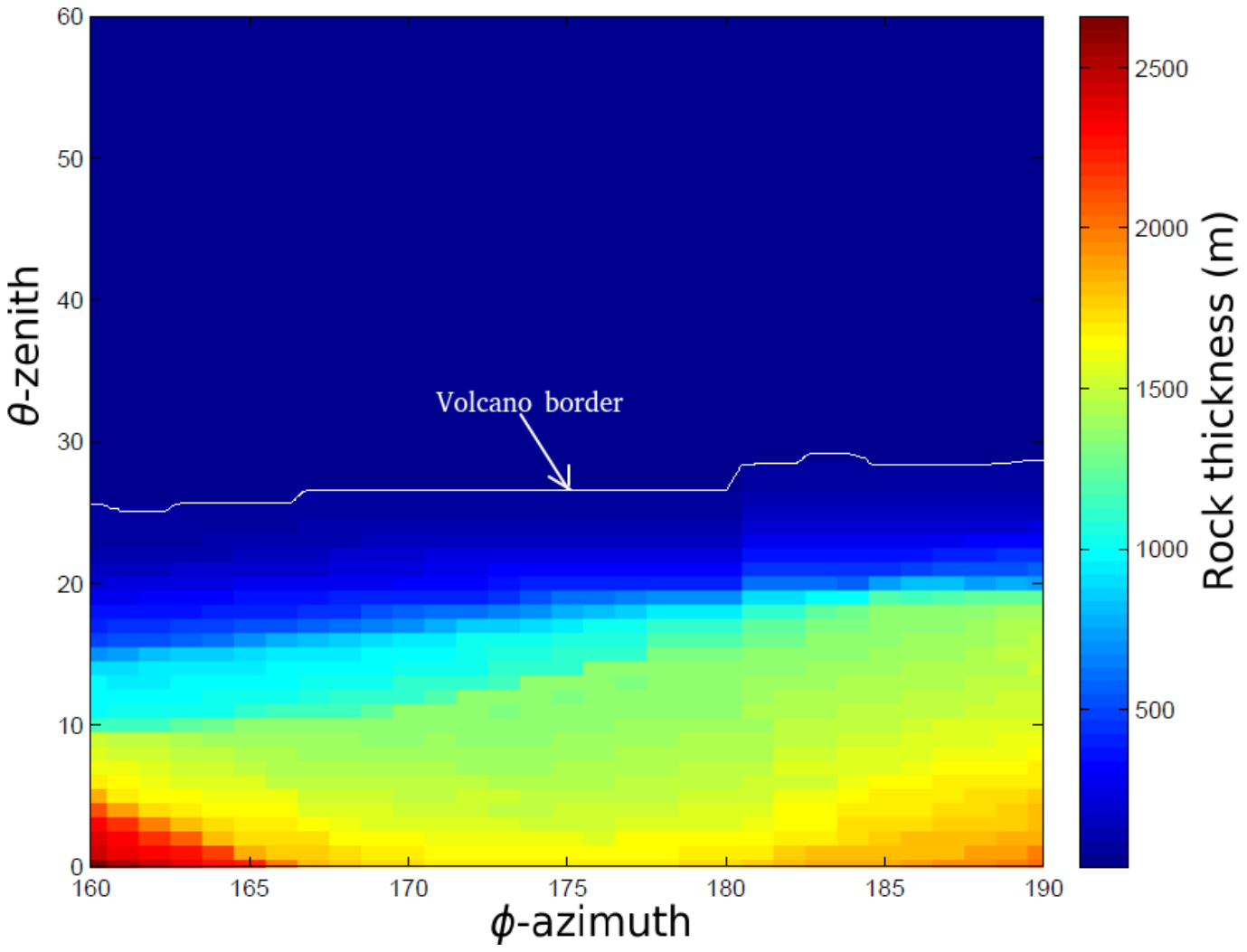}
\includegraphics[width=0.245\textwidth]{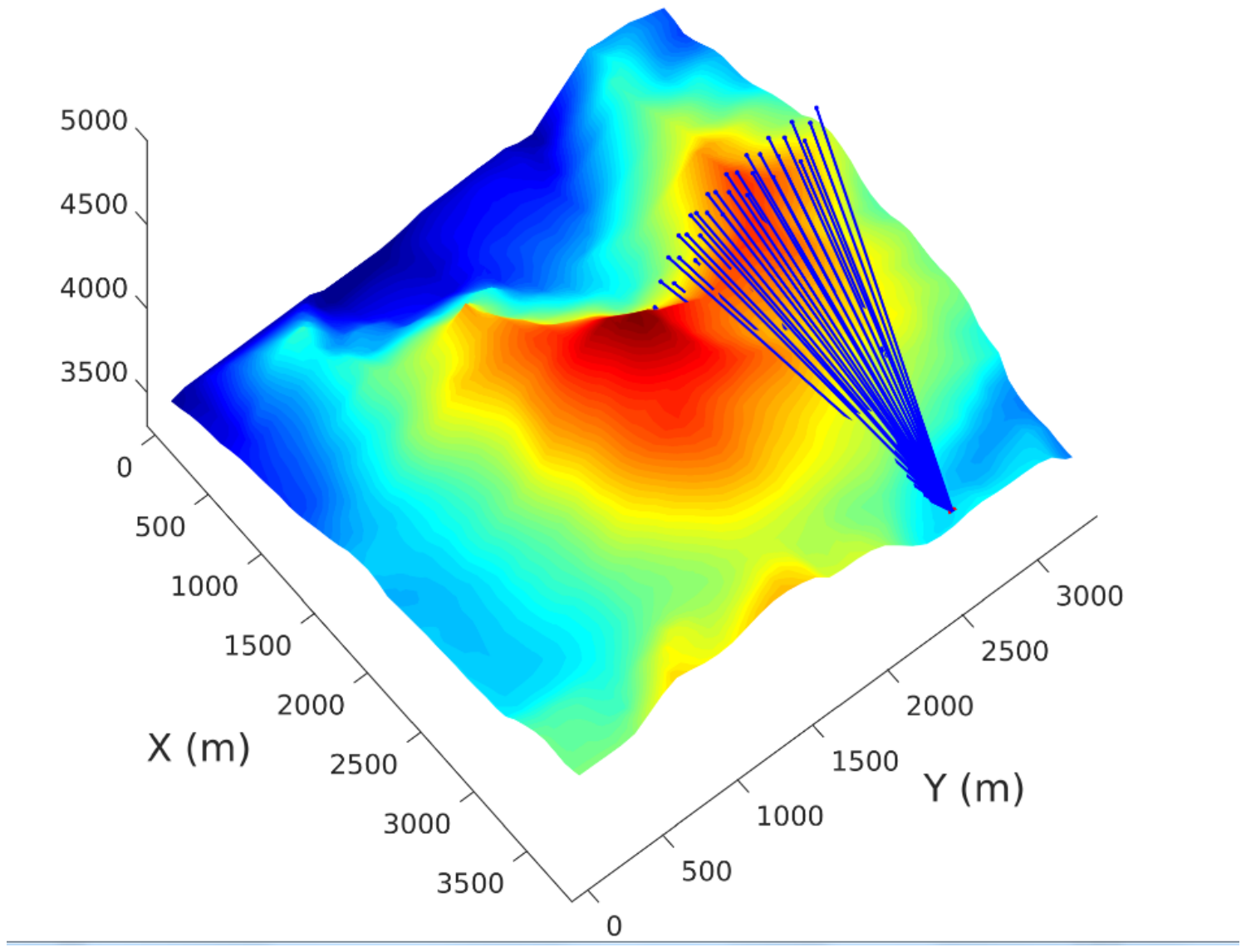}
\includegraphics[width=0.245\textwidth]{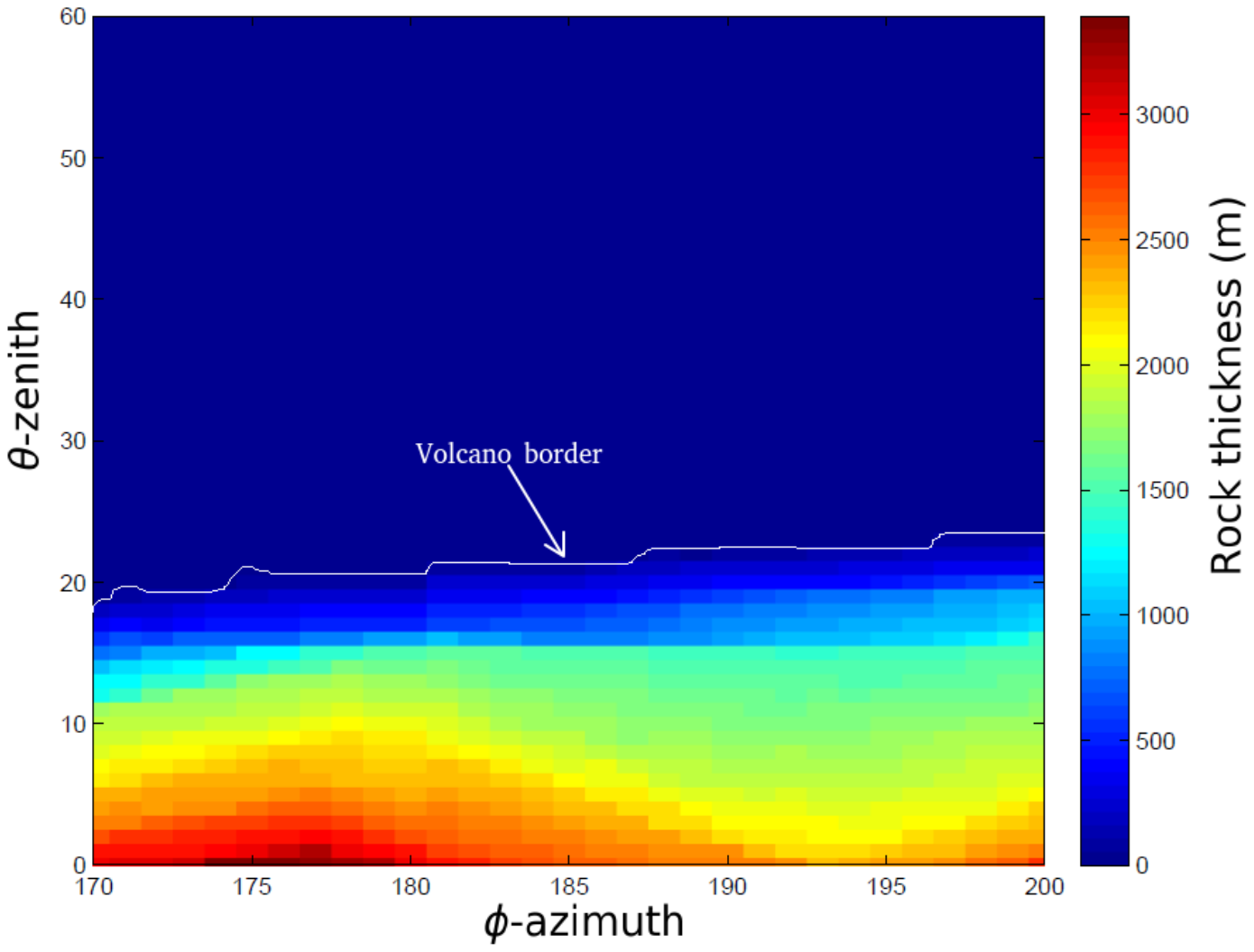}\\
\includegraphics[width=0.245\textwidth]{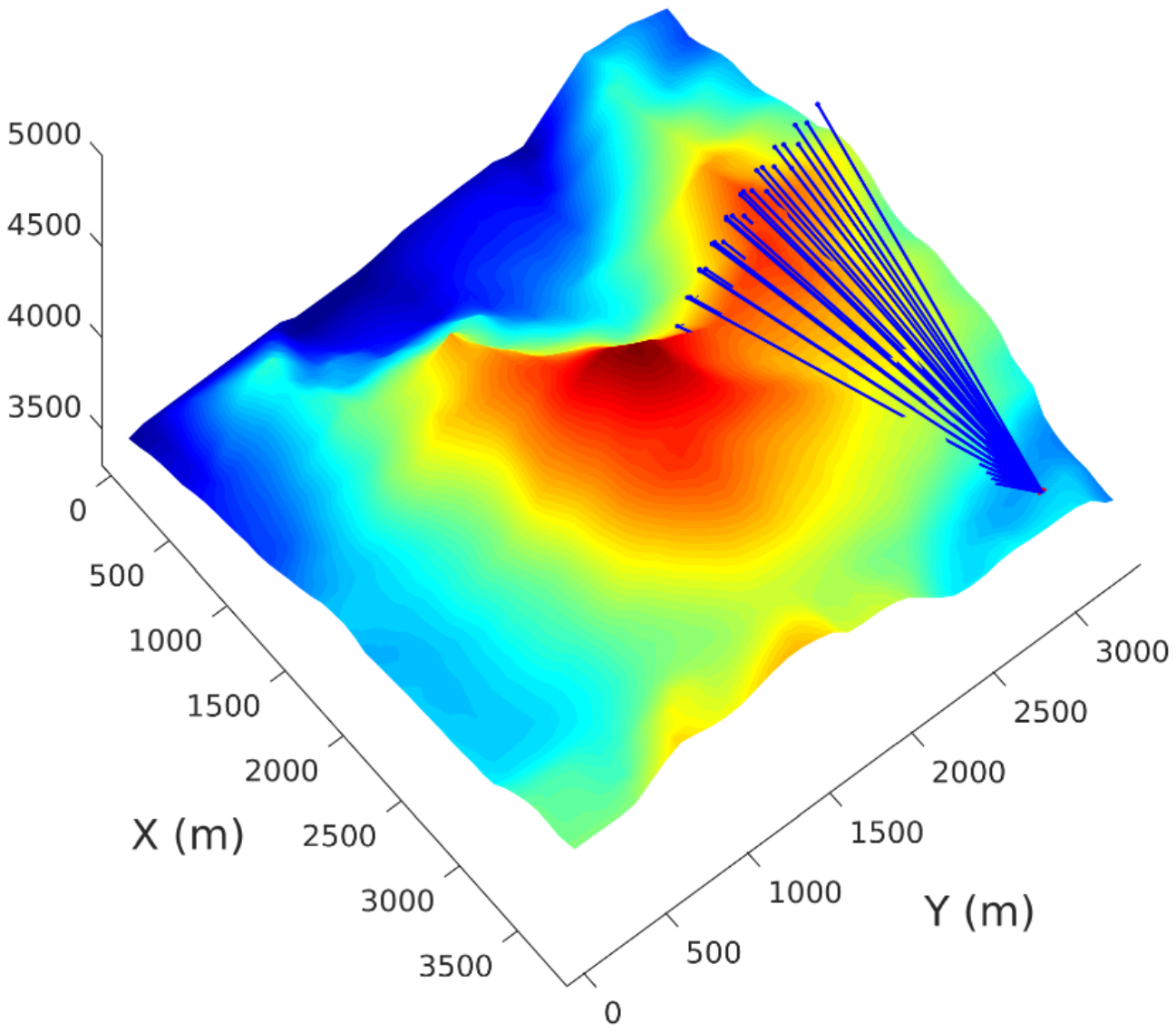}
\includegraphics[width=0.245\textwidth]{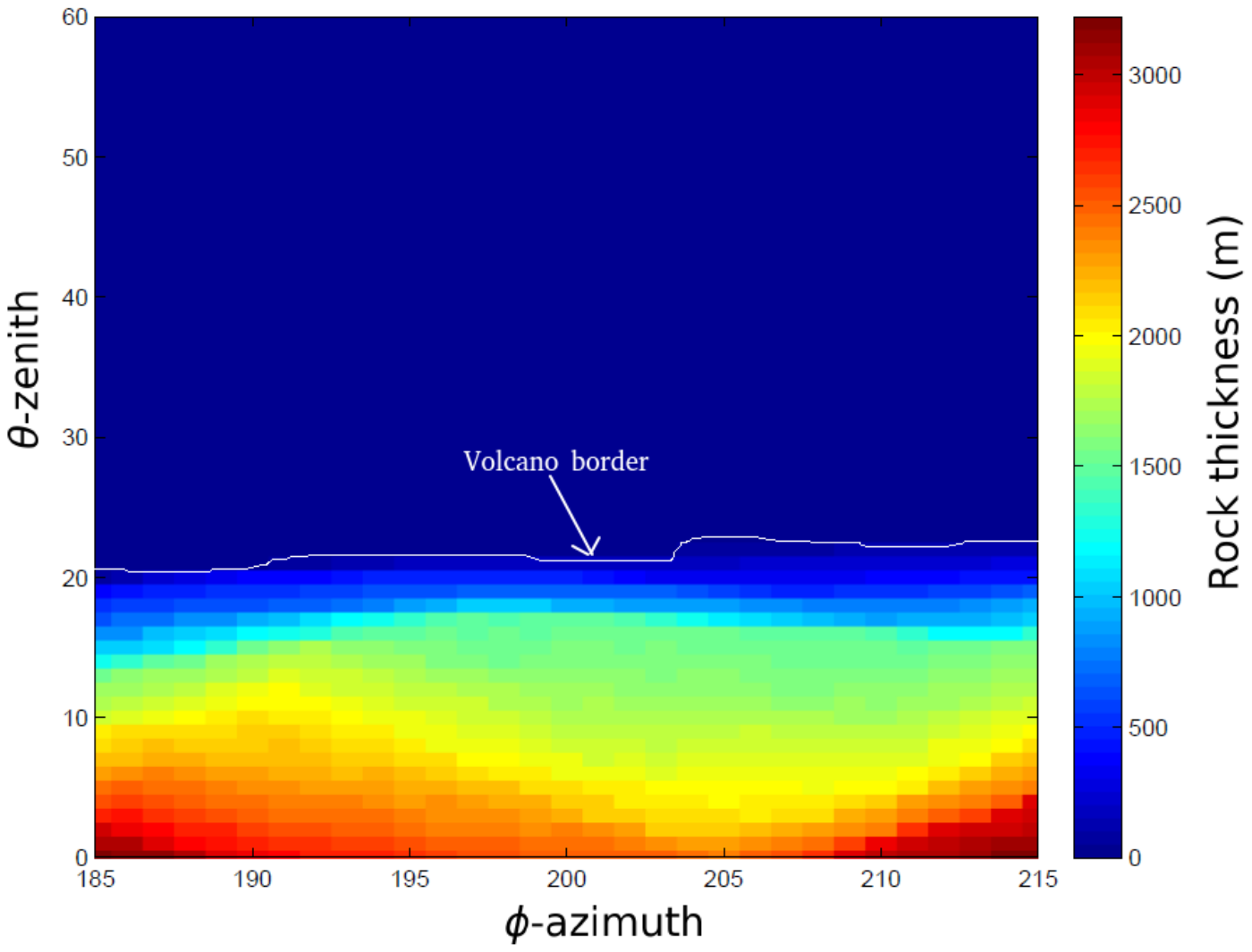}
\includegraphics[width=0.245\textwidth]{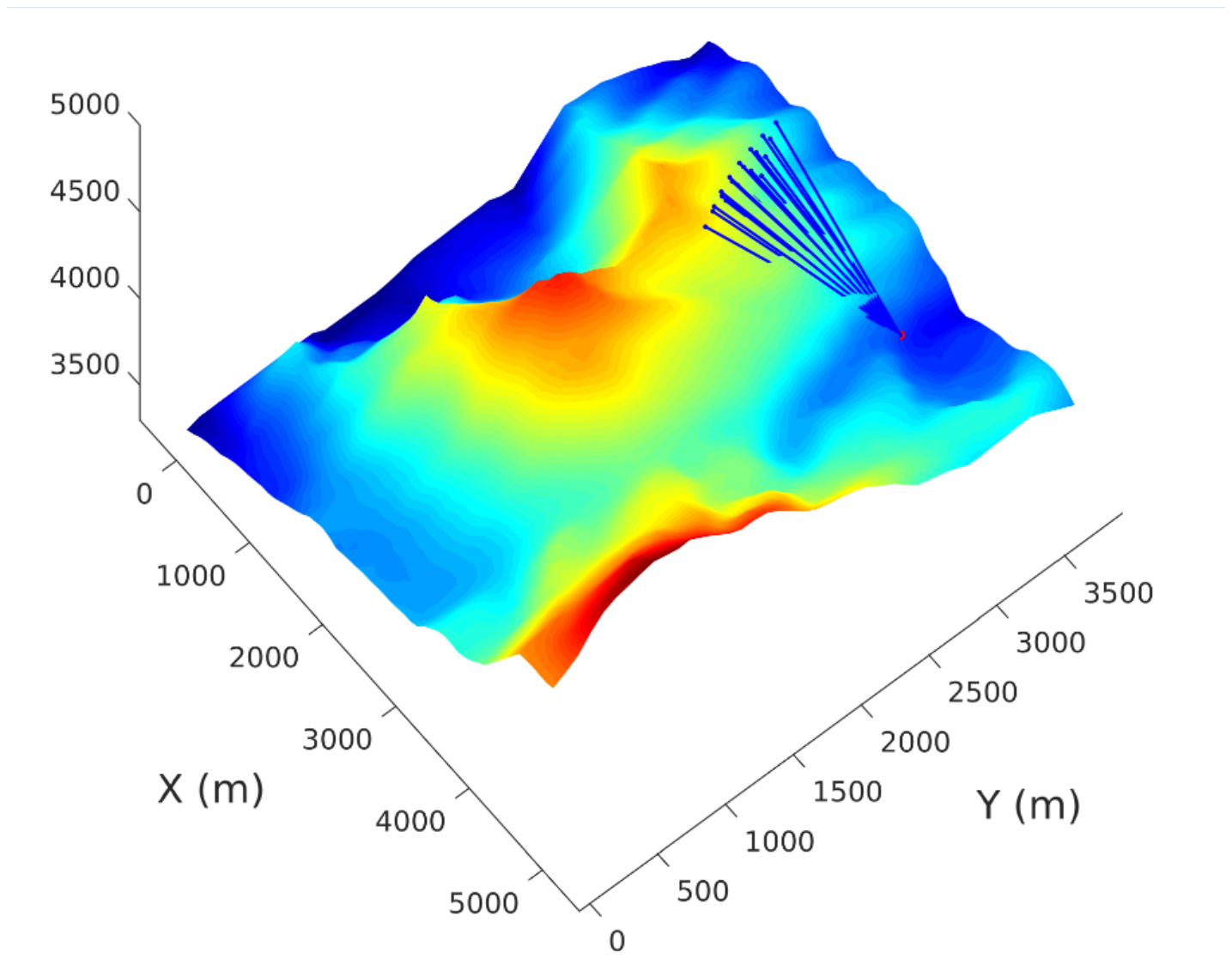}
\includegraphics[width=0.245\textwidth]{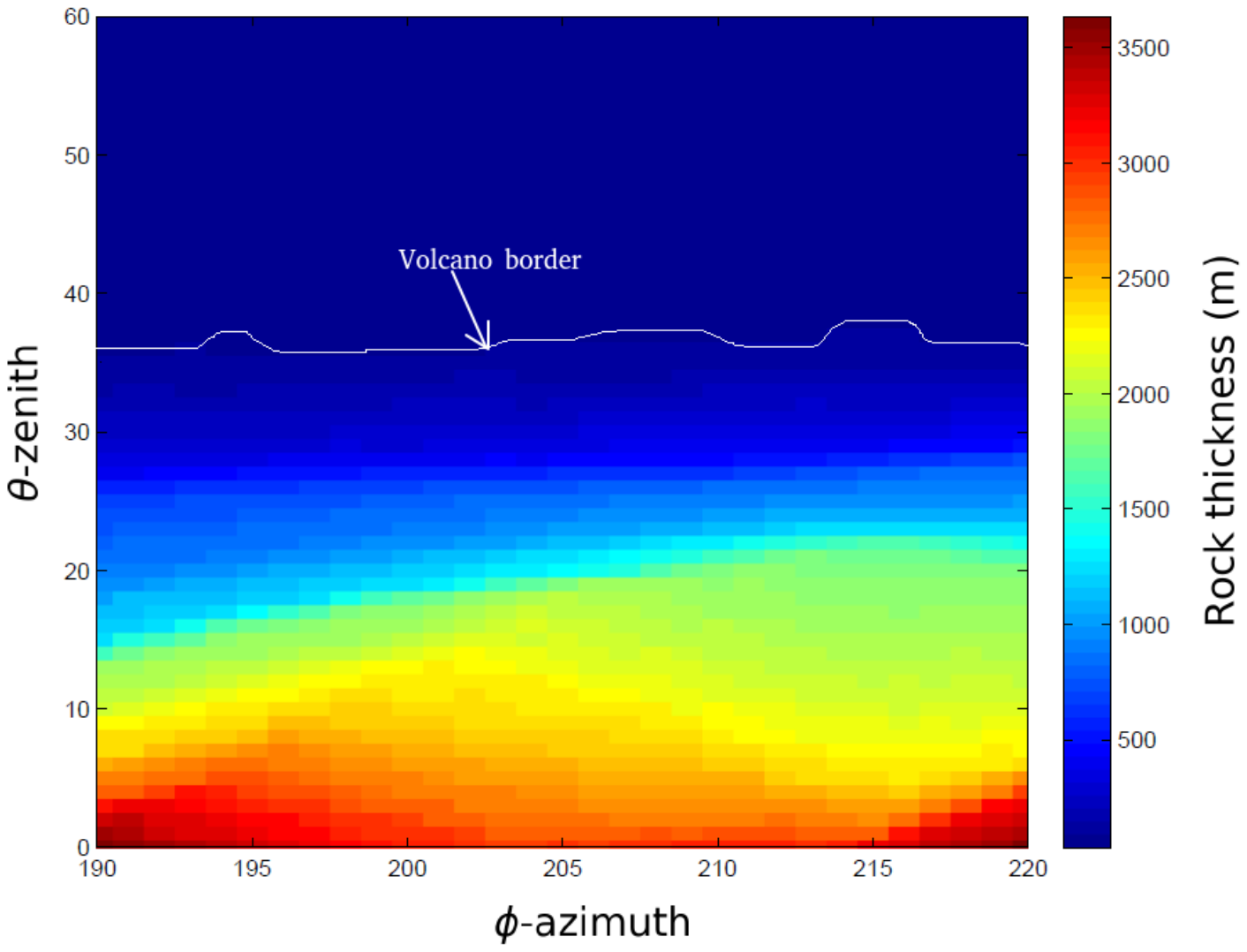}
	\caption{Particle trajectories crossing the Cerro Negro volcanic structure to the observation point {\bf P$_{1CN}$}, {\bf P$_{2CN}$}, {\bf P$_{3CN}$} and {\bf P$_{4CN}$}.}
  \label{DistanceTrajectoriesCerroNegro}
\end{figure}

In the case of Cerro Negro we have identified four points around it  (see Table \ref{TableCerroNegro}).
\begin{table}[!h]
\centering
\begin{tabular}{lllll}
\hline
\textbf{Cerro Negro points}        & \textbf{P$_{1CN}$}& \textbf{P$_{2CN}$} & \textbf{P$_{3CN}$} & \textbf{P$_{4CN}$} \\ \hline
\textbf{Latitude  ($^{\circ}$N)}        & 0.826250           & 0.832924         & 0.837090        & 0.840811     \\
\textbf{Longitude ($^{\circ}$W)}        & -77.954136         & -77.951177       & -77.952412       & -77.954454    \\
\textbf{Distance to edifice center (m)} & 1514               & 1912              & 1961              & 1982       \\ 
\hline
\end{tabular}
	\caption{Feasible observation points at Cerro Negro volcano (4$^{\circ}$29'23.08"N,$\;$ -75$^{\circ}$23'15.39"W). }
\label{TableCerroNegro}
\end{table}
In figure \ref{DistanceTrajectoriesCerroNegro} we can see the ray-tracing technique implemented for {\bf P$_{1CN}$}, {\bf P$_{2CN}$}, {\bf P$_{3CN}$} and {\bf P$_{4CN}$} to determine the distances of muon propagation through the Cerro Negro volcano, as well as the angular distribution of these rays around the upper part of the volcano. Notice that to obtain a reasonable muon flux the telescope should be tilted about seven degrees making the depth of investigation very short. 

Although the distances traveled by the muons measured from points P$_{1CN}$-P$_{4CN}$ are less than $1,500$m, this volcano, as well as Chiles, are discarded due to the difficult access to their potential observation points. The Chiles and Cerro Negro volcanoes are on the Colombia-Ecuador border, an intricate zone, where the safety of the scientific personnel and equipment is not guaranteed by either country. 

\subsubsection{Observation points at Galeras volcano}
Galeras volcano has been active for more than a million years. Its most recent phase of activity began about 4500 years ago and included six significant eruptions. After quiescence dating from 1948, Galeras renewed its activity in 1988 with the emplacement of a dome, which was destroyed in an explosion in 1992\cite{OrdonezCepeda1997}.
On January 14, 1993, an explosion in Galeras crater killed six visiting scientists and three tourist\cite{SeidlEtal2003}. 

The Galera's intense activity and its proximity to an important city, generate a significant interest to study this Volcano with all techniques available and muography do not escape from this pressure. Several local researchers have made a notable effort in modeling muon transport properties across this dangerous volcano\cite{TapiaEtal2016,GuerreroEtal2019,TorresEtal2019}, and some of them even proposed a design for a muon telescope\cite{GuerreroEtal2019}. It is not clear which are the geographical coordinates where they intend to place the instrument and what will be the typical exposures times so as to have statistically reasonable counts.

\begin{figure}[!ht]
\centering
{\includegraphics[width=0.6\textwidth]{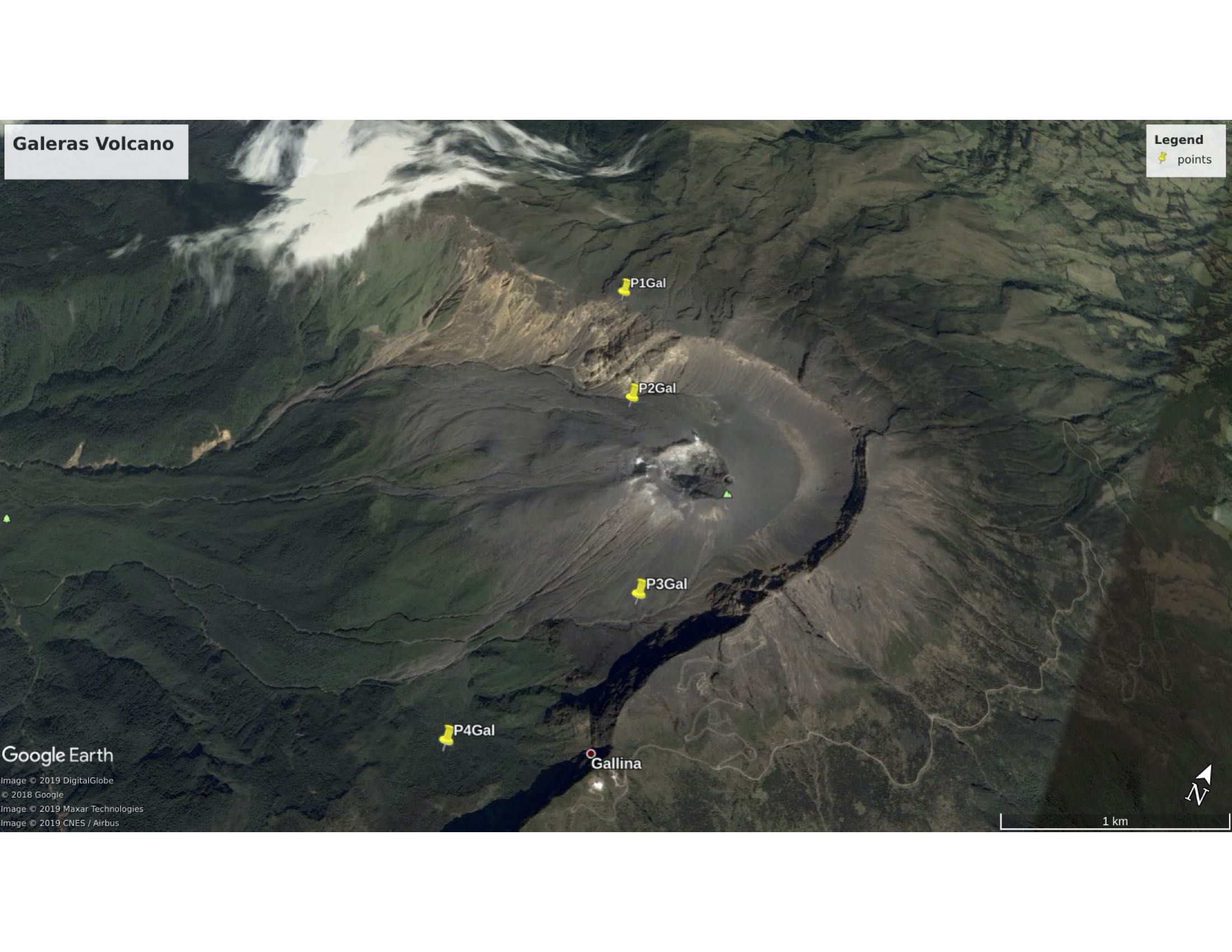}}
	\caption{Feasible observation points at Galeras volcano (1$^{\circ}$13'16.02"N,$\;$ 77$^{\circ}$21'33.09"W).} 
  \label{galerasgeo}
\end{figure}

\begin{figure}[!ht]
\centering
\includegraphics[width=0.245\textwidth]{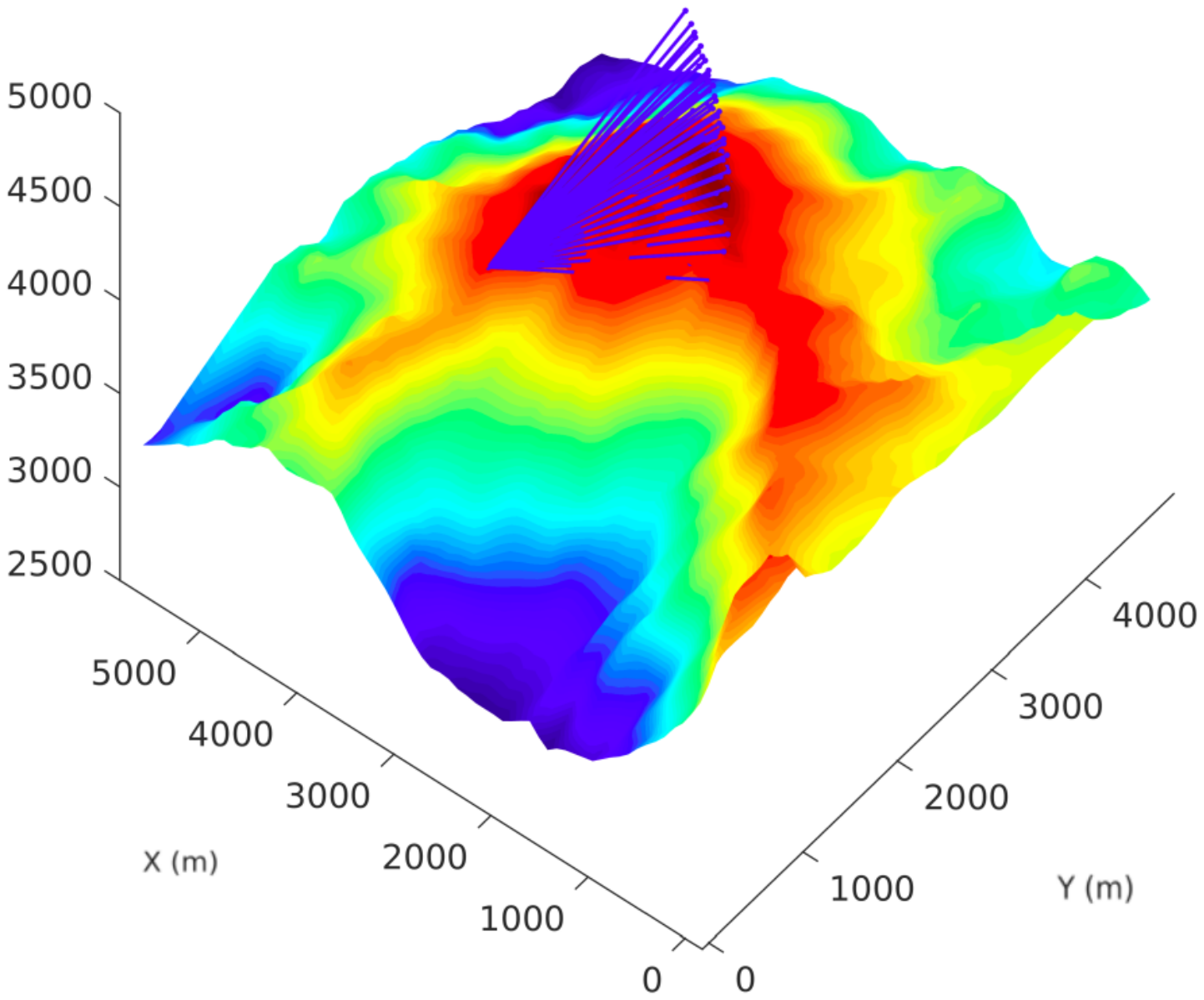}
\includegraphics[width=0.245\textwidth]{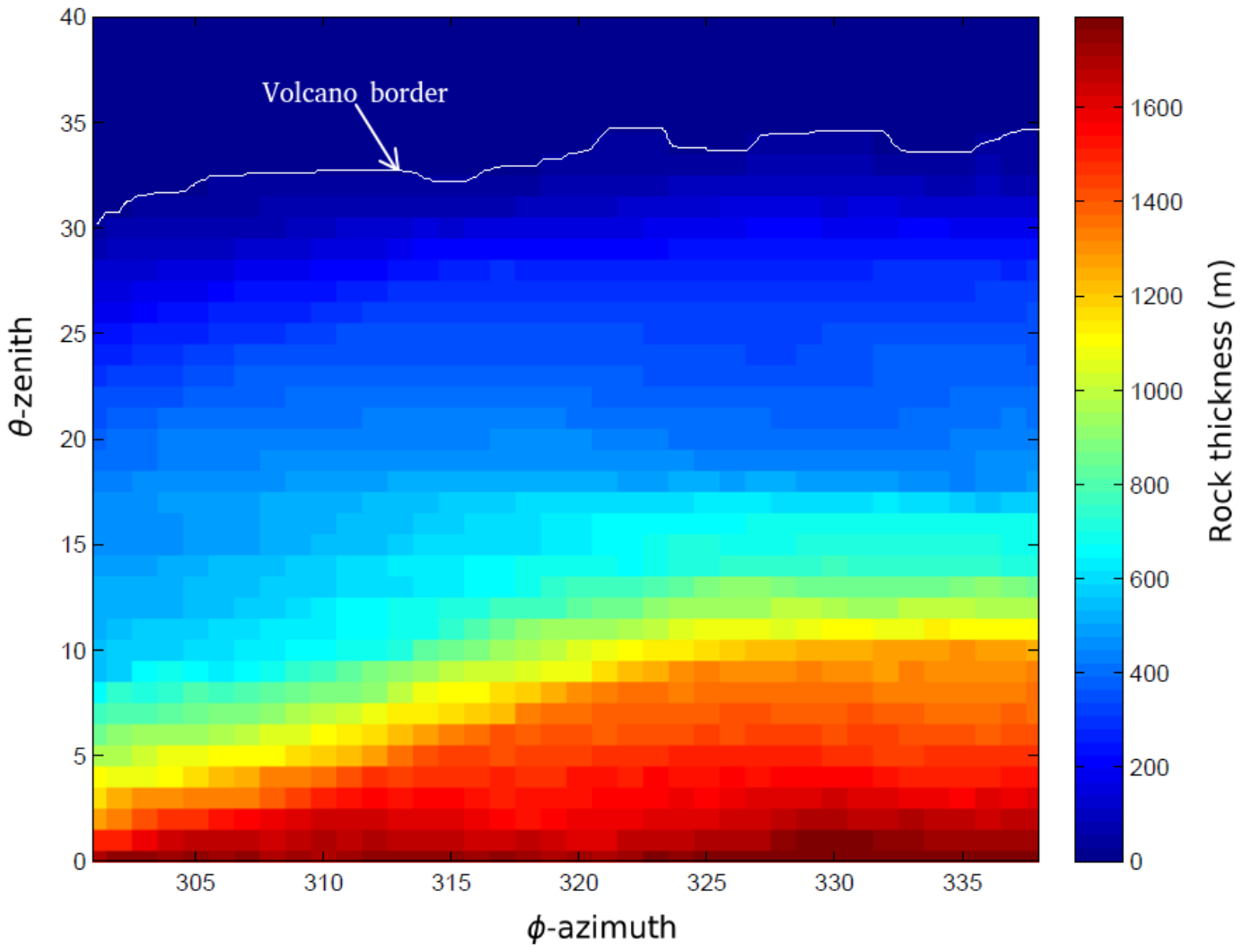}
\includegraphics[width=0.245\textwidth]{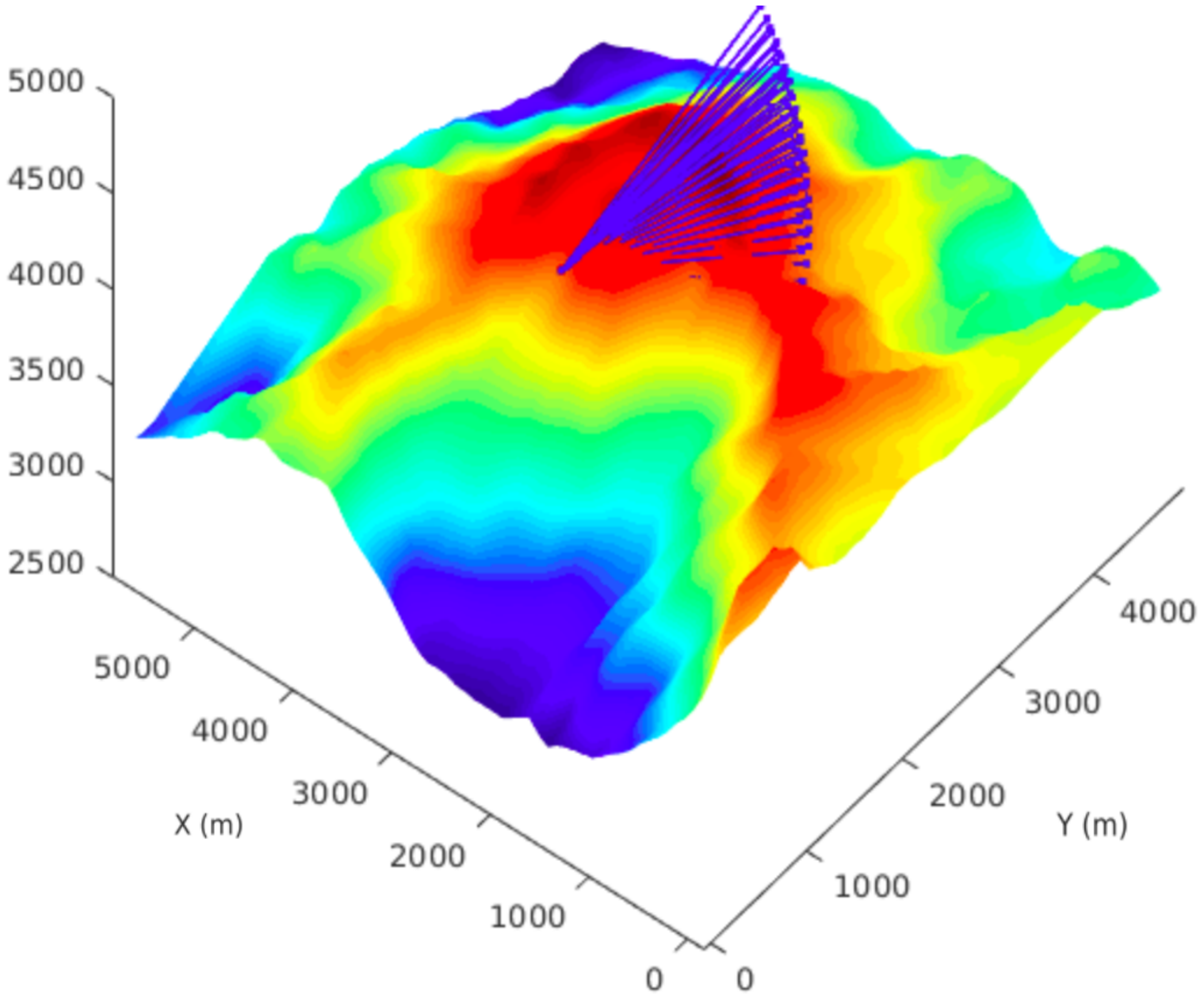}
\includegraphics[width=0.245\textwidth]{Figures/Distance2GAL}\\
\includegraphics[width=0.245\textwidth]{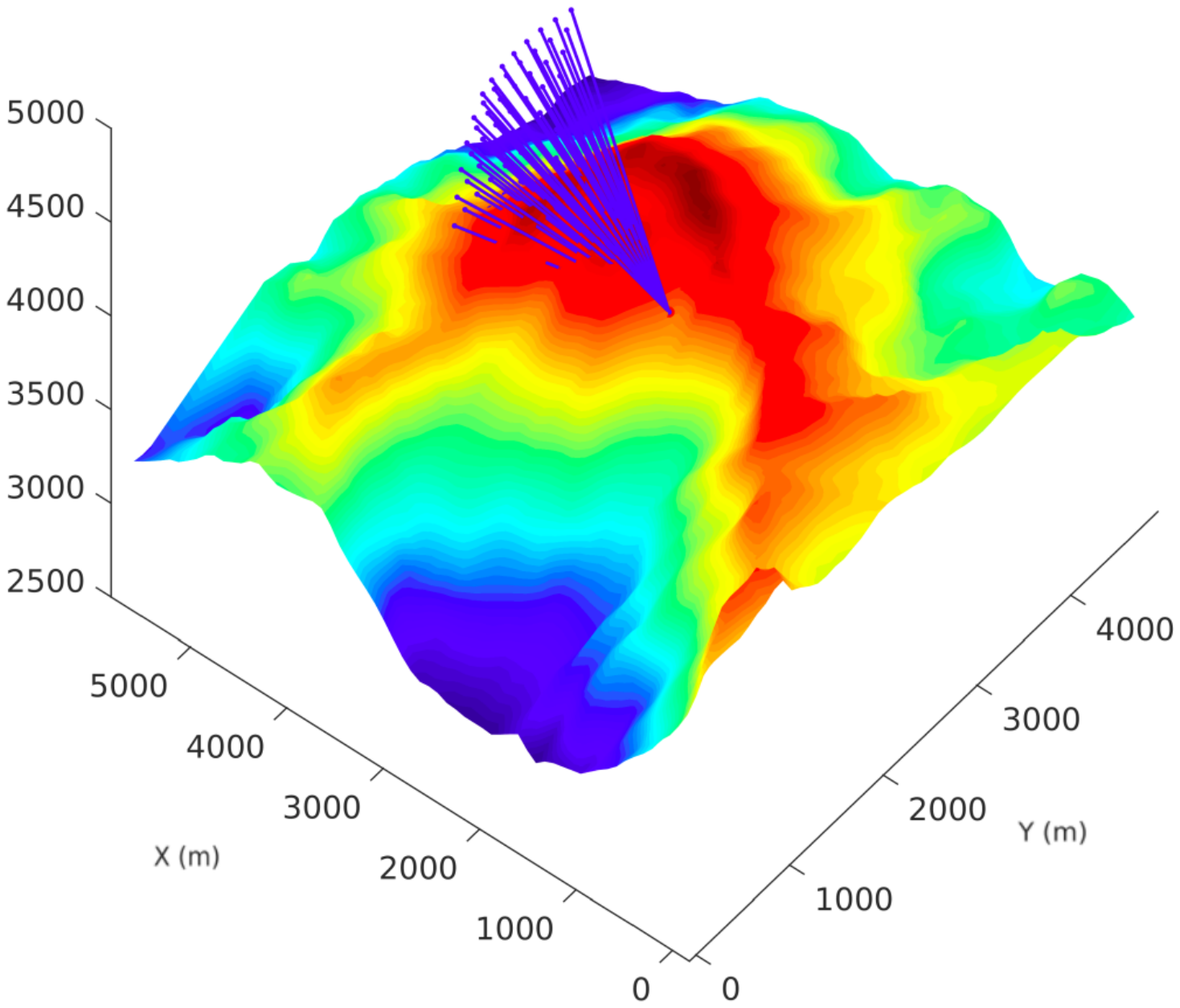}
\includegraphics[width=0.245\textwidth]{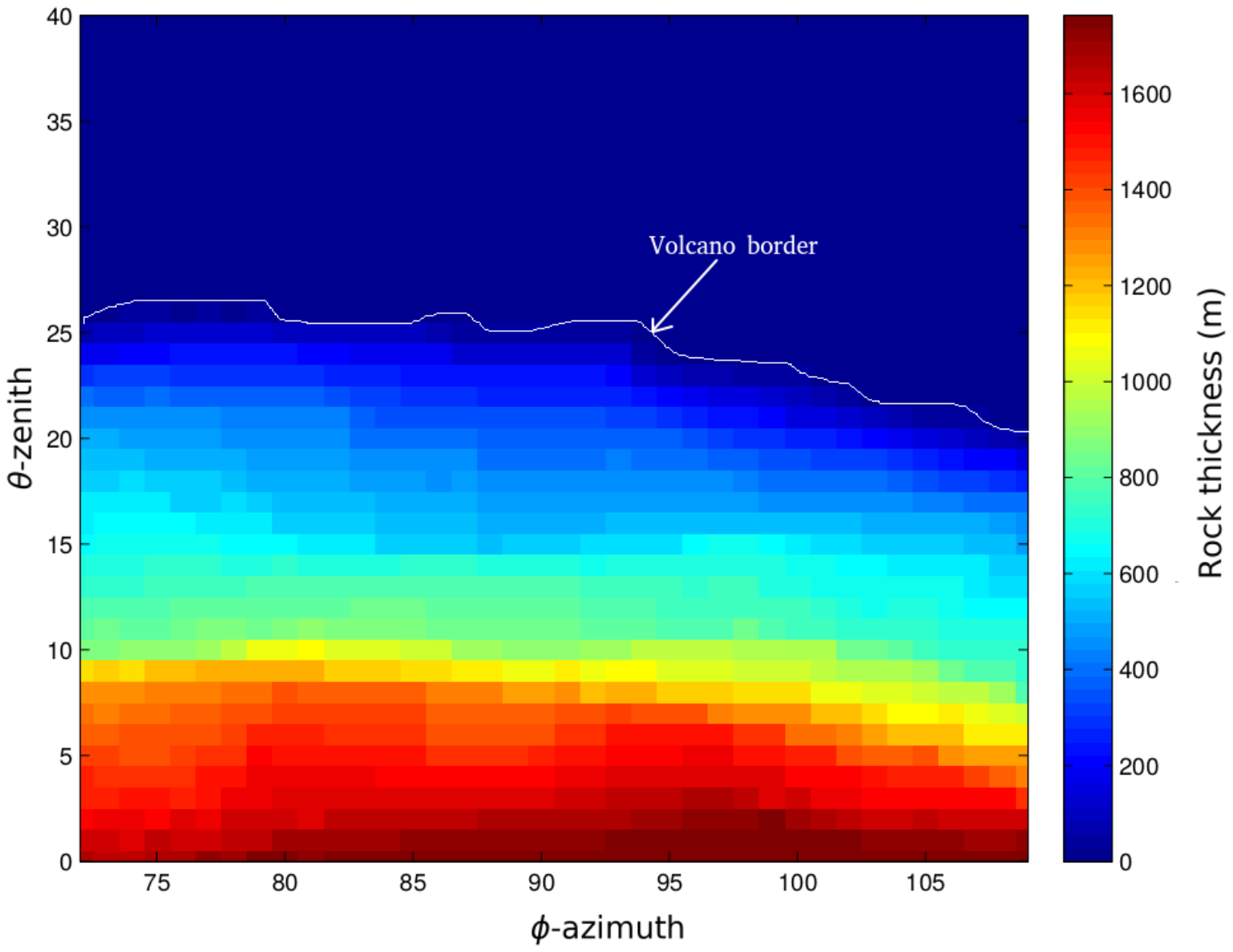}
\includegraphics[width=0.245\textwidth]{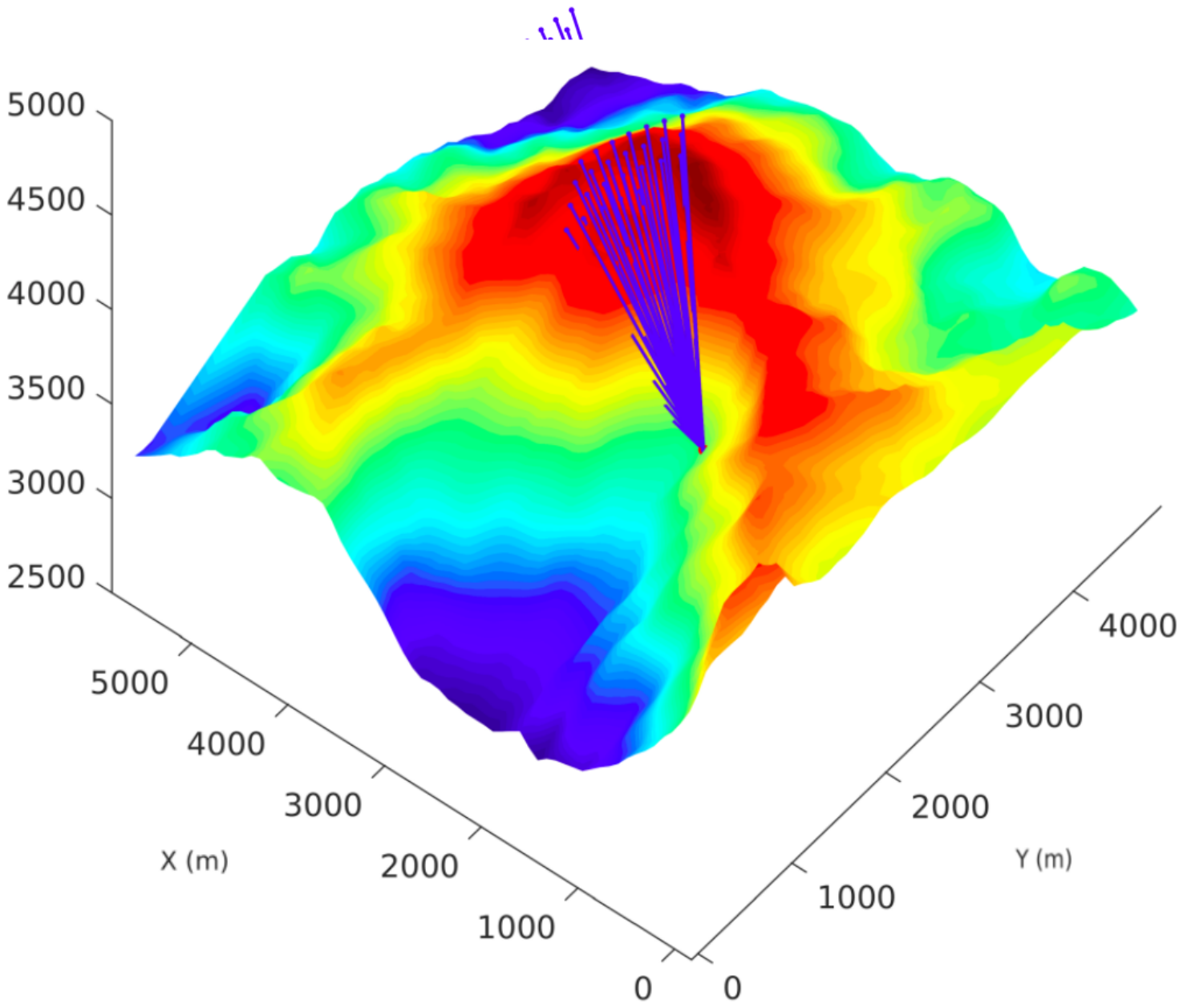}
\includegraphics[width=0.245\textwidth]{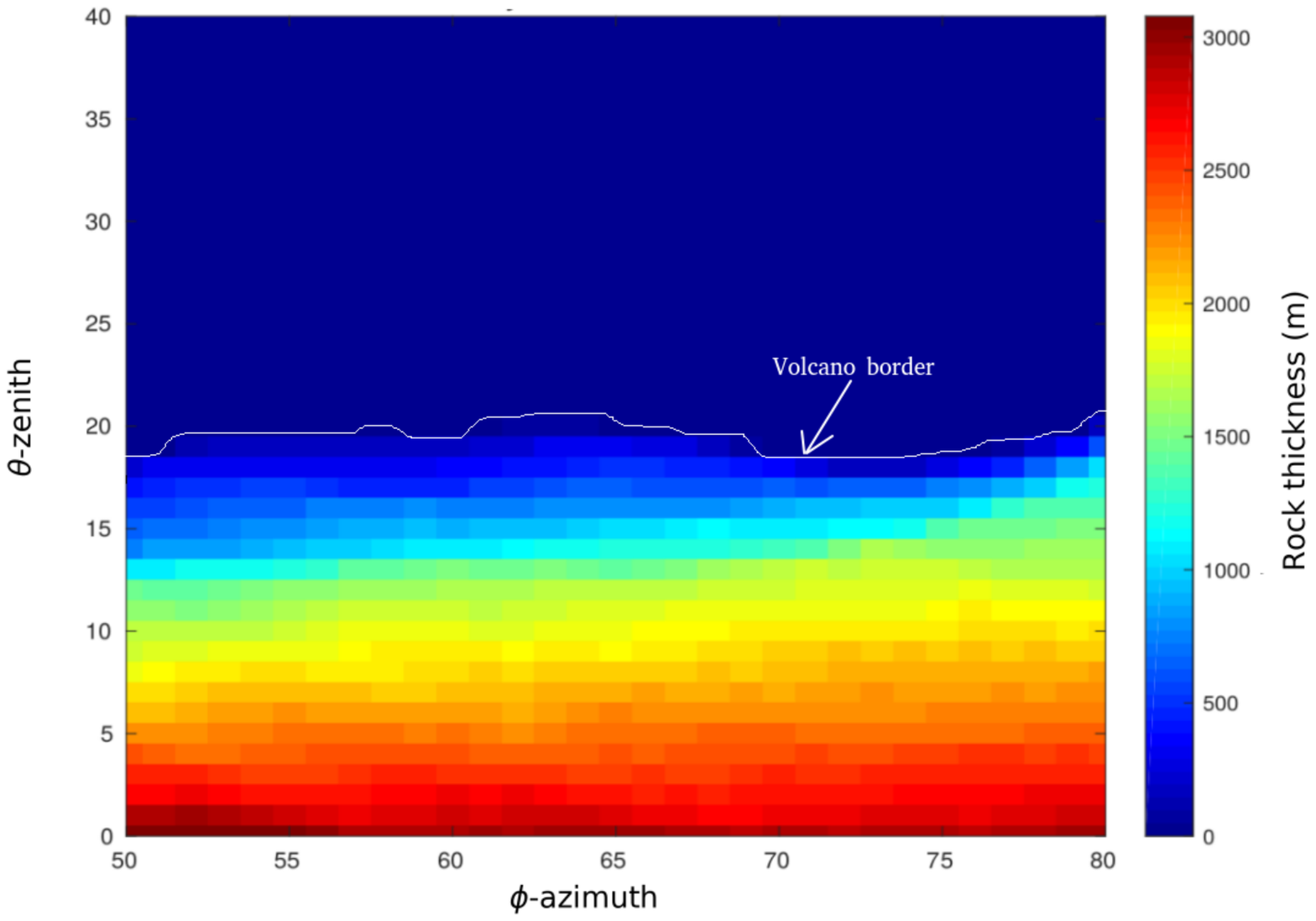}
	\caption{Particle trajectories crossing Galeras volcanic structure to the observation point {\bf P$_{1G}$}, {\bf P$_{2G}$}, {\bf P$_{3G}$} and {\bf P$_{4G}$}. }
  \label{DistanceTrajectoriesGaleras}
\end{figure}

We have proceeded to identify where safe observation points might be to install our MuTe, and how long would it takes to obtain a valuable density distribution. Following this procedure, we have identified four points around Galera Volcano  (see Table \ref{TableGaleras} and figure \ref{galerasgeo}). However, all are in the high volcanic risk area, and so located that the detected muons through the amphitheater and crater overlap, making difficult the analysis (see figure \ref{DistanceTrajectoriesGaleras}.

\label{GaleraObservationPoints}
\begin{table}[!ht]
\centering
\begin{tabular}{lllll}
\hline
\textbf{Galeras points}        & \textbf{P$_{1G}$}& \textbf{P$_{2G}$} & \textbf{P$_{3G}$} & \textbf{P$_{4G}$} \\ \hline
\textbf{Latitude  ($^{\circ}$N)}    & 1.228158           & 1.223786         & 1.216328        & 1.209212     \\
\textbf{Longitude ($^{\circ}$W)}    & -77.365237         & -77.362778       & -77.358964       & -77.363168    \\
\textbf{Distance to edifice center (m)} & 975              & 494              & 540              & 1482       \\ 
\hline
\end{tabular}
	\caption{Feasible observation points at Galeras volcano (1$^{\circ}$13'16.02"N,$\;$ 77$^{\circ}$21'33.09"W). }
\label{TableGaleras}
\end{table}

\section{Volcano observation site determination criteria}
\label{sec:Criteria}
Based on the detailed analysis of the Mach\'{\i}n volcano, and the ray-tracing studies of the Cerro Negro-Chiles complex and Galeras, we have devised a ``rule of thumb'' criteria to select the tentative muography observation sites of Colombian volcanoes.

These criteria for determination of the muon observation sites for the Colombian mainland volcanoes, --surrounded by other geological structures that could screen the atmospheric muon flux and having insecure access to various regions because of the internal conflict-- is qualitatively different from those made on peaceful islands with volcanoes free from the screening of other mountain systems. Thus, to determine muon observation points for active volcanoes in Colombia, we have devised a mix of technical and logistic criteria, which we call the ``rule of thumb''  criteria that should be fulfilled by the potential sites and which are listed below:  
\begin{itemize}
\item \textbf{Criterion 1: At the observational level, is the volcano base width less than $1,500$\,m?} This criterion is necessary because the energy of atmospheric horizontal muons are two orders of magnitude lower than vertical muons and, these incident particles need two more orders of magnitude to cross the volcanic structure \cite{NishiyamaEtal2016}. Therefore, most energetic horizontal muons can only cross $1,500$m of standard rock, and the horizontal line of sight from the tentative observation point should be less than this distance. This criterion can be appreciated quantitatively for trajectories/flux in the case of Mach\'in volcano (figures \ref{ParticleTrajectories} and  \ref{flux_volcano}), where only few (the most energetic and horizontal) muons can cross the structure while traveling distances greater than $1,500$m. 

\item \textbf{Criterion 2: Are there tentative observation points where the surrounding topography does not interfere with  the target?} Muons impacting the telescope should cross only the structure under study. Nearby mountains and any other geological formations neighboring the target volcano, must not contribute to the opacity. This requisite imposes a severe restriction on the tentative view points for the few observational places where a small window is present, with no mountains or other screening geological structures.

\item \textbf{Criterion 3: Are the sites accessible and secure?} Sites must be easily accessible, and the telescope should be securely transported and placed on the field. It is essential to consider: the weight and size of the assembled telescope and its parts; also the quality and accessibility of water resources in the area. Additionally, the volcano to be studied should not be cataloged in a situation of abundant activity due to the danger of volcanic products and processes associated with eruptions such as ashfall, pyroclastic materials, lahars, floods, among other risks, as well as earthquakes and landslides that may cause severe damage to instrument and personnel. Last but not least the --vanishing-- internal conflict persisting in several regions of the Colombia countryside and safe access should be taken into account. 
\end{itemize}

We devised these criteria based on the detailed study of Mach\'{\i}n volcano and the ray-tracing analysis for 13 Colombian volcanoes. The results of the application of the above criteria are summarized in table \ref{TableCriteria} and lead to the conclusion that the only Colombian volcano that could be studied through muography is Cerro Mach\'{\i}n \cite{AsoreyEtal2017}. We hope that shortly with the end of the internal conflict, Cerro Negro and Chiles will be safely accessible because they are good candidates to be studied with muography, but today they are not yet freely reachable. 

\begin{table}[ht!]
  \centering 
\begin{tabular}{|l|c|c|c|} \hline
 Volcano                & \textbf{Criterion 1:} & \textbf{Criterion 2:} & \textbf{Criterion 3:} \\ \hline 
{Azufral}               & N  					& Y 				    & N				        \\
{Cerro Negro$^{*}$}     & Y						& Y     			    & N 				    \\
{Chiles$^{*}$}  	    & Y						& Y    			        & N  				    \\
{Cumbal}  			    & N 					& Y  				    & N				        \\
{Dona Juana}		    & N 					& Y    			        & N				        \\ 
{Galeras}  		        & Y 					& N     			    & Y 				    \\
\textbf{Mach\'{\i}n}  	    & {\textbf{Y}} 			& {\textbf{Y}}		    & {\textbf{Y}}		    \\
{Nevado del Huila}      & N						& Y    			        & N				        \\
{Nevado del Ru\'{\i}z} 	& N 					& Y    			        & Y				        \\
{Nevado Santa Isabel}   & N 					& Y    			        & Y				        \\
{Nevado del Tolima}     & N 					& N    			        & Y				        \\
{Purac\'e}  		    & N 					& Y 				    & Y				        \\
{Sotar\'a}  			& N 					& Y  				    & N				        \\
\hline
\end{tabular}
  \caption{Which Colombian volcano can be studied by muography? The criteria discussed in section \ref{sec:Criteria} has been applied to 13 Colombian volcanoes. The three questions can be summarized as: \textbf{1-} Is the volcano base less than  $1,500$\,m?; \textbf{2-} The surrounding topography does not interfere with the observation point? and, \textbf{3-} Are the sites accessible and secure? Our studies suggest that only Cerro Mach\'{\i}n obtains three definite answers. We hope that shortly with the end of the internal conflict Cerro Negro and Chiles will be safely accessible.}
  \label{TableCriteria}
\end{table}

\section{Discussion and Conclusions}
\label{sec:disc}
In this work, we present the first comprehensive simulation of muon flux for the Mach\'{\i}n volcano dome. We have also carried out a ray-tracing analysis for 12 other inland volcanoes in Colombia surrounded by complex topographic environments. After a detailed study from the topography, we have identified the best volcano to be studied, spotted the finest points to place a muon telescope and estimated its time exposures for a significant statistics of muon flux. 

The rationale of our new approach stems from a four-step methodology:
\begin{enumerate}
\item \textbf{A ``rule of thumb'' criteria.} We have devised a mix of technical and logistic criteria --the ``rule of thumb'' criteria-- and applied them to 13 Colombian volcanoes. We have found that only Cerro Mach\'{\i}n, located at the Cordillera Central (4$^{\circ}$29'N 75$^{\circ}$22'W), can be feasibly studied today through muography. Cerro Negro and Chiles could be good candidates shortly.

\item \textbf{A unabridged simulation of open sky particle spectrum and composition.} {\sc Corsika} has calculated the energy spectrum and composition of open sky secondaries at Cerro Mach\'{\i}n and filtered them within {\sc{Magnetocosmics}} framework, providing a detailed description of different types of particles in the MeVs to TeVs secondary energy range\cite{AsoreyEtal2015B,SuarezENG2015}.

\item \textbf{A detailed calculation of the emerging muon flux.} With the above open sky particle flux and precise topographical information surrounding the volcano, we have simulated the muon propagation inside the geological edifice and estimated the emerging muon flux at four different points around Cerro Mach\'{\i}n. This was carried out integrating the energy loss equation \ref{lostenergy}, with the coefficients $a(E)$ (ionization) and $b(E)$ (radiative losses) from reference \cite{OliveEtal2014}.

\item \textbf{An estimation of the telescope time exposure.} With the emerging muon flux, we have calculated the time exposures of the instrument for an arbitrary statistics of $100$\, events at each pixel and, different values of the telescope acceptance (see Figure \ref{times}).  
\end{enumerate}

As we have mentioned before, most of the previous muography studied volcanoes --Mount Asama \cite{TanakaEtal2007} in Japan; Puy de D\^ome \cite{NoliEtal2017} in France; Mount Etna \cite{CarboneEtal2014} in Italy; La Soufuriere\cite{LesparreEtal2012,LesparreEtal2012B} in Guadalupe, to mention the most relevant studies-- are topographically isolated with a relatively good and accessible observation points. None are surrounded by geological structures screening the scarce high energy horizontal muons. However, Colombia --and surely all other Andean volcanoes-- is very different; most of the active volcanoes are along the Cordillera Central, surrounded by higher altitudes shielding cosmic ray flux. Therefore, we developed a methodology to identify the most feasible candidates, and only Cerro Mach\'{\i}n emerged as a possibility.    

Instead of using phenomenological and pseudo-empirical formulas to estimate the background flux at the volcano site (see \cite{TanakaEtal2007} and references therein), we proceeded to simulate its spectrum and composition, at each particular geographical location, with two standard astroparticle tools: {\sc Corsika} and {\sc Magnetocosmics}. We found that incident muons range from $0.1$ GeV/c to $10$ TeV, and the flux of high energy muons is very feeble:$\approx$ 10 muons per square meter per day at zenith angles $\theta \approx 82^\circ-84^\circ$.  

With the above simulation as an input, and including precise topographical information, we calculate the propagation of muons through the geological edifice and determine the emerging muon flux that could be detected at several particular observation points around the volcano (see figure \ref{ParticleTrajectories}). 

We have simulated more than $50$ days of muon flux to estimate the minimum time needed to obtain a flux of $100$ muons per pixel and found out that we require at least three months of data acquisition to explore the Mach\'{\i}n dome (depending on the observation point) at a depth of $\approx 110$m. 

Then, to discriminate density variations of $10$\%, we evaluated time exposures for our hybrid instrument as function of the acceptance. With these preliminary simulation results and by considering the standard configuration of our telescope, we have estimated time intervals between $100$ to $125$ days for the upper end ($114-150$\,m) of the volcano edifice. These results were recently reconfirmed using more precise muon underground propagation codes \cite{Kudryavtsev2009,MossEtal2018}. 

Muography can not image deep volcano structures, but it seems to be useful in determining shallow phenomena with an excellent spatial resolution. This technique can not give direct information on when a volcano will erupt, but it could provide significant insights about possible eruption processes, in the upper end of the edifice. This emerging technique requires significant progress in data analysis, treatment and interpretation of the experimental data obtained. For a bright future, it depends on the synergy between two active international communities: particle physicists and geophysicists \cite{Tanaka2016}.
\section*{Acknowledgments}
The authors acknowledge the financial support of  Departamento Administrativo de Ciencia, Tecnolog\'{\i}a e Innovaci\'on of Colombia (ColCiencias) under contract FP44842-082-2015 and to the Programa de Cooperaci\'on Nivel II (PCB-II) MINCYT-CONICET-COLCIENCIAS 2015, under project CO/15/02.  We are also very thankful to LAGO and to the Pierre Auger Collaboration for their continuous support.  The simulations in this work were partially possible thanks to The Red Iberoamericana de Computaci\'on de Altas Prestaciones (RICAP, 517RT0529), co-funded by the Programa Iberoamericano de Ciencia y Tecnolog\'{\i}a para el Desarrollo (CYTED) under its Thematic Networks Call. We also thank the computational support from the Universidad Industrial de Santander (SC3UIS) High Performance and Scientific Computing Centre. Finally, we would like to thank Vicerrector\'{\i}a Investigaci\'on y Extensi\'on Universidad Industrial de Santander for its permanent sponsorship. One of us, DSP, wants to thank the Escuela de F\'{\i}sica, the Grupo de Investigaci\'on en Relatividad y Gravitaci\'on, Grupo Halley de Astronom\'{\i}a y Ciencias Aeroespaciales and Vicerrector\'{\i}a Investigaci\'on y Extensi\'on of Universidad Industrial de Santander for the hospitality during his post-doctoral fellowship.

Our simulation codes can be found at \texttt{https://github.com/AstroparticulasBucaramanga} and other pertinent data can be also obtained from \texttt{https://zenodo.org/record/807741\#.WUMdlMaZPex }


\end{document}